\documentclass[%
reprint,
superscriptaddress,
nofootinbib,
showpacs,
citeautoscript,
 amsmath,amssymb,
 aps,
 prb
]{revtex4-2}

\usepackage[compat=1.1.0]{tikz-feynhand}
\usepackage{amsmath,amssymb,bm,braket,url,dsfont}
\usepackage{mathtools}
\usepackage{graphicx}
\usepackage{xcolor}
\usepackage{longtable}
\usepackage{multirow}
\usepackage{array}
\usepackage{booktabs}
\usepackage{physics}
\usepackage{here}
\usepackage[version=3]{mhchem}
\usepackage{siunitx}

\usepackage[
draft=false,
bookmarks=true,
bookmarksnumbered=false,
bookmarksopen=false,
bookmarksopenlevel=4,
colorlinks=true,
anchorcolor=black,
citecolor=blue,
filecolor=magenta,
linkcolor=blue,
linkbordercolor={1 0 0},
urlcolor=blue,
pdfborder={0 0 1},
pdftitle={},
pdfauthor={},
pdfsubject={},
pdfkeywords={},
pdfpagemode=UseThumbs,
]{hyperref}
\DeclareUnicodeCharacter{3000}{\hspace{1em}}
\DeclareUnicodeCharacter{2212}{\ensuremath{-}}

\newcommand{\expecval}[1]{\left \langle {#1} \right \rangle}
\newcommand{\secref}[1]{Sec.~\ref{#1}}
\newcommand{\figref}[1]{FIG.~\ref{#1}}
\newcommand{\tabref}[1]{TABLE~\ref{#1}}
\newcommand{\appref}[1]{Appendix~\ref{#1}}
\newcommand{\Eqref}[1]{Eq.~\eqref{#1}}

\begin{document}
\title{
        Nonlinear Hall effect driven by spin-charge-coupled motive force
}
\author{Kohei Hattori}
\email{hattori-kohei053@g.ecc.u-tokyo.ac.jp}
\affiliation{Department of Applied Physics, The University of Tokyo, {Bunkyo}, Tokyo 113-8656, Japan}

\author{Hikaru Watanabe} 
\email{hikaru-watanabe@g.ecc.u-tokyo.ac.jp}
\affiliation{Department of Physics, University of Tokyo, {Hongo}, Tokyo 113-0033, Japan}

\author{Ryotaro Arita} 
\affiliation{Department of Physics, University of Tokyo, {Hongo}, Tokyo 113-0033, Japan}
\affiliation{Center for Emergent Matter Science, RIKEN, {Wako}, Saitama 351-0198, Japan}

\begin{abstract}
Parity-time-reversal symmetric ($\mathcal{PT}$-symmetric) magnets have garnered much attention due to their spin-charge coupled dynamics enriched by the parity-symmetry breaking.
By real-time simulations, we study how localized spin dynamics can affect the nonlinear Hall effect in $\mathcal{PT}$-symmetric magnets.
To identify the leading-order term, we derive analytical expressions for the second-order optical response and classify the contributions by considering their transformation properties under $\mathcal{PT}$ symmetry.
Notably, our results reveal that the sizable contribution is attributed to the mixed dipole effect, which is analogous to the Berry curvature dipole term.
\end{abstract}

\maketitle
\section{introduction}\label{introduction}
The nonlinear Hall effect (NHE) is a nonlinear transverse current response induced by an electric field \cite{Du2021}.
Several mechanisms underlying the NHE have been proposed, exhibiting distinct behaviors depending on system parameters, particularly the relaxation time.
In time-reversal symmetric ($\mathcal{T}$-symmetric) systems, the Berry curvature dipole term is well known to play a crucial role, which scales linearly with the relaxation time \cite{Moore2010, Sodemann2015}. 
The Berry curvature dipole reflects the intrinsic properties of the electronic band structure. 
In fact, the topology of the electronic structure of nonmagnetic systems has been explored by measuring the nonlinear Hall conductivity experimentally \cite{Xu2018, Ma2019,He2021,Sinha2022}.
In addition to the Berry curvature dipole term, impurity effects also play a significant role in the NHE \cite{Kang2019,Du2019, Isobe2020}. 
On the other hand, in $\mathcal{T}$-breaking systems, the NHE can arise from the Drude term, which scales quadratically with the relaxation time \cite{Ideue2017, Watanabe2020, Holder2020, Wang2021,Ma2023}.
Another contribution is the positional shift effect, which is independent of the relaxation time and can dominate in dirty systems where the electron relaxation time is short \cite{Yao2014,Watanabe2020,Liu2021, Wang2021,MichishitaN2022, Kamal2023, Kaplan2024}. 
The NHE associated with this term has also been experimentally observed \cite{Wang2023, Gao2023}.
In these studies, the mechanisms of the NHE are primarily limited to the effects of an electric field on the electronic system.

In recent years, the nonlinear response generated by the dynamics of order parameters has garnered significant attention. 
The photocurrents arising from collective excitations have been theoretically investigated through perturbative analysis \cite{Morimoto2016_exciton, Morimoto2019_electromagnon, Morimoto2021, Morimoto2024}. 
Several studies have also examined the optical responses originating from the coupled dynamics of the charge and order parameter by using real-time simulations for excitonic insulators \cite{Kaneko2021, Chan2021} and magnetic insulators \cite{Iguchi2024, hattori_AIAO}.
Moreover, photocurrent responses driven by order parameter dynamics have been experimentally observed in ferroelectric materials \cite{Okumura2021}, excitonic insulators \cite{Sotome2021, Nakamura2024}, and magnetic insulators \cite{Ogino2024}.
The dynamics of order parameters enriches the nonlinear optical response, offering new insights into the mechanism underlying the NHE.

In this paper, we study the effect of localized spin dynamics on the NHE in $\mathcal{PT}$-symmetric collinear antiferromagnetic metals. This phenomenon is particularly intriguing due to recent developments in antiferromagnetic spintronics \cite{Jungwirth_review2016, Baltz2018, Manchon2019}. In parity-breaking antiferromagnets, the N\'eel vector can be manipulated by an electric field or current via the Edelstein effect \cite{Levitov1985, Edelstein1990, Yanase2014,Zelezny2014}.
Experimentally, electrical control of antiferromagnetic domains has been demonstrated in $\mathcal{PT}$-symmetric collinear antiferromagnetic metals such as CuMnAs \cite{Wadley2016,Godinho2018} and $\text{Mn}_2\text{Au}$ \cite{Bodnar2018}. 
Since the Edelstein effect can arise from the Fermi surface effect, spin-charge coupled dynamics can be induced by an external electric field in these magnets, particularly at low frequencies.
Given these considerations, collinear antiferromagnetic metals represent a promising platform for uncovering new mechanisms underlying the NHE driven by localized spin dynamics.

To clarify the role of the Edelstein effect in the NHE, we adopt a simple model representing a two-sublattice collinear antiferromagnet. 
We explore the impact of the localized spin dynamics on the NHE by calculating the time evolution of the charge and spin degrees of freedom simultaneously.
We employ the real-time simulation scheme for spin-charge coupled systems developed in the previous works \cite{Ono2021, Ono2023, Iguchi2024, hattori_AIAO}. 
This approach allows us to examine the effects of localized spin dynamics on current responses, providing a clear understanding of the NHE from the perspective of spin dynamics.

Using symmetry analysis, we identify an optically active mode in the localized spin system and investigate its role in both the linear response function and the NHE. In the NHE spectrum, we observe an enhancement of the peak in the low-frequency regime and the emergence of a new resonant peak associated with collective spin excitations.
To further elucidate the mechanism of the NHE, we analytically decompose the NHE into several components.
Through symmetry analysis and this decomposition, we identify a significant contribution, which we call the mixed dipole term analogous to the Berry curvature dipole term.

\begin{figure}
        \centering
        \includegraphics[width=\linewidth]{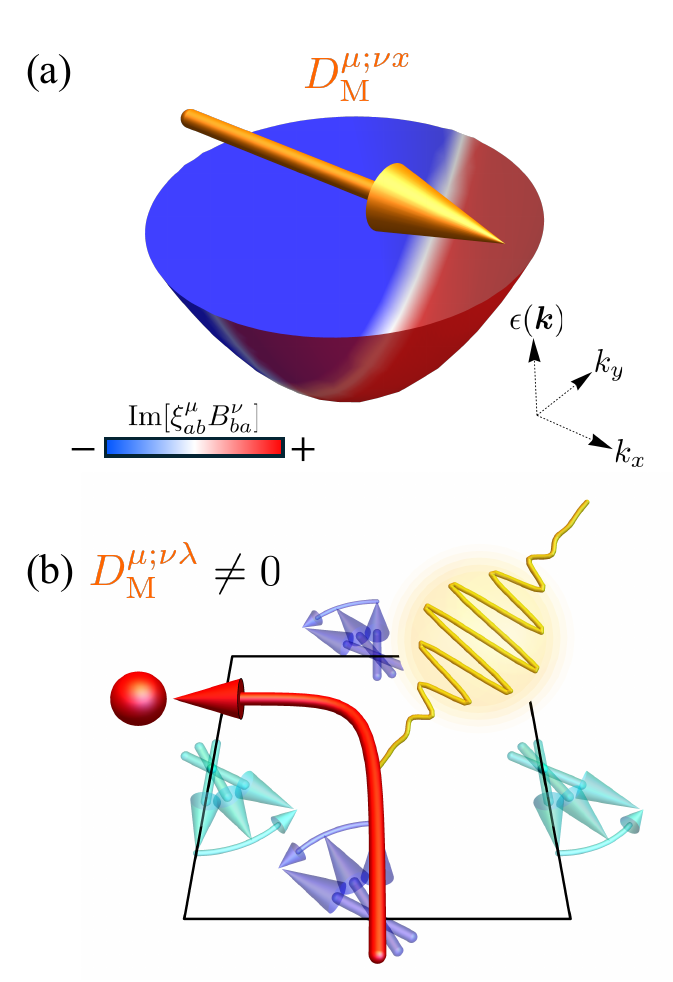}
        \caption{(a) Schematic representation of the band structure $\epsilon(\bm{k})$ colored by the distribution of $\operatorname{Im}[\xi_{ab}^{\mu}B_{ba}^{\nu}]$. The dipolar distribution of $\operatorname{Im}[\xi_{ab}^{\mu}B_{ba}^{\nu}]$ along the $k_x$ direction indicates the mixed dipole $D_{\text{M}}^{\mu;\nu x}$. (b) Illustration of the nonlinear Hall effect induced by localized spin dynamics. The blue and cyan arrows represent localized spins perturbed by the irradiating light (yellow curve), while the red arrow indicates the electric current $J$. When $D^{\mu;\nu\lambda}_{\text{M}} \neq 0$, the nonlinear Hall effect emerges from localized spin dynamics through the mixed dipole $D^{\mu;\nu\lambda}_{\text{M}}$.}
        \label{nonlinear_Hall}
\end{figure}
The Berry curvature dipole term is described as 
\begin{align}
&\sigma_{\mathrm{BCD;L}}^{\mu;\nu\lambda}=\frac{1/\tau}{\omega^2+1/\tau^2}\nonumber\\
    &\times\int\frac{d\bm{k}}{(2\pi)^d}\sum_{a\neq b}(\partial_{\lambda}\operatorname{Im}[\xi_{ab}^{\mu}\xi_{ba}^{\nu}]+\partial_{\nu}\operatorname{Im}[\xi_{ab}^{\mu}\xi_{ba}^{\lambda}])f_{a},\\
    &\sigma_{\mathrm{BCD;C}}^{\mu;\nu\lambda}=\frac{i\omega}{\omega^2+1/\tau^2}\nonumber\\
    &\times\int\frac{d\bm{k}}{(2\pi)^d}\sum_{a\neq b}(\partial_{\lambda}\operatorname{Im}[\xi_{ab}^{\mu}\xi_{ba}^{\nu}]-\partial_{\nu}\operatorname{Im}[\xi_{ab}^{\mu}\xi_{ba}^{\lambda}])f_{a},
\end{align}
where $\omega$ denotes the frequency of the external light field, $\xi_{ab}$ is the Berry connection, $\partial_{\nu}$ represents the partial derivative with respect to the $\nu$-axis in $\bm{k}$-space, and $f_a$ denotes the Fermi-Dirac distribution for the eigenenergy $\epsilon_{\bm{k}a}$ labeled by $\bm{k}$ and band index $a$.
Here, $\sigma_{\mathrm{BCD;L}}^{\mu;\nu\lambda}$ ($\sigma_{\mathrm{BCD;C}}^{\mu;\nu\lambda}$) is induced by linearly (circularly) polarized light.
This term is characterized by the Berry curvature dipole expressed as
\begin{align}
    D_{\text{BCD}}^{\mu;\nu\lambda}=\int\frac{d\bm{k}}{(2\pi)^d}\sum_{a}\partial_{\lambda}\Omega_{a}^{\mu\nu}f_a,
\end{align}
where $\Omega_{a}^{\mu\nu}=-2\sum_{b\neq a}\operatorname{Im}[\xi_{ab}^{\mu}\xi_{ba}^{\nu}]$ is the Berry curvature for the band $a$.
The Berry curvature dipole represents the dipole moment of the Berry curvature in momentum space and determines the amplitude of the NHE. 

On the other hand, the mixed dipole term is expressed as
\begin{align}
    \sigma_{\text{MD}}^{\mu;\nu\lambda}=\frac{J(1/\tau+i\omega)}{\omega^2+1/\tau^2}\int\frac{d\bm{k}}{(2\pi)^d}\sum_{a\neq b}\partial_{\lambda}\operatorname{Im}[\xi_{ab}^{\mu}B_{ba}^{\nu}]f_{a}.
\end{align}
Here, $J$ represents the exchange coupling between the spin moment of the itinerant electrons and the localized spin moment, and $B_{ab}$ is the spin operator.
This term is obtained by replacing the Berry connection $\xi_{ba}^{\nu}$ with the spin operator $B_{ba}^{\nu}$ in the Berry curvature dipole term.
This term is proportional to the dipole moment of the imaginary component of the product of the Berry connection and the spin operator over the occupied states as 
\begin{align}\label{Mixed_dipole}
    D^{\mu;\nu\lambda}_{\text{M}}=\int\frac{d\bm{k}}{(2\pi)^d}\sum_{a\neq b}\partial_{\lambda}\operatorname{Im}[\xi_{ab}^{\mu}B_{ba}^{\nu}]f_{a}.
\end{align}
We define this physical quantity as the mixed dipole.
The mechanism of the NHE driven by the mixed dipole closely resembles that of the NHE induced by the Berry curvature dipole.
A schematic illustration of the NHE originating from the mixed dipole is presented in \figref{nonlinear_Hall}.
The electron trajectory is bent by the localized spin dynamics through the mixed dipole.
This mixed dipole mechanism is important in the $\mathcal{PT}$-symmetric AFM metals, where the localized spin dynamics can be induced by an electric field via the Edelstein effect.
These findings highlight the importance of localized spin dynamics in understanding and engineering nonlinear current responses in complex magnetic systems.

The outline of this paper is as follows. In \secref{method}, we explain the details of the model and the computational scheme for the spin-charge coupled system. Section \ref{symmetry_analysis} discusses the symmetry classification of the localized spin dynamics coupled to the electric field. We elucidate the linear electromagnetic susceptibility of the localized spin systems in \secref{linear_response_functions}. 
\secref{photocurrent_spectra} shows the results of the nonlinear Hall conductivity obtained by real-time simulation.
We present the analytical formulas for the photocurrent conductivity under the external light field and localized spin dynamics in \secref{formulas_for_pc}.
\secref{decomposition_of_photocurrent}, and \secref{spin-motive} discuss the effects of localized spin dynamics on the NHE. Specifically, in \secref{spin-motive}, we decompose the NHE into several components and identify the dominant contribution: the mixed dipole term arising from the interference between the electric field and localized spin dynamics.
Finally, we summarize this work in \secref{summary}.

\section{method}\label{method}
\subsection{Model}
\begin{figure}[t]
        \centering
        \includegraphics[width=\linewidth]{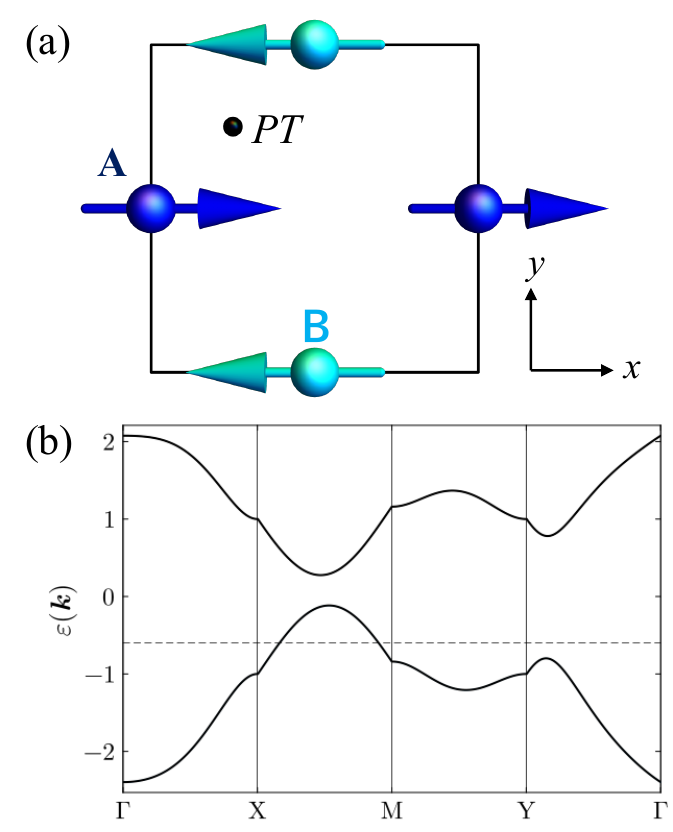}
        \caption{(a) Square lattice model with collinear antiferromagnetic order. The $\mathcal{PT}$ center is marked by the black dot. (b) Band dispersion of the system with the parameters $t_{1} = 1.0, t_2=0.08, \lambda = 0.8, J = 1.0, K_{x} = 0.05$ along the high-symmetry path $\Gamma$-X-M-Y-$\Gamma$. We set the chemical potential $\mu=-0.6$ at the horizontal dashed line.}
        \label{model_pic}
\end{figure}
This study focuses on a $\mathcal{PT}$-symmetric system in two dimensions, where conduction electrons are coupled to localized spin moments arranged in a collinear antiferromagnetic structure (see \figref{model_pic}). 
This model is minimal to consider the $\mathcal{PT}$-symmetric collinear antiferromagnet, as introduced in \cite{Smejkal2017}.

The Hamiltonian of the model reads
\begin{align}
        \hat{\mathcal{H}} = \hat{\mathcal{H}}_{\mathrm{ele}} + \hat{\mathcal{H}}_{\mathrm{exc}} + \mathcal{H}_{\mathrm{spin}} + \hat{\mathcal{H}}_{E}.
\end{align}
The first term,
\begin{align}
        \hat{\mathcal{H}}_{\mathrm{ele}} = \sum_{\bm{k}}\bm{\hat{c}^{\dag}}(\bm{k})\bm{\mathcal{H}}_{\mathrm{ele}}(\bm{k})\bm{\hat{c}}(\bm{k}),
\end{align}
is the Hamiltonian of the electronic system, where $\bm{\hat{c}_{k}}=(\hat{c}_{A\uparrow}(\bm{k}),\hat{c}_{A\downarrow}(\bm{k}),\hat{c}_{B\uparrow}(\bm{k}),\hat{c}_{B\downarrow}(\bm{k}))^{T}$ is a vector representation of annihilation operators.
$\hat{c}_{\alpha\sigma}^{\dagger}(\bm{k})$ ($\hat{c}_{\alpha\sigma}(\bm{k})$) is the creation (annihilation) operator of the electron on sublattice $\alpha$ ($\alpha = A, B$) having spin $\sigma$ ($\sigma=\uparrow, \downarrow$), and $\bm{k}$ is the wave vector. 
The matrix of the Hamiltonian is expressed as
\begin{align}
    \bm{\mathcal{H}}_{\mathrm{ele}}(\bm{k})=\begin{pmatrix}
        \epsilon_0(\bm{k})+\bm{g}_A(\bm{k})\cdot\bm{\sigma}&V_{AB}(\bm{k})\\
        V_{AB}(\bm{k})&\epsilon_0(\bm{k})+\bm{g}_B(\bm{k})\cdot\bm{\sigma}
    \end{pmatrix},
\end{align}
where $\bm{\sigma}$ are the Pauli matrices representing the spin degree of freedom.
The components are defined as
\begin{align}
    \epsilon_0(\bm{k})&=-t_2(\mathrm{cos}k_x+\mathrm{cos}k_y),\\
    V_{AB}(\bm{k})&=-2t_1\mathrm{cos}\frac{k_x}{2}\mathrm{cos}\frac{k_y}{2},\\
    \bm{g}_{A}(\bm{k})&=\lambda\begin{pmatrix}
        -\mathrm{sin}k_y&\mathrm{sin}k_x&0
    \end{pmatrix},
\end{align}
and $\bm{g}_{B}(\bm{k})=-\bm{g}_{A}(\bm{k})$.
This Hamiltonian consists of the nearest neighbor hopping $t_{1}$, the next-nearest neighbor hopping $t_{2}$, and the sublattice-dependent antisymmetric spin-orbit coupling $\lambda$ (sASOC). 
sASOC is essential for enhancing magnetoelectric coupling \cite{Yanase2014, Zelezny2014, Hayami2014}.
The second term
\begin{align}
        \hat{\mathcal{H}}_{\mathrm{exc}} = -J\sum_{\bm{k}}\sum_{\alpha}\sum_{\sigma\sigma^{\prime}}\qty(\vb*{\sigma}\cdot\vb{S}_{\alpha}(t))^{\sigma\sigma^{\prime}}\hat{c}_{\alpha\sigma}^{\dagger}(\bm{k})\hat{c}_{\alpha\sigma^{\prime}}(\bm{k})
        \label{Hund_Hamiltonian}
\end{align}
corresponds to the interaction between the electronic system and the localized spin system. 
Here, we assume that the localized spins $\vb{S}_{\alpha}$ are classical spins with a fixed magnitude, $\abs{\vb{S}_{\alpha}} = 1$. 
In \Eqref{Hund_Hamiltonian}, we consider a uniform spin dynamics across the unit cells.
This assumption is consistent with the dipole approximation, which we will employ in the following.
In \figref{model_pic}(b), we show the band dispersion of the electronic system along the high-symmetry path $\Gamma$-X-M-Y-$\Gamma$. The energy levels of the electronic system exhibit double degeneracy due to the $\mathcal{PT}$ symmetry.
The third term 
\begin{align}
        \mathcal{H}_{\mathrm{spin}} = -\sum_{\alpha}K_{x}\qty(\mathrm{S}_{\alpha}^{x})^{2}
\end{align}
describes the Hamiltonian for the localized spins. To stabilize the antiferromagnetic order along the $x$ axis, we consider the easy-axial anisotropy $K_{x}$. 

The last term 
\begin{align}
        \hat{\mathcal{H}}_{E} = -\bm{E}(t)\sum_{\bm{k}\bm{k}^{\prime}}\sum_{\alpha}\sum_{\sigma}\qty[i\pdv{\bm{k}}\delta(\bm{k}-\bm{k}^{\prime})]\hat{c}_{\bm{k}\alpha\sigma}^{\dagger}\hat{c}_{\bm{k}^{\prime}\alpha\sigma} \label{light-matter-coupling}
\end{align}
represents the light-matter coupling in the length gauge \cite{Aversa1995}, where $\bm{E}(t)$ is a time-dependent electric field. Here, we set lattice constant $a = 1$ and elementary charge $e=1$.
In \Eqref{light-matter-coupling}, we assume that the Wannier state of a conducting electron is well-localized at a given site, neglecting the light-matter coupling arising from its spatial dispersion. This light-matter coupling corresponds to the celebrated Peierls substitution in the dipolar gauge \cite{Murakami2022, Ono2024}.
However, in numerical calculations, the relaxation term in the velocity gauge violates the charge conservation law in a metallic system at zero temperature. 
To avoid this issue, we employ the length gauge in our calculation rather than the velocity gauge.
The challenge associated with handling the $k$-derivative of the delta function in \Eqref{light-matter-coupling} is addressed in the following subsection.

\subsection{Calculation scheme}
Following Ref.\cite{Iguchi2024}, here we introduce the calculation scheme of real-time simulation. To investigate the spin-charge coupled system, we solve the von Neumann equation and Landau-Lifshitz-Gilbert (LLG) equation simultaneously.
Firstly, the time evolution of the itinerant electrons can be described by the single-particle density matrix (SPDM) $\rho_{\alpha \beta}^{\sigma \sigma^{\prime}}(\bm{k}) = \expecval{\hat{c}_{\beta\sigma^{\prime}}^{\dagger}(\bm{k})\hat{c}_{\alpha\sigma}(\bm{k})}$. 
SPDM satisfies the von Neumann equation \cite{Yue2022} expressed as,
\begin{align}
        \begin{split}
            \pdv{\vb*{\rho}(\bm{k}, t)}{t} = -i\qty[\vb*{H}(\bm{k},t),\, \vb*{\rho}(\bm{k}, t)] - \bm{E}(t)\cdot\frac{\partial\vb*{\rho}(\bm{k}, t)}{\partial \bm{k}} \\
        - \frac{1}{\tau}(\vb*{\rho}(\bm{k}, t) - \vb*{\rho}_{\text{eq}}(\bm{k})).
        \label{vonNeumann}
        \end{split}
\end{align}
Secondly, the time evolution of the localized spin system is governed by the LLG equation described as 
\begin{align}
        \dfrac{d\boldsymbol{\mathrm{S}}_{\alpha}}{dt} &= \dfrac{1}{1 + \alpha_{G}^{2}}\left( \boldsymbol{\mathrm{h}}_{\alpha}^{\mathrm{eff}} \times \boldsymbol{\mathrm{S}}_{\alpha} + \alpha_{G} \boldsymbol{\mathrm{S}}_{\alpha} \times \left( \boldsymbol{\mathrm{S}}_{\alpha} \times\boldsymbol{\mathrm{h}}_{\alpha}^{\mathrm{eff}} \right)\right) \label{LLG}, \\
        \boldsymbol{\mathrm{h}}_{\alpha}^{\mathrm{eff}} &= -J\langle \boldsymbol{\sigma}_{\alpha} \rangle + \dfrac{\delta \mathcal{H}_{\text{spin}}}{\delta \boldsymbol{\mathrm{S}}_{\alpha}}.\label{LLG_sigma}
\end{align}
In \Eqref{vonNeumann}$, \vb*{H}(\bm{k},t)$ represents the time-dependent electronic Hamiltonian at each $\bm{k}$ point defined as follows
\begin{align}
        \hat{\mathcal{H}}_{\mathrm{ele}} + \hat{\mathcal{H}}_{\mathrm{exc}} = \sum_{\bm{k}}\sum_{\alpha\beta}\sum_{\sigma\sigma^{\prime}}\qty[\vb*{H}(\bm{k},t)]_{\alpha\beta}^{\sigma\sigma^{\prime}}\hat{c}_{\alpha\sigma}^{\dagger}(\bm{k})\hat{c}_{\beta\sigma^{\prime}}(\bm{k}). \label{time_dependent_hamiltonian}
\end{align}
The $k$-derivative of the delta function in \Eqref{light-matter-coupling} is handled as the $k$-derivative of the SPDM, making it computationally feasible.
In the LLG equation \Eqref{LLG_sigma}, $\expecval{\vb*{\sigma}_{\alpha}}$ represents the sublattice-resolved spin density of itinerant electrons, which can be calculated from SPDM. 
This method enables us to capture the dynamics of both itinerant electrons and localized spins.
We set the chemical potential $\mu=-0.6$, placing it below the bandgap, as indicated by the horizontal dashed line in \figref{model_pic}.
For the parameters $t_1 = 1.0$, $t_2 = 0.08$, $\lambda = 0.8$, $J = 1.0$, and $K_x = 0.05$, the collinear antiferromagnetic order characterized by $\bm{\mathrm{S}}_A = (1, 0, 0)$ and $\bm{\mathrm{S}}_B = (-1, 0, 0)$ is stable.

In real materials, excited carriers undergo relaxation processes due to electron-electron correlations, electron-phonon interactions, and impurity scattering. To model a physically realistic response to light, we incorporate these effects phenomenologically. Specifically, we apply the relaxation time approximation in the von Neumann equation as $1/\tau(\vb*{\rho}(\bm{k}, t) - \vb*{\rho}_{\text{eq}}(\bm{k}))$ in \Eqref{vonNeumann}, and introduce the Gilbert damping term $\alpha_{G}$ in \Eqref{LLG}. Here, $\vb*{\rho}_{\text{eq}}(\bm{k})$ denotes the SPDM at equilibrium for a temperature of $T=0$, as shown below.

The SPDM in equilibrium $\vb*{\rho}_{\text{eq}}(\bm{k})$ denotes the SPDM in the initial state at zero temperature ($T=0$). SPDM in the band basis $\tilde{\vb*{{\rho}}}_{\text{eq}}(\bm{k})$ is given by
\begin{align}
        \qty[\tilde{\vb*{{\rho}}}_{\text{eq}}(\bm{k})]_{nn^{\prime}} = \delta_{nn^{\prime}}\Theta(\mu - \epsilon_{\bm{k}n}),
\end{align}
where $\Theta(\mu - \epsilon_{\bm{k}n})$ represents the occupation number, where $\epsilon_{\bm{k}n}$ being the eigenvalue of the Hamiltonian $\vb*{H}(\bm{k})$.
By using this expression, we can calculate the SPDM in the original basis $\vb*{\rho}_{\text{eq}}(\bm{k})$ as
\begin{align}
        \vb*{\rho}_{\text{eq}}(\bm{k}) = \vb*{U}(\bm{k})\tilde{\vb*{\rho}}_{\text{eq}}(\bm{k})\vb*{U}^{\dagger}(\bm{k})
\end{align}
with $\vb*{U}$ is a unitary matrix which diagonalizes the Hamiltonian as, 
\begin{align}
        \vb*{U}^{\dagger}(\bm{k})\vb*{H}(\bm{k})\vb*{U}(\bm{k}) = \vb*{\mathcal{E}}(\bm{k}),\\
        \qty(\vb*{\mathcal{E}}(\bm{k}))_{nn^{\prime}} = \delta_{nn^{\prime}}\epsilon_{\bm{k}n}.
\end{align}

We can evaluate the current density from SPDM in each time step, as follows.
The current operator in the length gauge is described as 
\begin{align}
        \bm{\hat{J}}(t) &= \sum_{\bm{k}}\sum_{\alpha\beta}\sum_{\sigma\sigma^{\prime}}\pdv{\qty[\vb*{H}(\bm{k},t)]_{\alpha\beta}^{\sigma\sigma^{\prime}}}{\bm{k}}\hat{c}_{\alpha\sigma}^{\dagger}(\bm{k})\hat{c}_{\beta\sigma^{\prime}}(\bm{k}) \\
        &\equiv \sum_{\bm{k}}\sum_{\alpha\beta}\sum_{\sigma\sigma^{\prime}}\qty[\vb*{J}(\bm{k})]_{\alpha\beta}^{\sigma\sigma^{\prime}}\hat{c}_{\alpha\sigma}^{\dagger}(\bm{k})\hat{c}_{\beta\sigma^{\prime}}(\bm{k}).
\end{align}
It is noted that since the $\hat{\mathcal{H}}_{\text{exc}}$ term in the Hamiltonian is independent of the momentum $\bm{k}$, the current operator does not depend on the dynamics of the localized spins.

We solve the coupled equations of motion \Eqref{vonNeumann} and \Eqref{LLG} by using the fourth-order Runge-Kutta method.
In each time step, we evaluate the sublattice-resolved spin density $\langle\bm{\sigma}_{\alpha}\rangle$ using the SPDM and update the exchange Hamiltonian based on the newly determined spin configurations. Furthermore, we can evaluate the expectation value of the current operator using SPDM as
\begin{align}
        \bm{J}(t) = \sum_{\bm{k}}\operatorname{Tr}\qty[\vb*{J}(\bm{k})\vb*{\rho}(\bm{k}, t)].
\end{align}
To obtain the $\partial\vb*{\rho}(\bm{k}, t)/\partial \bm{k}$ in the von Neumann equation \Eqref{vonNeumann}, we use symmetric derivative
\begin{align}
        \pdv{\vb*{\rho}(\bm{k}, t)}{\bm{k}} = \dfrac{\vb*{\rho}(\bm{k}+d\bm{k}, t) - \vb*{\rho}(\bm{k}-d\bm{k}, t)}{2|d\bm{k}|},
\end{align}
where $|d\bm{k}| = 2\pi/N$, with $N = 1000$.

Based on this calculation scheme, we evaluate the nonlinear Hall conductivity from the calculated current response, which will be discussed in the next section. In the following calculations, we use the parameters $\alpha_{G} = 0.05$ and $\tau = 25.0$, otherwise explicitly mentioned.

\section{Real-time simulation}\label{result}
We present the result of the spin-charge coupled dynamics of the system under the external light field. Starting with the symmetry analysis, we identify the optically active collective modes in \ref{symmetry_analysis}. Secondly, in \ref{linear_response_functions}, we investigate the linear electromagnetic susceptibility of the localized spin system.
In \ref{photocurrent_spectra}, we focus on the nonlinear Hall conductivity arising from the spin-charge-coupled dynamics.

\subsection{Symmetry analysis}\label{symmetry_analysis}
\begin{figure}
        \centering
        \includegraphics[width=\linewidth]{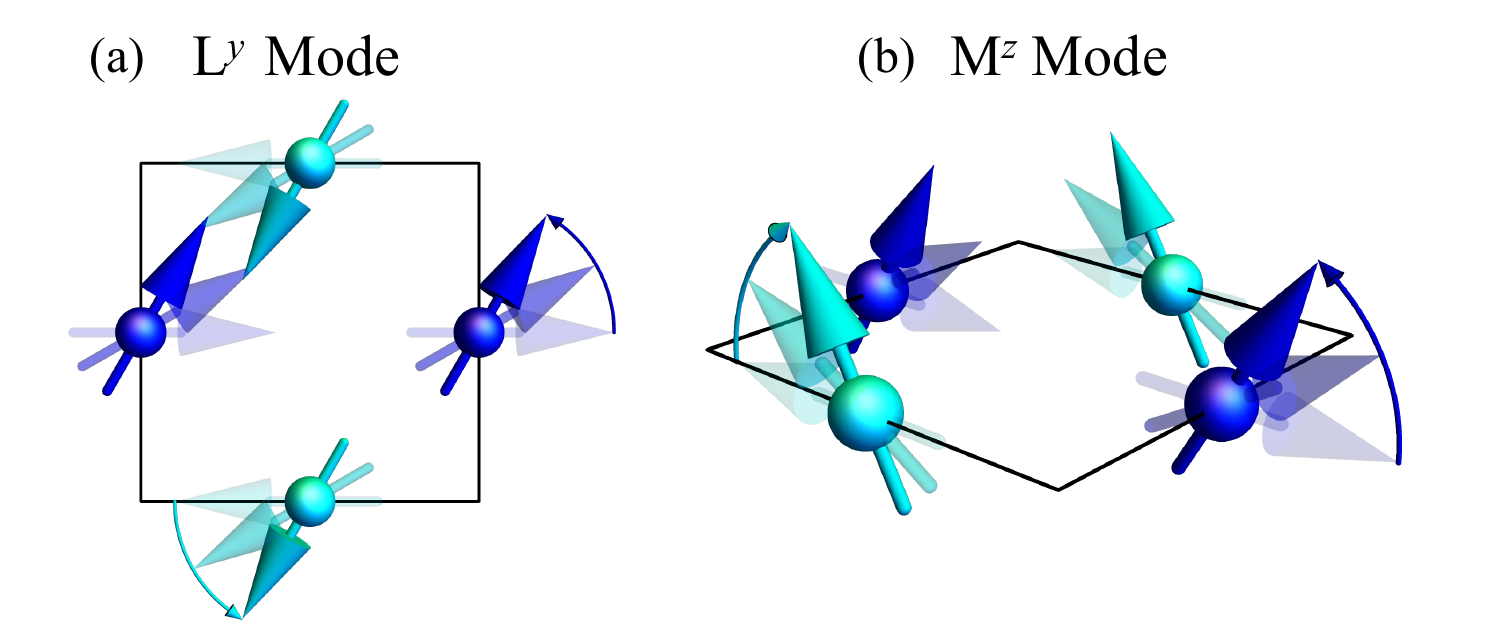}
        \caption{Collective mode of collinear antiferromagnetic moment, which is coupled to the electric field along the $x$-axis linearly. The blue (cyan) arrow indicates the spin moment in the $A$ ($B$) sublattice. (a) Picture of the symmetry-adapted basis $\text{L}^y$. (b) Picture of the symmetry-adapted basis $\text{M}^z$.}
        \label{Col_mode}
\end{figure}

\begin{table}
        \caption{Parity under symmetry operations of the magnetic point group $\bm{G}$. The symmetry operators listed in the first row are $\mathcal{G}_x = \left\{M_x \middle| \frac{1}{2}00\right\}$, $\mathcal{G}_z = \left\{M_z \middle| \frac{1}{2} \frac{1}{2} 0\right\}$, $\mathcal{S}_y = \left\{C_{2y} \middle| 0 \frac{1}{2} 0\right\}$, and $\mathcal
        {PT}$. $E^{x}$ is the light field along the $x$ direction, and $J^{\mu}$ is the electric current along the $\mu$ direction, respectively. $\mathrm{M}^{a}$ means spin magnetization along the $a$ direction, while $\mathrm{L}^{a}$ means staggered spin moments along the $a$ direction. $+$ means the observable does not change its sign, while $-$ means the observable flip its sign under the operation.}
        \begin{ruledtabular}
        \begin{tabular}{ccccc}
                & $\mathcal{G}_x$ & $\mathcal{G}_z$ & $\mathcal{S}_y$ & $\mathcal{PT}$ \\\colrule
                $E^{x}$ & $-$ & $+$           & $-$       & $-$ \\
                $J^{x}$ & $-$ & $+$           & $-$       & $+$  \\
                $J^{y}$ & $+$ & $+$           & $+$       & $+$  \\
        $\mathrm{M}^{x}$& $+$ & $-$           & $-$       & $-$  \\
        $\mathrm{L}^{x}$& $+$ & $+$           & $+$       & $+$ \\
        $\mathrm{M}^{y}$& $-$ & $-$           & $+$       & $-$ \\
        $\mathrm{L}^{y}$& $-$ & $+$           & $-$       & $+$ \\
        $\mathrm{M}^{z}$& $-$ & $+$           & $-$       & $-$ \\
        $\mathrm{L}^{z}$& $-$ & $-$           & $+$       & $+$ \\
        \end{tabular}
        \label{tab:symmetry}
        \end{ruledtabular}
\end{table}

In this subsection, we analyze the symmetry-adapted basis of the localized spins that is linearly coupled to the external light field. 
Owing to the stringent constraints on the photocurrent response by the symmetry of the system and the external field, the symmetry analysis of localized spin dynamics plays a crucial role in this study.

The system has the following symmetries and belongs to the magnetic point group $G$, explicitly given by
\begin{align}
        &\bm{G}=\bm{H}\oplus\mathcal{PT}\bm{H},\\
    &\bm{H}=\left\{\mathcal{I},\mathcal{G}_x,\mathcal{G}_z,\mathcal{S}_y\right\}.
\end{align}
Here, $\bm{G}$ represents the combined symmetry group comprising $\bm{H}$ and $\mathcal{PT}\bm{H}$. The group $\bm{H}$ consists of four symmetry operators: the identity $\mathcal{I}$, the nonsymmorphic glide operation $\mathcal{G}_x = \left\{M_x \middle| \frac{1}{2}00\right\}$ and $\mathcal{G}_z = \left\{M_z \middle| \frac{1}{2} \frac{1}{2} 0\right\}$, and the screw operation $\mathcal{S}_y = \left\{C_{2y} \middle| 0 \frac{1}{2} 0\right\}$.
The operator $\mathcal{G}_x$ combines the mirror symmetry $M_x$, which reflects across the (100) plane, with a half-unit-cell translation along the [100] direction.
The operator $\mathcal{G}_z$ combines the mirror symmetry $M_z$, which reflects across the (001) plane, with a half-unit-cell translation along the $[\frac{1}{2} \frac{1}{2} 0]$ direction.
The operator $\mathcal{S}_y$ combines the two-fold rotation symmetry $C_{2y}$ about the [010] axis with a half-unit-cell translation along the $[0 \frac{1}{2} 0]$ direction.

Let us identify the optically-active modes in collinear antiferromagnetic moments. In this discussion, we restrict our consideration to $k = 0$ magnons or antiferromagnetic resonance modes.
To clarify which mode is optically active, we analyze the parity of observables under a symmetry operation of $\left\{\mathcal{G}_x,\mathcal{G}_z,\mathcal{S}_y,\mathcal{PT}\right\}$. For example, the $\mathcal{G}_x$ operation is explicitly described as \footnote{Here, $\mathcal{G}_{z} = \qty(-i\sigma_{z}) \otimes \tau_{z}$, $\mathcal{S}_{y} = \qty(-i\sigma_{y}) \otimes \tau_{z}$ and $\mathcal{PT} = \qty(-i\sigma_y\mathcal{K}) \otimes \tau_{z}$ hold, where $\sigma_0=\bm{1}$ and $\mathcal{K}$ is the anti-unitary operator.}
\begin{align}
        \mathcal{G}_x = \qty(-i\sigma_{x}) \otimes \tau_{0}, 
\end{align}
where $\sigma$ and $\tau$ are Pauli matrices representing the spin and sublattice degrees of freedom and $\tau_0=\bm{1}$.
The transformation property of the staggered spin moment along the $x$ direction $\sigma_{x} \otimes \tau_{z}$ under $\mathcal{G}_x$ is described as
\begin{align}
        \mathcal{G}_x \qty(\sigma_{x} \otimes \tau_{z}) \mathcal{G}_x^{-1} = \sigma_{x}\otimes\tau_{z}.
\end{align}
Therefore, $\sigma_{x} \otimes \tau_{z}$ remains invariant under the $\mathcal{G}_x$ operation.
In \tabref{tab:symmetry}, we summarize the parity of the physical quantities under the operation in $\left\{\mathcal{G}_x,\mathcal{G}_z,\mathcal{S}_y,\mathcal{PT}\right\}$. 
Here, $E^{x}$ is the light field along the $x$ direction, and $J^{\mu}$ is the electric current along the $\mu$ direction, respectively. 
$\mathrm{M}^{a}$ represents ferroic configurations of spin moment along the $a$ direction, while $\mathrm{L}^{a}$ denotes staggered spin moments along the $a$ direction.
Owing to the incompatibility of the modes $\mathrm{M}^{x}, \mathrm{L}^{x}, \mathrm{M}^{y}, \mathrm{L}^{z}$ with the external light field $E^{x}$, under $\mathcal{G}_x$, $\mathcal{G}_z$ and $\mathcal{S}_y$ operation, these modes are not linearly excited by the light field.
Although the symmetry of $\mathrm{L}^{y}$ and $E^{x}$ differ under the $\mathcal{PT}$ operation, $\mathrm{L}^{y}$ can be linearly coupled to the electric current $J^{x}$.
This is because, in non-equilibrium conditions, the dissipation process effectively breaks time-reversal symmetry.
This can be interpreted as an extension of the dynamical response associated with the magnetoelectric effect and the Edelstein effect \cite{Freimuth2014,Watanabe2017, Zelezny2017,Hayami2018,Watanabe2024}. 
As a result, only the components associated with $\mathrm{L}^{y}$ and $\mathrm{M}^{z}$ can be linearly excited by light.
The symmetry argument can be applied to the case with the low-frequency light and with the high-frequency light as large as it induces interband transitions.
\begin{figure}[H]
        \centering
        \includegraphics[width=\linewidth]{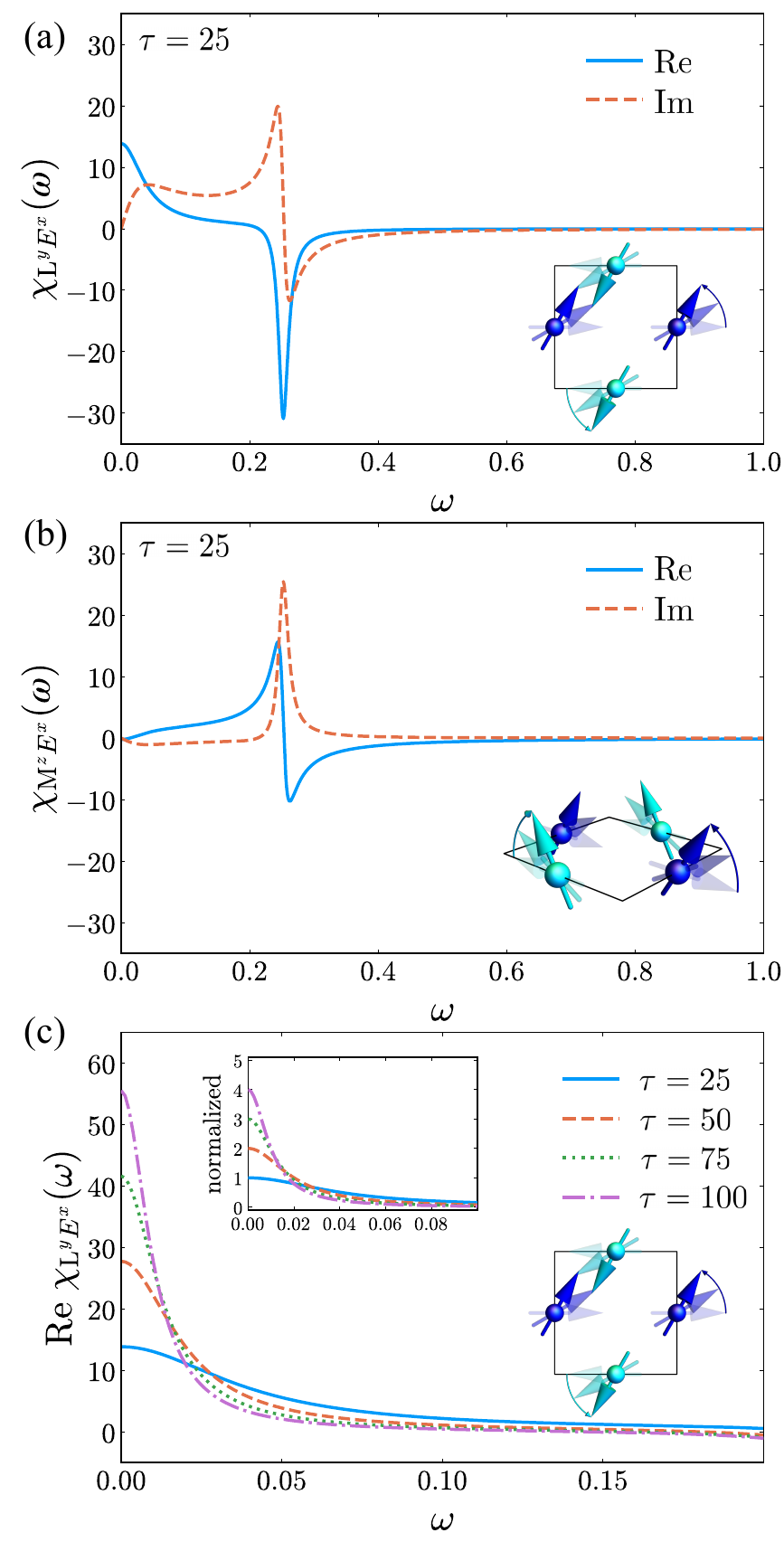}
        \caption{(a), (b) Linear electromagnetic susceptibility of the $\text{L}^y$ and $\text{M}^z$ modes to an external light field for $\tau = 25$. The blue solid lines and orange dashed lines represent the real and imaginary parts of the linear electromagnetic susceptibility, respectively. (c) Relaxation time dependence of the linear electromagnetic susceptibility for the $\text{L}^y$ mode. The inset shows the electromagnetic susceptibility normalized by the value with $\tau = 25$. This inset demonstrates that the electromagnetic susceptibility is proportional to $\tau^1$ in the limit $\omega\to0$.} 
        \label{linear_response_function}
\end{figure}

\subsection{Linear response functions}\label{linear_response_functions}
Before we investigate the nonlinear optical response, we examine the linear optical response.
We investigate the electromagnetic susceptibility, which offers direct insights into the localized spin dynamics induced by the external light field.
Electromagnetic susceptibility of localized spin system is defined as
\begin{align}
        \chi_{\text{M}^{a}E^{\mu}}(\omega)&= \dfrac{\Delta \mathrm{M}^{a}(\omega)}{E^{\mu}(\omega)}, \\
        \chi_{\text{L}^{a}E^{\mu}}(\omega)&=\dfrac{\Delta \mathrm{L}^{a}(\omega)}{E^{\mu}(\omega)}.
\end{align}
Here, we define $\Delta \mathrm{M}^{a}(\omega)$ and $\Delta \mathrm{L}^{a}(\omega)$ as the Fourier components of the modulation of the localized spins, given by
\begin{align}
        \Delta \mathrm{M}^{a}(t) &= \frac{1}{2}\qty(\Delta \mathrm{S}_{A}^{a}(t) + \Delta \mathrm{S}_{B}^{a}(t)), \\
        \Delta \mathrm{L}^{a}(t) &= \frac{1}{2}\qty(\Delta \mathrm{S}_{A}^{a}(t) - \Delta \mathrm{S}_{B}^{a}(t)).
\end{align}

We define the localized spin dynamics and linear electromagnetic susceptibilities related to the optically active modes as 
\begin{align}
        \Delta\vb{S}(\omega) &= \qty(\Delta\mathrm{L}^{y}(\omega),\Delta\mathrm{M}^{z}(\omega) ), \label{alpha_mode_dynamics}\\
        \vb*{\chi}_{\text{S}E^{\mu}}(\omega) &= \qty(\chi_{\mathrm{L}^{y}E^{\mu}}(\omega), \chi_{\mathrm{M}^{z}E^{\mu}}(\omega))^{T} \label{electromagnetic_susceptibilities}.
\end{align}
Using these quantities, we can express spin dynamics linearly coupled to the light field as
\begin{align}
        \Delta\vb{S}(\omega) = \vb*{\chi}_{\text{S}E^{\mu}}(\omega)E^{\mu}(\omega).
\end{align}

To calculate these response functions, we apply a light field with a Gaussian profile described as
\begin{align}
        E^{\mu}(t) = \dfrac{E_{0}}{\sqrt{2\pi\sigma^{2}}} \exp(-\dfrac{(t-t_{0})^{2}}{2\sigma^{2}})
        \label{gaussianpulse}.
\end{align}
We can calculate the linear response functions by the Fourier transform of the resulting response in the time domain.
For example, the linear electromagnetic susceptibility is obtained as 
\begin{align}
        \chi_{\mathrm{S}E^{\mu}}(\omega) = \dfrac{1}{E_{0}}e^{\sigma^{2}\omega^{2}/2}e^{i\omega t_{0}}\int_{0}^{\infty}e^{i\omega t}\Delta\mathrm{S}(t)dt .
\end{align}
In this calculation, we use $E_0=1.0\times10^{-5},\ t_0=0.2,$ and $\sigma=0.03$.

In \figref{linear_response_function}, we show the frequency dependence of the linear response functions $\bm{\chi}_{\text{S}E^x}(\omega)$. Specifically, we plot the electromagnetic susceptibility $\bm{\chi}_{\text{S}E^x}(\omega)$ for the optically active modes $\text{L}^y$ and $\text{M}^z$. The electromagnetic susceptibility of the localized spin system is computed by simultaneously solving the time evolution of both the electronic and localized spin systems.
The $\chi_{\text{M}^zE^x}$ shows the peak structure only at $\omega = 0.25$ corresponding to the collective excitations of the localized spin system. 
In contrast, the peak structure of $\chi_{\text{L}^yE^x}$ at $\omega = 0$ is attributed to the Edelstein effect. This behavior of $\chi_{\text{L}^yE^x}$ is consistent with the results obtained from the symmetry analysis.
In \figref{linear_response_function}, we also illustrate the dependence of the peak structure of $\operatorname{Re}\chi_{\text{L}^yE^x}$ at $\omega = 0$ on the relaxation time $\tau$. 
The $\tau$ dependence is consistent with the mechanism of the Edelstein effect \cite{Freimuth2014,Watanabe2017, Zelezny2017,Hayami2018}.

\subsection{Photocurrent Spectra}\label{photocurrent_spectra}
We show the result of NHE spectra, revealing the effect of localized spin dynamics.
We focus on the nonlinear Hall current under the electric field along the $x$-direction.
$J_y$ is transformed in the same manner as $(E^x)^2$ under the symmetry operators of the magnetic point group $ \bm{G}$.
Consequently, the nonlinear Hall conductivity, $\sigma^{yxx}$, can be nonzero in this system. 
The longitudinal photocurrent conductivity $\sigma^{xxx}$ vanishes owing to the mirror symmetry $\mathcal{G}_x$.
Thus, the photocurrent is transverse to the polarization direction of the applied light field, implying the nonlinear Hall response.
The second-order photocurrent can be described as
\begin{align}
        \begin{split}
            J^{yxx}(\omega=0,\omega_{p}) = \sigma^{yxx}(0;\omega_{p}, -\omega_{p})E^{x}(\omega_{p})E^{x}(-\omega_{p})\\+ \sigma^{yxx}(0;-\omega_{p},\omega_{p})E^{x}(-\omega_{p})E^{x}(\omega_{p}).
        \end{split}
\end{align}
Here, $J^{y}(\omega=0,\omega_{p}) $ is the DC component of current response along the $y$-direction induced by the light field with frequency $\omega_{p}$. 
To calculate the NHE spectra, we apply the continuous light field $E^{x}(t) = E_{0}\sin\omega_{p} t$. 
Here, we define the nonlinear Hall conductivity $\sigma^{yxx}(0;\omega_{p})$ as 
\begin{align}
        \sigma^{yxx}(\omega=0;\omega_{p}) &= \dfrac{1}{2}\qty(\sigma^{yxx}(0; \omega_{p}, -\omega_{p}) + \sigma^{yxx}(0; -\omega_{p}, \omega_{p})).
\end{align}
In this definition, the nonlinear Hall conductivity is calculated by the following formula \cite{Kaneko2021, Iguchi2024}
\begin{align}
\sigma^{yxx}(\omega=0; \omega_{p}) = \dfrac{2}{E_{0}^{2}NT_{p}}\int_{t_{\text{sat}}}^{t_{\text{sat}} + NT_{p}}J^{y}(t)dt,
\end{align}
where $T_{p} = 2\pi/\omega_{p}$ is the period of the external light field.
We choose $t_{\text{sat}}$ to be sufficiently large to ensure that the system reaches the steady state and use $N > 20$ to calculate the average.

\begin{figure}[H]
        \centering
        \includegraphics[width=\linewidth]{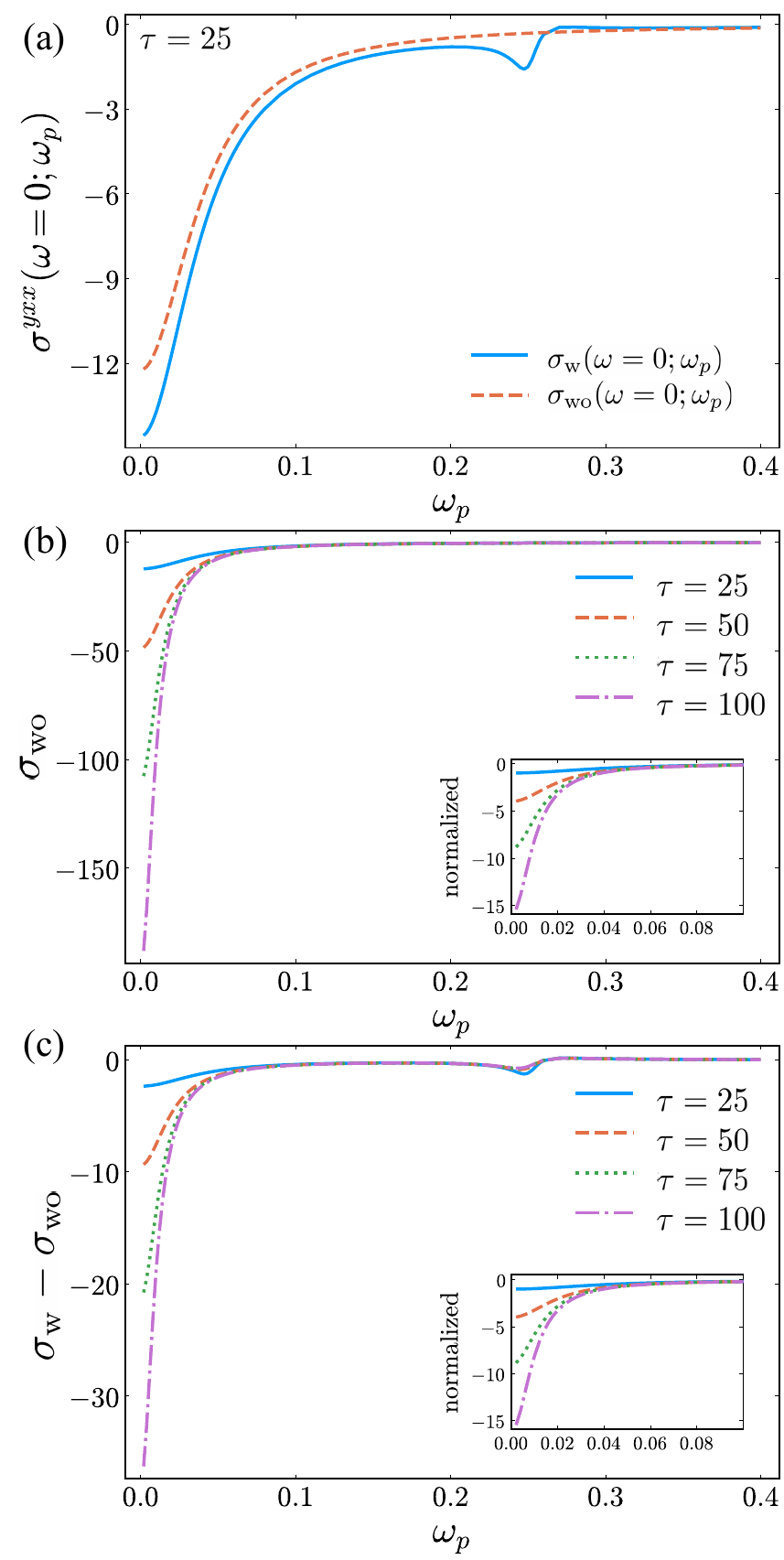}
        \caption{(a) Photocurrent spectra for $\tau=25$. The orange dashed and blue solid lines indicate the independent particle approximation calculation and calculation including the spin-dynamics effect, respectively. (b) Relaxation time dependence of the nonlinear Hall conductivity $\sigma_{\text{wo}}(\omega=0;\omega_p)$ without the effect of the localized spin dynamics. The inset shows $\sigma_{\text{wo}}(\omega=0;\omega_p)$ normalized by the value with for $\tau = 25$. 
        This inset demonstrates that $\sigma_{\text{wo}}(\omega=0;\omega_p)$ is proportional to $\tau^2$ in the limit $\omega_p\to0$. (c) Relaxation time dependence of the nonlinear Hall conductivity $\sigma_{\text{w}}(\omega=0;\omega_p)-\sigma_{\text{wo}}(\omega=0;\omega_p)$. The inset shows $\sigma_{\text{w}}(\omega=0;\omega_p)-\sigma_{\text{wo}}(\omega=0;\omega_p)$ normalized by the value with $\tau = 25$. 
        This inset demonstrates that $\sigma_{\text{w}}(\omega=0;\omega_p)-\sigma_{\text{wo}}(\omega=0;\omega_p)$ is proportional to $\tau^2$ in the limit $\omega_p\to0$.}
        \label{PVE}
\end{figure}

In \figref{PVE}(a), we present the NHE spectra with and without contributions from localized spin dynamics, denoted as $\sigma_{\text{w}}(\omega=0;\omega_p)$ and $\sigma_{\text{wo}}(\omega=0;\omega_p)$, respectively. The blue solid line represents the photocurrent spectrum, including the effects of spin dynamics, while the orange dashed line corresponds to the calculation based on the independent particle approximation.
A pronounced NHE peak is observed at $\omega_p = 0$, which corresponds to the NHE originating from the Fermi surface contribution. Additionally, in the spectrum incorporating localized spin dynamics, a peak emerges around $\omega_p = 0.25$. 
This peak arises from the collective excitations of the localized spin system and is absent in $\sigma_{\text{wo}}$.

In \figref{PVE}(b), we illustrate the relaxation-time dependence of the photocurrent spectrum, $\sigma_{\text{wo}}(\omega=0;\omega_p)$, in the absence of localized spin dynamics. The photocurrent response at $\omega_p = 0$ exhibits a quadratic dependence on the relaxation time, scaled with $\tau^2$. This result indicates that the leading contributions to the photocurrent response, driven by the external electric field, are proportional to $\tau^2$. We find that the Drude term, which is proportional to $\tau^2$, is dominant in $\sigma_{\text{wo}}(\omega=0;\omega_p)$.
In \figref{PVE}(c), we show the relaxation-time dependence of the photocurrent spectrum difference, $\sigma_{\text{w}}(\omega=0;\omega_p) - \sigma_{\text{wo}}(\omega=0;\omega_p)$, which includes the effect of localized spin dynamics. This term also exhibits a $\tau^2$ dependence in the low-frequency regime.
$\sigma_{\text{w}}(\omega=0;\omega_p) - \sigma_{\text{wo}}(\omega=0;\omega_p)$ has a sub-leading contribution to the Drude term for larger $\tau$.
In the next subsection, we investigate the origin of the relaxation time dependence of the NHE induced by localized spin dynamics.

\section{Mechanism of Nonlinear Hall effect}\label{decomposition_of_NHE}
We focus on the decomposition of the NHE arising from the spin-charge-coupled motive force.
We show the analytical expression applicable to general photocurrent response in the presence of an external light field and localized spin dynamics in \ref{formulas_for_pc}.
In \ref{decomposition_of_photocurrent} and \ref{spin-motive}, we study the effect of localized spin dynamics on photocurrent conductivity. Here, we decompose the photocurrent into different processes, where the linear coupling between spin and light obtained in  \ref{linear_response_functions} plays a crucial role.
In \ref{spin-motive}, we discuss the relaxation time dependence of the NHE driven by the spin-motive force.

\subsection{Analytical formula for photocurrent conductivity}\label{formulas_for_pc}
We elucidate mechanisms for the photocurrent conductivity under the external electric field and localized spin dynamics.
Firstly, we focus on the photocurrent generated by the optical excitation, discussed in the previous studies \cite{Watanabe2021,Ahn2020}.
Photocurrent response arising from the external field can be induced by the Fermi surface effect and the interband effect.
The Fermi-surface effect is a photocurrent generation from the light-induced shift of the Fermi surface, and it can occur only in the metallic system.
On the other hand, the interband effect is governed by the electronic transitions between the conduction and valence bands. 
The photocurrent response associated with the interband effect exhibits a resonant peak structure at the optical gap energy separating the conduction and valence bands.

In the previous paragraph, we discuss the photocurret response arising only from the electric field effect.
In this study, we discuss the dynamics of both the electronic system and the localized spin system in the spin-charge-coupled system by a real-time simulation, and therefore the calculated responses also include that coming from the spin-mediated force.
Thus, we consider the photocurrent generation from the external light field and the localized spin dynamics in the spin-charge-coupled system.
Localized spin dynamics can induce both Fermi-surface and interband mechanisms.
In particular, for the Fermi-surface contribution, the interference between the electric field and the localized spin dynamics can give rise to the mixed dipole effect.
The detailed derivation of the photocurrent conductivity is summarized in \appref{pve_conductivity} and \appref{U(2)_formula}.

First, we show the expressions for the photocurrent conductivity originating from the Fermi surface effect as
\begin{align}
    \sigma^{\mu;\nu\lambda}_{\mathrm{D}}&=\frac{1}{\omega^2+(1/\tau)^2}\int\frac{d\bm{k}}{(2\pi)^d}\sum_{a}\partial_{\mu}\epsilon_{\bm{k}a}\partial_{\nu}\partial_{\lambda}f_a,\\
    \sigma_{\mathrm{BCD;C}}^{\mu;\nu\lambda}&=\frac{i\omega}{\omega^2+1/\tau^2}\nonumber\\
    &\times\int\frac{d\bm{k}}{(2\pi)^d}\sum_{a\neq b}(\partial_{\lambda}\operatorname{Im}[\xi_{ab}^{\mu}\xi_{ba}^{\nu}]-\partial_{\nu}\operatorname{Im}[\xi_{ab}^{\mu}\xi_{ba}^{\lambda}])f_{a},\\
    \sigma_{\mathrm{BCD;L}}^{\mu;\nu\lambda}&=\frac{1/\tau}{\omega^2+1/\tau^2}\nonumber\\
    &\times\int\frac{d\bm{k}}{(2\pi)^d}\sum_{a\neq b}(\partial_{\lambda}\operatorname{Im}[\xi_{ab}^{\mu}\xi_{ba}^{\nu}]+\partial_{\nu}\operatorname{Im}[\xi_{ab}^{\mu}\xi_{ba}^{\lambda}])f_{a},\\
    \sigma_{\text{MD,C}}^{\mu;\nu\lambda} &=\frac{Ji\omega}{\omega^2+1/\tau^2}\int\frac{d\bm{k}}{(2\pi)^d}\sum_{a\neq b}\partial_{\lambda}\operatorname{Im}[\xi_{ab}^{\mu}B_{ba}^{\nu}]f_{a},\\
    \sigma_{\text{MD,L}}^{\mu;\nu\lambda} &=\frac{J/\tau}{\omega^2+1/\tau^2}\int\frac{d\bm{k}}{(2\pi)^d}\sum_{a\neq b}\partial_{\lambda}\operatorname{Im}[\xi_{ab}^{\mu}B_{ba}^{\nu}]f_{a}.
\end{align}
Here, $\sigma^{\mu;\nu\lambda}_{\mathrm{D}}$ represents the Drude term, $\sigma_{\mathrm{BCD;C}}^{\mu;\nu\lambda}$ ($\sigma_{\mathrm{BCD;L}}^{\mu;\nu\lambda}$) denotes the Berry curvature dipole term induced by the circular (linear) polarized light, and $\sigma_{\mathrm{MD;C}}^{\mu;\nu\lambda}$ ($\sigma_{\mathrm{MD;L}}^{\mu;\nu\lambda}$) is the mixed dipole term induced by the circular (linear) polarized light.

The Drude term $\sigma^{\mu;\nu\lambda}_{\mathrm{D}}$ depends on the band dispersion, and its peak amplitude shows the $\tau^2$ dependence at the zero frequency \cite{Ideue2017,Holder2020,Watanabe2020}.
As explained in \secref{introduction}, the Berry curvature dipole terms $\sigma_{\mathrm{BCD;C}}^{\mu;\nu\lambda}$ and $\sigma_{\mathrm{BCD;L}}^{\mu;\nu\lambda}$ are the photocurrent conductivities closely related to the dipole moment of the Berry curvature in momentum space \cite{Moore2010,Sodemann2015}. 
Similarly, the mixed dipole terms $\sigma_{\text{MD,C}}^{\mu;\nu\lambda}$ and $\sigma_{\text{MD,L}}^{\mu;\nu\lambda}$ stem from the mixed dipole $D_{\text{M}}^{\mu;\nu\lambda}$ defined in \Eqref{Mixed_dipole}.
The Berry curvature dipole term $\sigma_{\mathrm{BCD;L}}^{\mu;\nu\lambda}$ and the mixed dipole term $\sigma_{\mathrm{MD;L}}^{\mu;\nu\lambda}$, which are induced by linearly polarized light, exhibit a peak structure with $\tau^1$ dependence at $\omega = 0$.
By partial integration, we can find that these terms are proportional to the partial derivative of the Fermi distribution function $\partial_{\mu}f_a$, and they can survive only in metallic systems. 
We note that the mixed dipole term arises solely from the interference between the external light field and localized spin dynamics, whereas the Drude and Berry curvature dipole terms are generated purely by the external electric field.

Second, we tabulate analytical formulas for the generalized photocurrent conductivity arising from the interband effect as 
\begin{align}
        \sigma_{XY,\text{shift}}^{\mu;\nu\lambda} &= \dfrac{\pi}{2}\int \dfrac{d\vb*{k}}{(2\pi)^{d}}\sum_{a\neq b}\operatorname{Im}\left[\qty[D^{\mu}X^{\nu}]_{ab}Y_{ba}^{\lambda}\right.\nonumber\\
        &\left.- \qty[D^{\mu}Y^{\lambda}]_{ba}X_{ab}^{\nu}\right]f_{ab}\delta(\Omega-\epsilon_{ba}), \\
        \sigma_{XY,\text{gyro}}^{\mu;\nu\lambda} &= -\dfrac{i\pi}{2}\int \dfrac{d\vb*{k}}{(2\pi)^{d}}\sum_{a\neq b}\operatorname{Re}\left[\qty[D^{\mu}X^{\nu}]_{ab}Y_{ba}^{\lambda} \right.\nonumber\\
        &\left.- \qty[D^{\mu}Y^{\lambda}]_{ba}X_{ab}^{\nu}\right]f_{ab}\delta(\Omega-\epsilon_{ba}), \\
        \sigma_{XY,\text{Inj;M}}^{\mu;\nu\lambda} &= \pi \tau\int \dfrac{d\vb*{k}}{(2\pi)^{d}}\sum_{a\neq b}\Delta_{ab}^{\mu}\operatorname{Re}\qty[X^{\nu}_{ab}Y^{\lambda}_{ba}]f_{ab}\delta(\Omega-\epsilon_{ba}),\\
        \sigma_{XY, \text{Inj;E}}^{\mu;\nu\lambda} &= i\pi \tau\int \dfrac{d\vb*{k}}{(2\pi)^{d}}\sum_{a\neq b}\Delta_{ab}^{\mu}\operatorname{Im}\qty[X^{\nu}_{ab}Y^{\lambda}_{ba}]f_{ab}\delta(\Omega-\epsilon_{ba}),\\
        \sigma_{XY,\mathrm{IFSI;M}}^{\mu;\nu\lambda}&=\frac{1}{2}\int\dfrac{d\vb*{k}}{(2\pi)^{d}}\sum_{a\neq b}\operatorname{Re}[X^{\nu}_{ab}Y^{\lambda}_{ba}]f_{ab}\partial_{\mu}\mathrm{P}\frac{1}{\Omega-\epsilon_{ba}},\\
        \sigma_{XY,\mathrm{IFSI;E}}^{\mu;\nu\lambda}&=\frac{i}{2}\int\dfrac{d\vb*{k}}{(2\pi)^{d}}\sum_{a\neq c}\operatorname{Im}[X^{\nu}_{ab}Y^{\lambda}_{ba}]f_{ab}\partial_{\mu}\mathrm{P}\frac{1}{\Omega-\epsilon_{ba}},\\
        \sigma_{XY,\text{IFSI\hspace{-1.2pt}I;M}}^{\mu;\nu\lambda}&=\frac{1}{2}\int \dfrac{d\vb*{k}}{(2\pi)^{d}}\sum_{a\neq b}\partial_{\mu}\operatorname{Re}[X^{\nu}_{ab}Y^{\lambda}_{ba}]f_{ab}\mathrm{P}\frac{1}{\Omega-\epsilon_{ba}},\\
        \sigma_{XY,\text{IFSI\hspace{-1.2pt}I;E}}^{\mu;\nu\lambda}&=\frac{i}{2}\int \dfrac{d\vb*{k}}{(2\pi)^{d}}\sum_{a\neq b}\partial_{\mu}\operatorname{Im}[X^{\nu}_{ab}Y^{\lambda}_{ba}]f_{ab}\mathrm{P}\frac{1}{\Omega-\epsilon_{ba}}.
\end{align}
Here, $\sigma_{XY}^{\mu;\nu\lambda}$ represents the photocurrent conductivity under the external fields $X$ and $Y$.
The quantities $\bm{X}_{ab}$ and $\bm{Y}_{ab}$ represent either the matrix elements of the Berry connection $\bm{\xi}$ or the product of the spin operator and the exchange coupling coefficient, $J\bm{B}$. 
We define the U(1)-gauge covariant derivative of physical quantity $O$ as $[D^{\mu}O]_{ab}=\partial_{\mu}O_{ab}-i(\xi^{\mu}_{aa}-\xi^{\mu}_{bb})O_{ab}$, and $\Delta_{ab}^{\mu}=\partial_{k_\mu}\epsilon_a-\partial_{k_\mu}\epsilon_b$ represents the group velocity difference between the bands $a$ and $b$.
$f_{ab}=f_a-f_b$ denotes the difference between the Fermi–Dirac distribution functions of the bands $a$ and $b$.

$\sigma_{XY,\text{shift}}^{\mu;\nu\lambda}$ ($\sigma_{XY,\text{gyro}}^{\mu;\nu\lambda}$) is called the shift current term (gyration current term), which includes the shift vector (gyration vector) in the case of $X=Y=E$ \cite{Morimoto2016_Floquet,Fregoso2017,Ahn2020,Watanabe2021}. 
The shift current and the gyration current can be interpreted as the photocurrent generation from the positional shift of wave packets caused by the interband transition of electrons.
The magnetic injection current term $\sigma_{XY,\text{Inj;M}}^{\mu;\nu\lambda}$ (electric injection current term $\sigma_{XY,\text{Inj;E}}^{\mu;\nu\lambda}$) can be induced by the linear polarized light (circular polarized light) \cite{Sipe2000,deJuan2017,Zhang2019}.
The injection current is related to the group velocity difference $\Delta_{ab}^{\mu}$ between the conduction band and the valence band.
The magnetic intrinsic Fermi surface effect I $\sigma_{XY,\mathrm{IFSI;M}}^{\mu;\nu\lambda}$ (electric intrinsic Fermi surface effect I $\sigma_{XY,\mathrm{IFSI;E}}^{\mu;\nu\lambda}$) can be induced by the linear polarized light (circular polarized light), the magnetic intrinsic Fermi surface effect I\hspace{-1.2pt}I $\sigma_{XY,\mathrm{IFSI\hspace{-1.2pt}I;M}}^{\mu;\nu\lambda}$ (electric intrinsic Fermi surface effect I\hspace{-1.2pt}I $\sigma_{XY,\mathrm{IFSI\hspace{-1.2pt}I;E}}^{\mu;\nu\lambda}$) can be induced by the linear polarized light (circular polarized light).
By combining the $\sigma_{XY,\mathrm{IFSI;E}}^{\mu;\nu\lambda}$, $\sigma_{XY,\mathrm{IFSI\hspace{-1.2pt}I;E}}^{\mu;\nu\lambda}$, $\sigma_{XY,\mathrm{IFSI;M}}^{\mu;\nu\lambda}$, and $\sigma_{XY,\mathrm{IFSI\hspace{-1.2pt}I;M}}^{\mu;\nu\lambda}$, we get the intrinsic Fermi Surface term $\sigma_{XY,\mathrm{IFS}}^{\mu;\nu\lambda}$ as
\begin{align}
    \sigma_{XY,\mathrm{IFS}}^{\mu;\nu\lambda}&=\sigma_{XY,\mathrm{IFSI};\text{E}}^{\mu;\nu\lambda}+\sigma_{XY,\mathrm{IFSI\hspace{-1.2pt}I};\text{E}}^{\mu;\nu\lambda}\nonumber\\
    &\ \ \ \ \ \ \ \ \ \ \ \ \     +\sigma_{XY,\mathrm{IFSI};\text{M}}^{\mu;\nu\lambda}+\sigma_{XY,\mathrm{IFSI\hspace{-1.2pt}I};\text{M}}^{\mu;\nu\lambda},\nonumber\\ 
    &=-\frac{1}{2}\int \dfrac{d\vb*{k}}{(2\pi)^{d}}\sum_{a\neq b}X^{\nu}_{ab}Y^{\lambda}_{ba}\partial_{\mu}f_{ab}\mathrm{P}\frac{1}{\Omega-\epsilon_{ba}}.
\end{align}
This term contains the partial derivative of the Fermi distribution function $\partial_{\mu}f_{ab}$ and is finite only in the metallic system \cite{deJuan2017prr,Gao2021}.
In this context, $\sigma_{XY,\mathrm{IFS}}^{\mu;\nu\lambda}$ is called as the intrinsic Fermi surface term.
These interband terms can be induced by the electric field effect, the interference between the electric field and the localized spin dynamics, and the purely localized spin dynamics effect.

Here, we note that these expressions for the photocurrent conductivity are valid only in the non-degenerate band systems.
We cannot apply these expressions to the spinful $\mathcal{PT}$-symmetric system due to the U(2)-gauge symmetry arising from the Kramers degeneracy at each $\bm{k}$-points.
To ensure the U(2)-gauge symmetry, we introduce the U(2)-gauge Berry connection, the U(2)-gauge covariant derivative, and the U(2)-gauge spin operator.
We summarize the analytical formulas for the U(2)-gauge symmetric invariant formulas for the photocurrent conductivity in \tabref{Table_photocurrent_classification} and the detailed discussion for the formulation of the photocurrent conductivity in U(2)-gauge symmetric system is summarized in \appref{U(2)_formula}.

\subsection{Decomposition of photocurrent}\label{decomposition_of_photocurrent}
\begin{figure}[t]
        \centering
        \includegraphics[width = \linewidth]{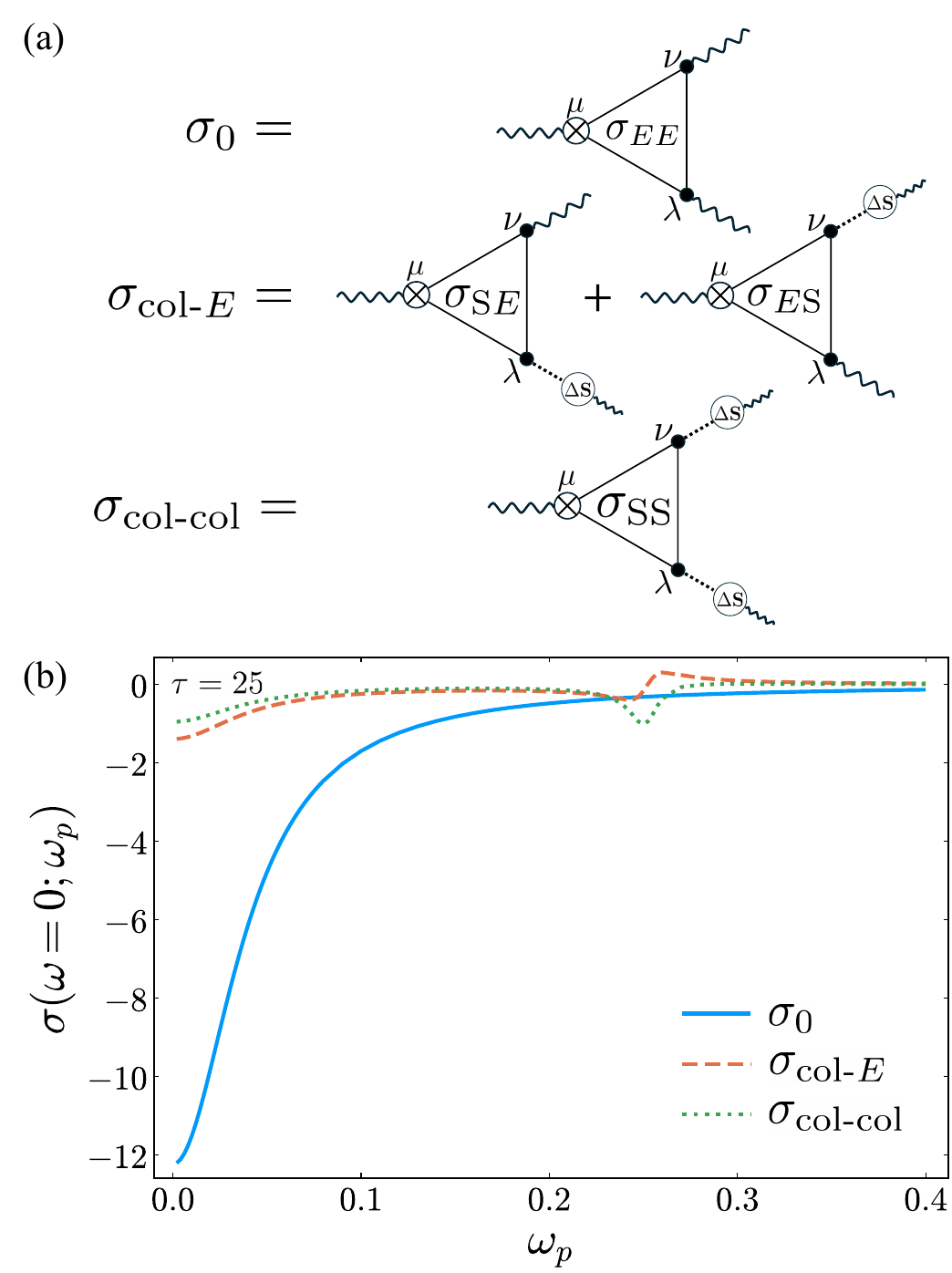}
        \caption{(a) Diagrams of three different processes of the photocurrent in the presence of collective spin dynamics. The wavy and dashed lines indicate the light field and interaction $J$, respectively. $\otimes$ represents the output photocurrent.
        The solid triangles describe the photocurrent susceptibility concerning only the electrons. $\Delta \vb{S}$ is the light induced spin dynamics.
        (b) Photocurrent spectra originating from the three different processes for $\tau=25$.}
        \label{diag_decomp}
\end{figure}

To gain deeper insight into the effect of spin dynamics on the NHE, we decompose the photocurrent contributions into three processes.
First, as illustrated in \figref{diag_decomp}(a), the photocurrent response arises from three contributing processes, described as
\begin{align}
        J^{y}(\omega=0;\omega_{p}) &= J^{y}_{0}(\omega_{p}) + J^{y}_{\text{col-E}}(\omega_{p}) +J^{y}_{\text{col-col}}(\omega_{p}),
\end{align}
where 
\begin{align}
        J_{0}^{y}(\omega_{p}) &= \sigma_{EE}^{y;xx}(0;-\omega_{p}, \omega_{p})E^{x}(-\omega_{p})E^{x}(\omega_{p}) \nonumber\\
        &+ \sigma_{EE}^{y;xx}(0;\omega_{p}, -\omega_{p})E^{x}(\omega_{p})E^{x}(-\omega_{p})\label{J_0},\\
        J_{\text{col-E}}^{y}(\omega_{p}) &= \sum_{\lambda}\sigma_{E\mathrm{S}}^{y;x\lambda}(0;-\omega_{p},\omega_{p})E^{x}(-\omega_{p})\Delta\mathrm{S}^{\lambda}(\omega_{p}) \nonumber\\ 
        &+\sum_{\lambda}\sigma_{E\mathrm{S}}^{y;x\lambda}(0;\omega_{p},-\omega_{p})E^{x}(\omega_{p})\Delta\mathrm{S}^{\lambda}(-\omega_{p})\nonumber\\
        &+\sum_{\nu}\sigma_{\mathrm{S}E}^{y;\nu x}(0;-\omega_{p},\omega_{p})\Delta\mathrm{S}^{\nu}(-\omega_{p})E^{x}(\omega_{p}) \nonumber\\ 
        &+\sum_{\nu}\sigma_{\mathrm{S}E}^{y;\nu x}(0;\omega_{p},-\omega_{p})\Delta\mathrm{S}^{\nu}(\omega_{p})E^{x}(-\omega_{p}), \label{J_colE}\\
        J_{\text{col-col}}^{y}(\omega_{p}) &= \sum_{\nu\lambda}\sigma_{\mathrm{S}\mathrm{S}}^{y;\nu\lambda}(0;-\omega_{p},\omega_{p})\Delta \mathrm{S}^{\nu}(-\omega_{p})\Delta \mathrm{S}^{\lambda}(\omega_{p}) \nonumber \\
        &+ \sum_{\nu\lambda}\sigma_{\mathrm{S}\mathrm{S}}^{y;\nu\lambda}(0;\omega_{p},-\omega_{p})\Delta \mathrm{S}^{\nu}(\omega_{p})\Delta \mathrm{S}^{\lambda}(-\omega_{p}) \label{J_colcol}.
\end{align}
In this study, we use the spin dynamics $\Delta\vb{S}(\omega)$, as defined in \Eqref{alpha_mode_dynamics}, with indices $\nu$ and $\lambda$ representing the components of $\Delta\vb{S}(\omega)$.
The first contribution, $J_{0}$, represents the photocurrent in the absence of collective spin dynamics and is described by the photocurrent conductivity $\sigma_{EE}$. 
This term arises in the independent particle approximation.
The second contribution, $J_{\text{col-E}}$, originates from the synergistic interaction between the external light field and light-induced spin dynamics, characterized by $\sigma_{E\mathrm{S}}$.
This term can be interpreted as the interference between the external light field and the collective spin dynamics.
The third contribution, $J_{\text{col-col}}$, involves photocurrent generation mediated by the light-induced spin motive force and is characterized by $\sigma_{\mathrm{SS}}$.
Since the spin dynamics are linearly coupled to the light field, as demonstrated in this study, all these components represent second-order responses to the light field.
\begin{table*}[htbp]
        \caption{
        Classification of photocurrent responses. The mechanisms with the superscript "$\ast$" originate from the Fermi surface effect. Each term of the photocurrent conductivity is expressed as $\sigma^{\mu;\nu\lambda} = \int d\bm{k}/(2\pi)^d \mathcal{I}^{\mu;\nu\lambda}$, where the integrand $\mathcal{I}^{\mu;\nu\lambda}$ is listed in the ``Integrand'' column. The generalized quantum metric is defined as $g_{ab}^{\mu\nu} = \operatorname{Re}[X^{\mu}_{ab}Y^{\nu}_{ba}]$, and the generalized Berry curvature as $\Omega_{ab}^{\mu\nu} = -2 \operatorname{Im}[X^{\mu}_{ab}Y^{\nu}_{ba}]$, where $X^{\mu}$ and $Y^{\nu}$ correspond to the interband component of the Berry connection $\mathcal{A}^{\mu}$ or spin operator $\mathcal{S}^{\mu}$ in the U(2) gauge. Owing to the double degeneracy of the electronic bands, we introduce the U(2) gauge description of the photocurrent response (see \appref{U(2)_formula}).
        When $X = Y = E$, the generalized quantum metric and Berry curvature are the conventional quantum metric and Berry curvature, respectively \cite{Ahn2022}. The quantities $A_{ab}^{\mu;\nu\lambda}$ are defined as $A_{ab}^{\mu;\nu\lambda} = \big[ [\mathcal{D}^{\mu}X^{\nu}]_{ab} Y_{ba}^{\lambda} - [\mathcal{D}^{\mu}Y^{\lambda}]_{ba} X_{ab}^{\nu} \big]$. The ``Intra/Inter'' column indicates whether each photocurrent response arises from the intraband or interband component of the output current. In the ``$\bm{e}$'' column, the symbol $\updownarrow$ marks photocurrents induced by the non-circularly polarized light, and $\circlearrowleft$ denotes circularly polarized light-induced photocurrents. The ``$\tau$'' column specifies the dependence on the relaxation time in the low-frequency regime, while the ``Field'' column classifies the external fields that induce each photocurrent response.} 
    \label{Table_photocurrent_classification}
    \centering
    \renewcommand{\arraystretch}{1.5}
    \vspace{5pt}
    \begin{tabular}{lccccc}\hline \hline
    Mechanism&Integrand&Intra/Inter&$\bm{e}$&$\tau$&Field\\ \hline
    Drude$^{\ast}$                               & \small $\mathcal{I}^{\mu;\nu\lambda}_{\mathrm{D}}=1/(\Omega^2+1/\tau^2)\sum_{a}\partial_{\mu}\partial_{\nu}\partial_{\lambda}\epsilon_{\bm{k}a}f(\epsilon_{\bm{k}a})$  & Intra & $\updownarrow$ & $O(\tau^2)$&$EE$ \\
    Berry curvature dipole$^{\ast}$ (L)              & \small $\mathcal{I}_{\mathrm{BCD;L}}^{\mu;\nu\lambda}=-1/(2\tau(\Omega^2+1/\tau^2))\sum_{a\neq b}(\partial_{\lambda}\Omega_{ab}^{\mu\nu}+\partial_{\nu}\Omega_{ab}^{\mu\lambda})f_{a}$  & Inter & \small $\updownarrow$ & $O(\tau^1)$&$EE$ \\
    Berry curvature dipole$^{\ast}$ (C)              &\small  $\mathcal{I}_{\mathrm{BCD;C}}^{\mu;\nu\lambda}=-i\Omega/2(\Omega^2+1/\tau^2)\sum_{a\neq b}(\partial_{\lambda}\Omega_{ab}^{\mu\nu}-\partial_{\nu}\Omega_{ab}^{\mu\lambda})f_{a}$  & Inter & $\circlearrowleft$ & $O(\tau^0)$&$EE$ \\
    Mixed dipole$^{\ast}$ (L)              & \small $\mathcal{I}_{\mathrm{MD;L}}^{\mu;\nu\lambda}=-1/(2\tau(\Omega^2+1/\tau^2))\sum_{a\neq b}\partial_{\lambda}\Omega_{ab}^{\mu\nu}f_{a}$  & Inter & \small $\updownarrow$ & $O(\tau^1)$&$\text{S}E$ \\
    Mixed dipole$^{\ast}$ (C)              &\small  $\mathcal{I}_{\mathrm{MD;C}}^{\mu;\nu\lambda}=-i\Omega/2(\Omega^2+1/\tau^2)\sum_{a\neq b}\partial_{\lambda}\Omega_{ab}^{\mu\nu}f_{a}$  & Inter & $\circlearrowleft$ & $O(\tau^0)$&$\text{S}E$ \\
    Intrinsic Fermi surface I$^{\ast}$ (electric)  &\small  $\mathcal{I}_{\mathrm{IFSI;E}}^{\mu;\nu\lambda}=-i/2\sum_{a\neq b}\Omega_{ab}^{\nu\lambda}f_{ab}\partial_{\mu}\mathrm{P}(1/(\Omega-\epsilon_{ba}))$  & Intra & $\circlearrowleft$ & $O(\tau^0)$&All\\
    Intrinsic Fermi surface I\hspace{-1.2pt}I$^{\ast}$ (electric)  & \small $\mathcal{I}_{\mathrm{IFSI\hspace{-1.2pt}I;E}}^{\mu;\nu\lambda}=-i/2\sum_{a\neq b}\partial_{\mu}\Omega_{ab}^{\nu\lambda}f_{ab}\mathrm{P}(1/(\Omega-\epsilon_{ba}))$  & Inter & $\circlearrowleft$ & $O(\tau^0)$&All\\
    Intrinsic Fermi surface I$^{\ast}$ (magnetic)  & \small $\mathcal{I}_{\mathrm{IFSI;M}}^{\mu;\nu\lambda}=1/2\sum_{a\neq b}g_{ab}^{\nu\lambda}f_{ab}\partial_{\mu}\mathrm{P}(1/(\Omega-\epsilon_{ba}))$  & Intra & $\updownarrow$ & $O(\tau^0)$&All \\
    Intrinsic Fermi surface I\hspace{-1.2pt}I$^{\ast}$ (magnetic)  & \small $\mathcal{I}_{\mathrm{IFSI\hspace{-1.2pt}I;M}}^{\mu;\nu\lambda}=1/2\sum_{a\neq b}\partial_{\mu}g_{ab}^{\nu\lambda}f_{ab}\mathrm{P}(1/(\Omega-\epsilon_{ba}))$  & Inter & $\updownarrow$ & $O(\tau^0)$&All \\
    Injection current (electric)       &\small  $\mathcal{I}^{\mu;\nu\lambda}_{\text{Inj;E}}=-(i\pi\tau/2)\sum_{a\neq b}\Delta_{ab}^{\mu}\Omega_{ab}^{\nu\lambda}f_{ab}\delta(\Omega-\epsilon_{ba})$ & Intra & $\circlearrowleft$ &$O(\tau^0)$&All    \\
    Injection current (magnetic)        & \small $\mathcal{I}^{\mu;\nu\lambda}_{\text{Inj;M}}=\pi\tau\sum_{a\neq b}\Delta_{ab}^{\mu}g_{ab}^{\nu\lambda}f_{ab}\delta(\Omega-\epsilon_{ba})$  & Intra & $\updownarrow$ &$O(\tau^0)$&All    \\
    Shift current                       & \small  $\mathcal{I}^{\mu;\nu\lambda}_{\text{shift}}=(\pi/2)\sum_{a\neq b}\operatorname{Im}[A_{ab}^{\mu;\nu\lambda}]f_{ab}\delta(\Omega-\epsilon_{ab})$& Inter & $\updownarrow$ & $O(\tau^{-1})$&All   \\
    Gyration current                    & \small $\mathcal{I}^{\mu;\nu\lambda}_{\text{gyro}}=-(i\pi/2)\sum_{a\neq b}\operatorname{Re}[A_{ab}^{\mu;\nu\lambda}]f_{ab}\delta(\Omega-\epsilon_{ab})$  & Inter & $\circlearrowleft$ &$O(\tau^{-1})$ &All    \\ 
    \hline \hline
    \end{tabular}
\end{table*}

In linear response theory, the total current response to multiple external fields, such as an electric field and localized spin dynamics, can be expressed as the superposition of the response to each field.
However, in nonlinear response, the total output is affected by interference between different fields.
Using the photocurrent classification outlined in \Eqref{J_0}, \Eqref{J_colE}, and \Eqref{J_colcol}, we can define the photocurrent conductivity in the presence of collective spin dynamics as
\begin{align}
        J_{0}^{y}(\omega_{p}) &= 2\sigma_{0}(0;\omega_{p})E^{x}(\omega_{p})E^{x}(-\omega_{p}), \\
        J_{\text{col-E}}^{y}(\omega_{p}) &= 2\sigma_{\text{col-E}}(0;\omega_{p})E^{x}(\omega_{p})E^{x}(-\omega_{p}), \\
        J_{\text{col-col}}^{y}(\omega_{p}) &= 2\sigma_{\text{col-col}}(0;\omega_{p})E^{x}(\omega_{p})E^{x}(-\omega_{p}),
\end{align}
where 
\begin{align}
        \sigma_{0}(0;\omega_{p}) &= \dfrac{1}{2}\qty(\sigma_{EE}^{y;xx}(0; -\omega_{p}, \omega_{p}) + \sigma_{EE}^{y;xx}(0; \omega_{p}, -\omega_{p})), \\
        \sigma_{\text{col-E}}(0;\omega_{p})
        &=\dfrac{1}{2}\sum_{\lambda}\sigma_{E\mathrm{S}}^{y;x\lambda}(0;-\omega_{p},\omega_{p})\chi_{\mathrm{S}^{\lambda}E^x}(\omega_{p})\nonumber \\
        &+ \dfrac{1}{2}\sum_{\lambda}\sigma_{E\mathrm{S}}^{y;x\lambda}(0;\omega_{p},-\omega_{p})\chi_{\mathrm{S}^{\lambda}E^x}(-\omega_{p})\nonumber\\
        &+\dfrac{1}{2}\sum_{\nu}\sigma_{\mathrm{S}E}^{y;\nu x}(0;-\omega_{p},\omega_{p})\chi_{\mathrm{S}^{\nu}E^x}(-\omega_{p})\nonumber \\
        &+ \dfrac{1}{2}\sum_{\nu}\sigma_{\mathrm{S}E}^{y;\nu x}(0;\omega_{p},-\omega_{p})\chi_{\mathrm{S}^{\nu}E^x}(\omega_{p}),\nonumber \\
        &=\frac{1}{2}\sum_{\nu}(\sigma_{\text{MD}}^{y;\nu x}(0;-\omega_{p},\omega_{p})+\tilde{\sigma}_{\text{S}E}^{y;\nu x}(0;-\omega_{p},\omega_{p}))\nonumber\\
        &\times \chi_{\mathrm{S}^{\nu}E^x}(-\omega_{p})\nonumber\\
        &+\frac{1}{2}\sum_{\nu}(\sigma_{\text{MD}}^{y;\nu x}(0;\omega_{p},-\omega_{p})+\tilde{\sigma}_{\text{S}E}^{y;\nu x}(0;\omega_{p},-\omega_{p}))\nonumber\\
        &\times \chi_{\mathrm{S}^{\nu}E^x}(\omega_{p}),\\
        \begin{split}
            \sigma_{\text{col-col}}(0;\omega_{p})
            &= \dfrac{1}{2}\sum_{\nu\lambda}\sigma_{\mathrm{SS}}^{y;\nu\lambda}(0;-\omega_{p},\omega_{p})\chi_{\mathrm{S}^{\nu}E^x}(-\omega_{p})\chi_{\mathrm{S}^{\lambda}E^x}(\omega_{p}) \\
        &+ \dfrac{1}{2}\sum_{\nu\lambda}\sigma_{\mathrm{SS}}^{y;\nu\lambda}(0;\omega_{p},-\omega_{p})\chi_{\mathrm{S}^{\nu}E^x}(\omega_{p})\chi_{\mathrm{S}^{\lambda}E^x}(-\omega_{p}).
        \end{split}
\end{align}
We employ electromagnetic susceptibilities defined in \Eqref{electromagnetic_susceptibilities} to analyze the system.
We summarize the expressions for the photocurrent conductivities in these terms in \tabref{Table_photocurrent_classification}. The detailed derivation of formulas for these conductivities is provided in \appref{pve_conductivity} and \appref{U(2)_formula}.

Our time-dependent calculations naturally decompose the photocurrent response as follows. First, $J_{0}(\omega_{p})$ is determined from calculations where the spin configuration is not updated. Second, $J_{\text{col-col}}(\omega_{p})$ is obtained by turning off the external light field $E^{x}(t)$ while updating the Hamiltonian using the localized spin dynamics $\vb{S}(t)$, which is derived from the simulations used to compute the total photocurrent spectrum $J(\omega_{p})$. Finally, $J_{\text{col-E}}(\omega_{p})$ is obtained by subtracting $J_{0}(\omega_{p})$ and $J_{\text{col-col}}(\omega_{p})$ from $J(\omega_{p})$.

Using this approach, we decompose the photocurrent response into three components. 
The spectrum of each term is shown in \figref{diag_decomp}(b).
The $\sigma_{0}(\omega)$ component corresponds to the orange dashed line in \figref{PVE}. 
$\sigma_{\text{col-}E}$ as well as $\sigma_{\text{col-col}}$ exhibits a peak at $\omega = 0.25$, which corresponds to the resonance frequency of the localized spin system. 
At the $\omega=0$, a significant contribution is obtained from $\sigma_{\text{col-}E}$, while $\sigma_{\text{col-col}}$ does not have a substantial effects.

Next, we can further decompose the photocurrent response into two components by separating the current operator at the output vertex into intraband and interband components, expressed as
\begin{align}
    J^{y}_{ab}=v_a^y\delta_{ab}+i\epsilon_{ab}\xi_{ab}^y.
\end{align}
The intraband component of the current operator corresponds to the group velocity of the band electron $v_a^y=\partial_{k_y}\epsilon_a$, while the interband component is associated with the Berry connection $\xi_{ab}^y$ and the energy difference between the bands $\epsilon_{ab}=\epsilon_{a\bm{k}}-\epsilon_{b\bm{k}}$.
In numerical calculations, this decomposition is obtained by evaluating the expectation values of the intraband and interband components of the current operator at each time step. By using this approach, the photocurrent responses can be expressed as
\begin{align}
    \sigma_0(0;\omega_p)&=\sigma^{\mathrm{intra}}_0(0;\omega_p)+\sigma^{\mathrm{inter}}_0(0;\omega_p),\\
    \sigma_{\text{col-E}}(0;\omega_p)&=\sigma^{\mathrm{intra}}_{\text{col-E}}(0;\omega_p)+\sigma^{\mathrm{inter}}_{\text{col-E}}(0;\omega_p),\\
    \sigma_{\text{col-col}}(0;\omega_p)&=\sigma^{\mathrm{intra}}_{\text{col-col}}(0;\omega_p)+\sigma^{\mathrm{inter}}_{\text{col-col}}(0;\omega_p),
\end{align}
where $\sigma^{\mathrm{intra}}$ and $\sigma^{\mathrm{inter}}$ denote the intraband and interband components of the nonlinear optical conductivity, respectively.

The Drude term $\sigma_{\mathrm{D}}$, injection current term $\sigma_{\mathrm{inj}}$, and parts of the intrinsic Fermi surface effect $\sigma_{\mathrm{IFSI}}$ originate from the intraband component of the current matrix. On the other hand, the Berry curvature dipole term $\sigma_{\mathrm{BCD}}$, shift current $\sigma_{\mathrm{shift}}$, and the rest of contributions to the intrinsic Fermi surface effect $\sigma_{\mathrm{IFSI\hspace{-1.2pt}I}}$ arise from the interband component of the current matrix \cite{Watanabe2021}.
We summarize the expressions for these photocurrent conductivities in \tabref{Table_photocurrent_classification}. Detailed discussions on whether each photocurrent response arises from the intraband or interband component of the output current are provided in \appref{pve_conductivity} and \appref{U(2)_formula}.

It is important to note that the phase degrees of freedom of the driving fields impose constraints on the types of photocurrents that can be generated. 
As a general case, let us consider the photocurrent induced by two external fields $X(\omega_{p})$ and $Y(\omega_{p})$. This process can be written by using the
photocurrent conductivity $\sigma_{XY}(0; \omega_{p}, -\omega_{p})$ as
\begin{align}
    \begin{split}
        J_{XY}(\omega_{p}) &= \sigma_{XY}(0;-\omega_{p}, \omega_{p})X(-\omega_{p})Y(\omega_{p})\\
     &+  \sigma_{XY}(0; \omega_{p}, -\omega_{p})X(\omega_{p})Y(-\omega_{p}).
    \end{split}\label{current_XY}
\end{align}
Owing to the fact that the time-domain external field is real, the external field in the frequency domain satisfies the relation
\begin{align}
    X(\omega)=X^{\ast}(-\omega).
\end{align}
Therefore, the product of the two external fields in \Eqref{current_XY} is transformed as
\begin{align}
    X(-\omega_p)Y(\omega_p)&=X^{\ast}(\omega_p)Y(\omega_p),\nonumber\\
    &=R^{XY}(\omega_p)+iI^{XY}(\omega_p).
\end{align}
Here, we decomposed the product of the external field into real and imaginary parts defined by 
\begin{align}
    R^{XY}(\omega_p)&=\operatorname{Re}\{[X(\omega_p)]^{\ast}Y(\omega_p)\},\\
    I^{XY}(\omega_p)&=\operatorname{Im}\{[X(\omega_p)]^{\ast}Y(\omega_p)\},
\end{align}
which are related to the Stokes parameters \cite{wolf2007introduction} in the case of $X=Y=E$.
By using this method, we can rewrite \Eqref{current_XY} as
\begin{align}
    \begin{split}
        J_{XY}(\omega_p)=2\operatorname{Re}[\sigma_{XY}(0;-\omega_p,\omega_p)]R^{XY}(\omega_p)\\
        -2\operatorname{Im}[\sigma_{XY}(0;-\omega_p,\omega_p)]I^{XY}(\omega_p).
    \end{split}
\end{align}
In this formulation, the real part of the photocurrent conductivity arises from the real part (in-phase part) of the product of the external fields, whereas the imaginary part of the photocurrent conductivity originates from the imaginary part (out-of-phase part) of the product of the external fields.
In this work, $\sigma_{XY, \text{D}}$, $\sigma_{XY, \text{BCD;L}}$, $\sigma_{XY, \text{shift}}$, $\sigma_{XY, \text{Inj;M}}$, and $\sigma_{XY, \text{IFS;M}}$ ($\sigma_{XY, \text{BCD;C}}$, $\sigma_{XY, \text{gyro}}$, $\sigma_{XY, \text{Inj;E}}$, $\sigma_{XY, \text{IFS;E}}$) represent the real (purely imaginary) components characterizing the photocurrent induced by the in-phase (out-of-phase) part of the product of the external fields.

The symmetry of fields also imposes further constraints on the types of photocurrent that can arise.
In the present case, the fields $X(\omega_{p})$ and $Y(\omega_{p})$ correspond to the electric field of light, $E(\omega_{p})$, and/or localized spin dynamics, $\Delta\mathrm{S}^{\nu}(\omega_{p})$. 
For example, we observe that the spin dynamics, specifically the product of the $\mathrm{L}^{y}(\omega_{p})$ and $\mathrm{M}^{z}(\omega_{p})$, exhibits both in-phase and out-of-phase components. 
Moreover, their parity under the $\mathcal{PT}$ transformation differs, indicating the mechanism of the photocurrent generation, such as the shift current, electric injection current, and the electric intrinsic Fermi surface effect.
These mechanisms are similar to those in $\mathcal{T}$-symmetric systems \cite{Baltz1981,Sipe2000,deJuan2017prr,Watanabe2021}.
A similar analysis applies to the photocurrent originating from the interference between the light field and collective spin dynamics, denoted by $\sigma_{E\mathrm{S}}$. The classification of intraband and interband components of the photocurrent is summarized in \tabref{photocurrent_classification_intra} and \tabref{photocurrent_classification_inter}, respectively, based on the symmetry analysis detailed in \appref{sym_analysis_photocurrent}.
It is important to note that the spin operator does not include derivative terms, such as $\partial_{\bm{k}}$, and therefore does not give rise to Fermi surface terms like $\sigma_{\mathrm{D}}$ or $\sigma_{\mathrm{BCD}}$ in $\sigma_{\text{col-col}}$.
\begin{table}[]
        \caption{Photocurrent mechanism related to the intraband components induced by the light field and spin dynamics.
        The external fields listed in the first column correspond to the index $\nu$ in the nonlinear optical conductivity $\sigma^{y;\nu\lambda}$, while those listed in the first row correspond to the index $\lambda$ in $\sigma^{y;\nu\lambda}$.
        For instance, light field $E^x$ and spin dynamics along the $y$ direction $\mathrm{L}^{y}$ induces $\sigma_{\text{Inj;E}}$ and $\sigma_{\text{IFSI;E}}$. $\sigma_{\text{Inj;E}}$ and $\sigma_{\text{IFSI;E}}$ are induced by out-of-phase component of the product of $E^x$ and $\mathrm{L}^{y}$, while the $\sigma_{\text{Inj;M}}$ and $\sigma_{\text{IFSI;M}}$ are induced by in-phase component. The detailed expressions for $\sigma_{\text{D}},\ \sigma_{\text{Inj;E}},\ \sigma_{\text{Inj;M}},\ \sigma_{\text{IFSI;M}},\ \text{and}\ \sigma_{\text{IFSI;E}}$ are given in \tabref{Table_photocurrent_classification}.}
        \begin{ruledtabular}
        \begin{tabular}{cccc}
$\nu\backslash\lambda$& $E^x$ & $\mathrm{L}^{y}$ & $\mathrm{M}^{z}$  \\ \hline
$E^x$&$\sigma_{\mathrm{D}},\sigma_{\mathrm{Inj;M}},\sigma_{\mathrm{IFSI;M}}$&$\sigma_{\mathrm{Inj;E}},\sigma_{\mathrm{IFSI;E}}$&$\sigma_{\mathrm{Inj;M}},\sigma_{\mathrm{IFSI;M}}$\\
$\mathrm{L}^y$&$\sigma_{\mathrm{Inj;E}},\sigma_{\mathrm{IFSI;E}}$&$\sigma_{\mathrm{Inj;M}},\sigma_{\mathrm{IFSI;M}}$&$\sigma_{\mathrm{Inj;E}},\sigma_{\mathrm{IFSI;E}}$\\
$\mathrm{M}^z$&$\sigma_{\mathrm{Inj;M}},\sigma_{\mathrm{IFSI;M}}$&$\sigma_{\mathrm{Inj;E}},\sigma_{\mathrm{IFSI;E}}$&$\sigma_{\mathrm{Inj;M}},\sigma_{\mathrm{IFSI;M}}$
\end{tabular}
\end{ruledtabular}   
\label{photocurrent_classification_intra}
\end{table} 
\begin{table}[]
        \caption{Photocurrent mechanism related to the interband components induced by the light field and spin dynamics. 
        The external fields listed in the first column correspond to the index $\nu$ in the nonlinear optical conductivity $\sigma^{y;\nu\lambda}$, while those listed in the first row correspond to the index $\lambda$ in $\sigma^{y;\nu\lambda}$.
        For instance, light field $E^x$ and uniform spin dynamics along the $z$ direction $\mathrm{M}^{z}$ induces $\sigma_{\text{gyro}}$ and $\sigma_{\text{IFSI\hspace{-1.2pt}I;M}}$. $\sigma_{\text{IFSI\hspace{-1.2pt}I;M}}$ is induced by in-phase component of the product of $E^x$ and $\mathrm{M}^{z}$, while the $\sigma_{\text{gyro}}$ is induced by out-of-phase component. The detailed expressions for $\sigma_{\text{MD}},\ \sigma_{\text{shift}},\ \sigma_{\text{gyro}},\ \sigma_{\text{IFSI\hspace{-1.2pt}I;M}},\ \text{and}\ \sigma_{\text{IFSI\hspace{-1.2pt}I;E}}$ are given in \tabref{Table_photocurrent_classification}.}
        \begin{ruledtabular}
        \begin{tabular}{cccc}
$\nu\backslash\lambda$& $E^x$ & $\mathrm{L}^{y}$ & $\mathrm{M}^{z}$  \\ \hline
$E^x$&$\sigma_{\mathrm{IFSI\hspace{-1.2pt}I;M}}$&$\sigma_{\mathrm{Shift}},\sigma_{\mathrm{IFSI\hspace{-1.2pt}I;E}}$&$\sigma_{\mathrm{gyro}},\sigma_{\mathrm{IFSI\hspace{-1.2pt}I;M}}$\\
$\mathrm{L}^y$&$\sigma_{\mathrm{MD}},\sigma_{\mathrm{Shift}},\sigma_{\mathrm{IFSI\hspace{-1.2pt}I;E}}$&$\sigma_{\mathrm{IFSI\hspace{-1.2pt}I;M}}$&$\sigma_{\mathrm{Shift}},\sigma_{\mathrm{IFSI\hspace{-1.2pt}I;E}}$\\
$\mathrm{M}^z$&$\sigma_{\mathrm{gyro}},\sigma_{\mathrm{IFSI\hspace{-1.2pt}I;M}}$&$\sigma_{\mathrm{Shift}},\sigma_{\mathrm{IFSI\hspace{-1.2pt}I;E}}$&$\sigma_{\mathrm{IFSI\hspace{-1.2pt}I;M}}$
\end{tabular}
\end{ruledtabular}   
\label{photocurrent_classification_inter}
\end{table} 
\subsection{Relaxation time dependence of nonlinear Hall effect driven by spin-motive force}\label{spin-motive}
\begin{figure}[t]
        \centering
        \includegraphics[width = \linewidth]{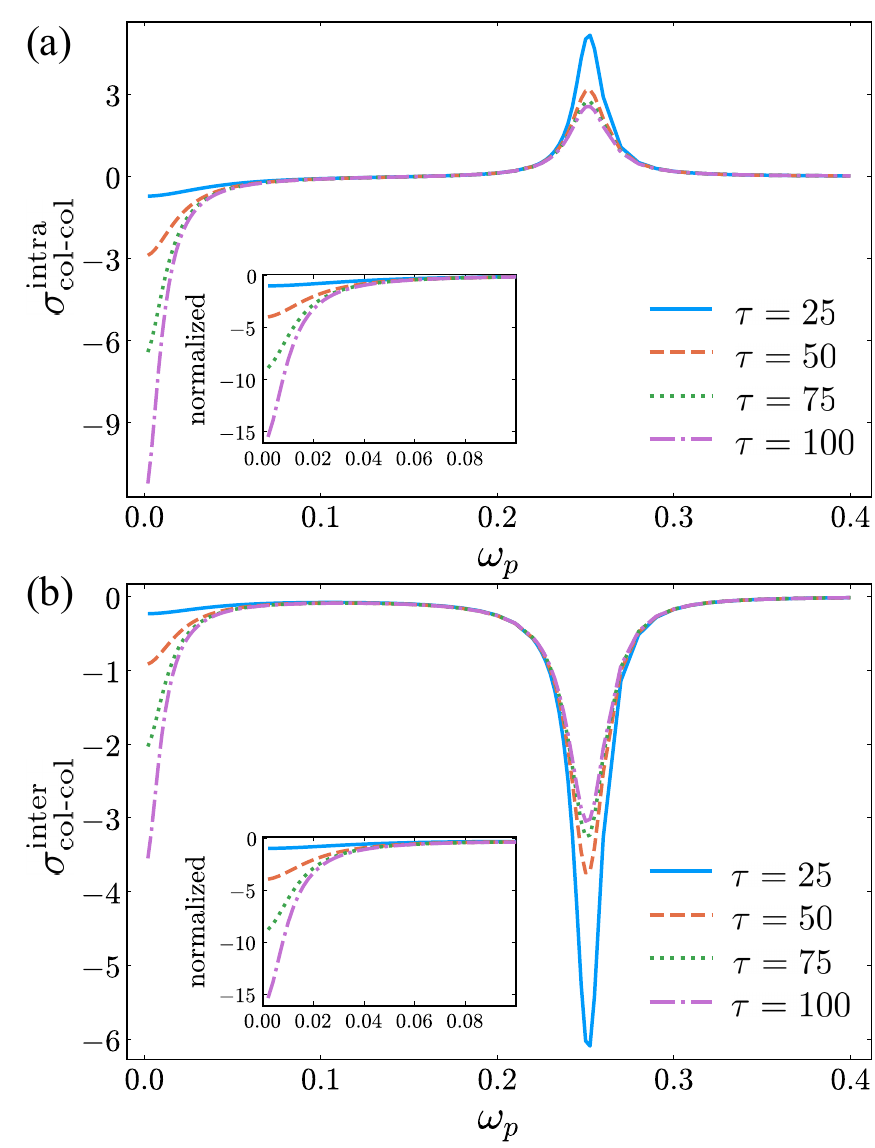}
        \caption{(a) Relaxation time dependence of $\sigma_{\text{col-col}}^{\text{intra}}$. The inset shows $\sigma_{\text{col-col}}^{\text{intra}}$ normalized by the value with $\tau = 25$. 
        This inset demonstrates that $\sigma_{\text{col-col}}^{\text{intra}}$ is proportional to $\tau^2$ in the limit $\omega_p\to0$.
        (b) Relaxation time dependence of $\sigma_{\text{col-col}}^{\text{inter}}$. The inset shows $\sigma_{\text{col-col}}^{\text{inter}}$ normalized by the value with $\tau = 25$. 
        This inset demonstrates that $\sigma_{\text{col-col}}^{\text{inter}}$ is proportional to $\tau^2$ in the limit $\omega_p\to0$.}
        \label{sigma_colcol}
\end{figure}

Let us focus on $\sigma_{\text{col-col}}^{\text{intra}}$, $\sigma_{\text{col-col}}^{\text{inter}}$ and $\sigma_{\text{col-}E}^{\text{inter}}$ near $\omega = 0$ and identify the dominant contributions in the clean limit by considering the relaxation-time dependence.
First, we plot the spectra of $\sigma_{\text{col-col}}^{\text{intra}}$ and $\sigma_{\text{col-col}}^{\text{inter}}$ in \figref{sigma_colcol}. We observe that both $\sigma_{\text{col-col}}^{\text{intra}}$ and $\sigma_{\text{col-col}}^{\text{inter}}$ exhibit a $\tau^2$ dependence in the low-frequency regime.
In \appref{freq_dep}, we demonstrate that $\sigma_{\text{Inj}}$ and $\sigma_{\text{IFS}}$ are proportional to $\tau^0$ in the low-frequency regime $\omega \ll \epsilon_g$, where $\epsilon_g$ represents the optical gap of the electronic system\footnote{Note that the relaxation-time dependence of injection current differs from that shown in \tabref{Table_photocurrent_classification} where the light frequency is as large as that producing particle-hole pairs.}.
Therefore, we can express the relaxation time dependence of the injection current (intrinsic Fermi surface effect) $\sigma_{\text{SS,Inj(IFS)}}$ term arising from the localized spin dynamics as
\begin{align}
    \sigma_{\text{SS,inj(IFS)}}(0;\mp\omega,\pm\omega)\propto\tau^0.
\end{align}
As shown in \figref{linear_response_function}, the real part of the linear electromagnetic susceptibility of the symmetry-adapted basis $\text{L}^y$ is proportional to $\tau^1$ in the low-frequency regime as
\begin{align}
    \operatorname{Re}\chi_{\text{L}^yE^x}(\omega)\propto\tau^1.
\end{align}
Consequently, the injection current (intrinsic Fermi surface effect) $\sigma^{\text{inj(IFS)}}_{\text{col-col}}$ from the light-induced spin motive force can exhibit a $\tau^2$ dependence at $\omega = 0$ as
\begin{align}
    \sigma^{\text{inj(IFS)}}_{\text{col-col}}(0;\omega)&=\frac{1}{2}\sigma_{\text{SS,Inj(IFS)}}(0;-\omega,\omega)\chi_{\text{L}^yE^x}(-\omega)\chi_{\text{L}^yE^x}(\omega)\nonumber\\
    &+\frac{1}{2}\sigma_{\text{SS,Inj(IFS)}}(0;\omega,-\omega)\chi_{\text{L}^yE^x}(\omega)\chi_{\text{L}^yE^x}(-\omega),\nonumber\\
    &=O(\tau^2),
\end{align}
using $\sigma_{\text{SS,inj(IFS)}} \propto \tau^0$ and $\chi_{\text{S}E^x} \propto \tau^1$.
From \tabref{photocurrent_classification_intra} and \tabref{photocurrent_classification_inter}, the injection current and intrinsic Fermi surface effects arising from two localized spin dynamics $\text{L}^y$ are not forbidden by the symmetry.
In contrast, we find that the shift current term $\sigma_{\text{SS,shift}}$ is proportional to $\tau^{-1}$ in the low-frequency regime $\omega \ll \epsilon_g$, as shown in \appref{freq_dep}. The shift current term arising from $\sigma_{\text{col-col}}$ does not exhibit $\tau^2$ dependence at $\omega = 0$, even when combined with the linear electromagnetic susceptibility.  
Thus, the shift mechanism plays a minor role in the clean limit.
These observations allow us to identify leading contributions from $\sigma_{\text{col-col}}^{\text{intra}}$ and $\sigma_{\text{col-col}}^{\text{inter}}$.

The peak of $\sigma_{\text{col-col}}^{\text{intra}}$ near $\omega = 0$ originates from $\sigma_{\text{Inj;M}}$ and $\sigma_{\text{IFSI;M}}$ enhanced by the linear electromagnetic susceptibility $\chi_{\text{L}^yE^x}(\omega)$. Similarly, the peak of $\sigma_{\text{col-col}}^{\text{inter}}$ near $\omega = 0$ arises from $\sigma_{\text{IFSI\hspace{-1.2pt}I;M}}$ and the same susceptibility $\chi_{\text{L}^yE^x}(\omega)$.
In contrast to the NHE behavior near $\omega = 0$, the NHE around $\omega = 0.25$ does not significantly depend on $\tau$. Instead, the $\tau$ dependence of the NHE in this regime is determined by the relaxation-time dependence of $\chi_{\text{S}E^x}$. This analysis shows that the NHE driven by the spin-motive force at the low-frequency regime proportional to $\tau^2$ can result from $\sigma_{\text{SS;Inj}}$ and $\sigma_{\text{SS;IFS}}$ when combined with a linear electromagnetic susceptibility proportional to $\tau^1$.
Finally, we note that the leading term from $\sigma_{\text{SS}}$ includes a factor of $1/(\omega - \epsilon_g)$, which produces a resonant structure near the optical gap $\epsilon_g$. In systems with a large optical gap, these terms can be significantly suppressed by this resonant factor at the low-frequency regime.
\begin{figure}[t]
        \centering
        \includegraphics[width = \linewidth]{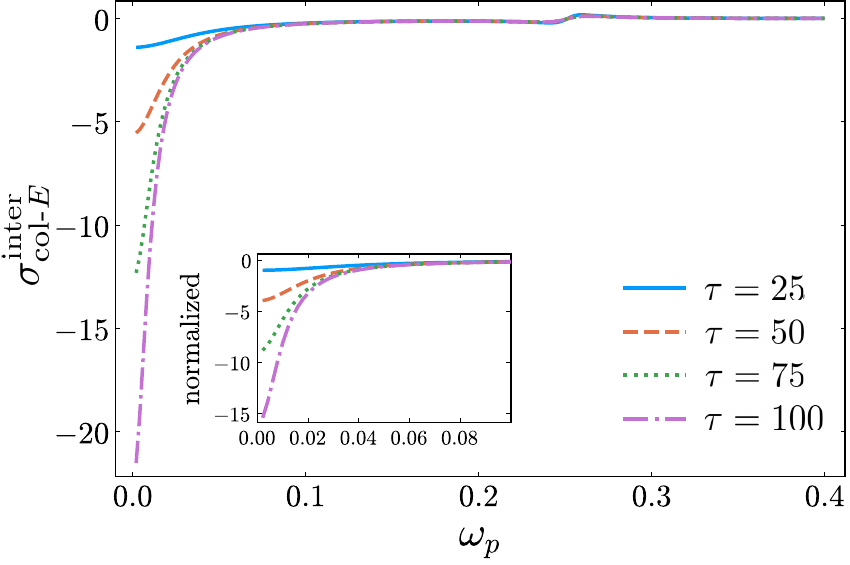}
        \caption{Relaxation time dependence of $\sigma_{\text{col-}E}^{\text{inter}}$. The inset shows $\sigma_{\text{col-}E}^{\text{inter}}$ normalized by the value with $\tau = 25$. 
        This inset demonstrates that $\sigma_{\text{col-}E}^{\text{inter}}$ is proportional to $\tau^2$ in the limit $\omega_p\to0$.}
        \label{sigma_colE}
\end{figure}

Next, we discuss the NHE arising from $\sigma_{\text{col-}E}^{\text{inter}}$, which contains the mixed dipole term. The $\tau$-dependence of $\sigma_{\text{col-}E}^{\text{inter}}$ is shown in \figref{sigma_colE}, where it is evident that $\sigma_{\text{col-}E}^{\text{inter}}$ near $\omega = 0$ is proportional to $\tau^2$.
As detailed in \appref{U(2)_formula}, $\sigma_{\text{col-}E}^{\text{inter}}$ contains a mixed dipole term $\sigma_{\text{MD}}$ expressed as
    \begin{align}
        \sigma_{\text{MD,C}}^{y;\nu x} &=\frac{Ji\omega}{\omega^2+(1/\tau)^2}\int\frac{d\bm{k}}{(2\pi)^d}\sum_{a\neq b}\partial_{x}\operatorname{Im}[\mathcal{A}_{ab}^{y}\mathcal{S}_{ba}^{\nu}]f_{a},\\
    \sigma_{\text{MD,L}}^{y;\nu x} &=\frac{J/\tau}{\omega^2+(1/\tau)^2}\int\frac{d\bm{k}}{(2\pi)^d}\sum_{a\neq b}\partial_{x}\operatorname{Im}[\mathcal{A}_{ab}^{y}\mathcal{S}_{ba}^{\nu}]f_{a}.
    \end{align}
Here, the $\bm{\mathcal{A}}_{ab}$ is the U(2) gauge Berry connection and the $\bm{\mathcal{S}}_{ab}$ is the interband spin operator in the U(2) gauge. 
Considering the double degeneracy, we introduce the U(2) gauge description of the photocurrent response (see \appref{U(2)_formula}).
This term originates from the second-order response perturbed by one photon and one localized spin dynamics, $\text{L}^y$, as summarized in \tabref{photocurrent_classification_inter}.
The former term, $\sigma_{\text{MD,C}}$, is induced by the out-of-phase component of $E^x$ and $\text{L}^y$ and vanishes at $\omega=0$. 
In contrast, the latter term, $\sigma_{\text{MD,L}}$, arises from the in-phase component of $E^x$ and $\text{L}^y$ and shows $\tau^1$-dependence in the low-frequency regime.
As discussed earlier, the linear electromagnetic susceptibility of the collective spin mode, $\operatorname{Re}\chi_{\text{L}^yE^x}$, is proportional to $\tau$. 
Consequently, the mixed dipole term $\sigma_{\text{MD,L}}$ is proportional to $\tau^2$ as
\begin{align}
    \sigma_{\text{col-}E}^{\text{MD}}(0;\omega)&=\sigma_{\text{MD;L}}(0;\omega)\operatorname{Re}\chi_{\text{L}^yE^x}(-\omega),\nonumber\\
    &\propto\tau^2.
\end{align}
In comparison, other terms in $\sigma_{\text{S}E}$ and $\sigma_{E\text{S}}$ are at most proportional to $\tau^0$, leading to an NHE with a linear dependence on $\tau$ in the highest order. These contributions are, therefore, negligible compared to the mixed dipole term.
Unlike the NHE contributions from $\sigma_{\text{SS}}$, the mixed dipole term $\sigma_{\text{MD;L}}$ does not include a suppression factor that reduces its amplitude $\propto1/(\omega-\epsilon_g)$. As a result, even in systems with a large optical gap, the contribution from the mixed dipole term can be significant.

\section{Summary}\label{summary}
In this study, we investigated the effect of localized spin dynamics on NHE in a $\mathcal{PT}$-symmetric collinear antiferromagnet by using real-time simulations. First, we identified the optically active mode of the localized spin system and identified the significant role of the Edelstein effect in the low-frequency regime. Our findings reveal that collective spin dynamics significantly enhance the NHE near $\omega = 0$.
To clarify the role of localized spin dynamics in the NHE, we decomposed the NHE into three components and found that $\sigma_{\text{col-}E}$ and $\sigma_{\text{col-col}}$ contribute significantly.
Furthermore, to determine the dominant terms in the interference contributions $\sigma_{\text{col-}E}$, we decomposed the NHE into intraband and interband components.
Through symmetry analysis, we classified the photocurrent response originating from the interplay of the external light field and localized spin dynamics.
Our results show that the mixed dipole term is dominant for the interference contributions and is proportional to $\tau^1$.
This is consistent with the analytical formulas. 
This term is further enhanced by the Edelstein effect, making its contribution to the NHE substantial as large as $\tau^2$ in the clean limit.
These insights highlight the pivotal role of the mixed-dipole mechanism in the NHE of $\mathcal{PT}$-symmetric antiferromagnetic metals, where the localized spin dynamics can be driven electrically via the Edelstein effect.
Experimental identification of the NHE induced by the Edelstein effect is an important challenge. 
The NHE from localized spin dynamics could be significant in electrically switchable antiferromagnets such as CuMnAs and $\text{Mn}_2\text{Au}$. 
Our findings reveal another avenue for understanding the mechanisms of the NHE and suggest a promising direction for future research in this area.

\section*{acknowledgement}
K.H. thanks T. Morimoto and S. Hayami for helpful discussion.
This work is supported by Grant-in-Aid for Scientific Research from JSPS, KAKENHI Grant No.~JP23K13058 (H.W.), No.~JP24K00581 (H.W.), No.~JP25H02115 (H.W.), No.~21H04990 (R.A.), No.~JP25H01246 (R.A.), No. 25H01252 (R.A.), JST-CREST No.~JPMJCR23O4(R.A.), JST-ASPIRE No.~JPMJAP2317 (R.A.), JST-Mirai No.~JPMJMI20A1 (R.A.). K.H. was supported by the Program for Leading Graduate Schools (MERIT-WINGS).
This work was supported by the RIKEN TRIP initiative (RIKEN Quantum, Advanced General Intelligence for Science Program, Many-body Electron Systems).

\appendix

\onecolumngrid

\section{Nonlinear optical response functions based on density matrix formalism}\label{pve_conductivity}
In this section, we formulate the perturbative expansion of the von Neumann equation \Eqref{vonNeumann} incorporating both the light field and light-induced spin dynamics.
The electronic Hamiltonian is expressed as 
\begin{align}
        \mathcal{H}(t) = \mathcal{H}_{0} + \Delta\mathcal{H}(t).
\end{align}
$\Delta\mathcal{H}(t)$ represents the perturbative Hamiltonian, comprising an external light field and the light-induced spin dynamics, defined as
\begin{align}
        \Delta\mathcal{H}(t) &= \Delta\mathcal{H}_{E}(t) + \Delta\mathcal{H}_{\text{spin}}(t), \\
        \Delta\mathcal{H}_{E} &= -r^{\mu}E^{\mu}, \label{lengthgauge}\\
        \Delta\mathcal{H}_{\text{spin}}(t) &= -JB^{\mu}\Delta\mathrm{S}^{\mu}(t), \\
        \Delta\vb{S}(t) &= \qty(\Delta\mathrm{L}^{y}(t),\Delta\mathrm{M}^{z}(t))^{T}, \\
        \vb*{B} &=  \qty(\sigma_{y}\tau_{z},\sigma_{z}\tau_{0})^{T}.
\end{align}
Here, $\sigma^{\mu}$ and $\tau^{\mu}$ denote the Pauli matrices representing the spin and sublattice degrees of freedom, respectively.
$\mathcal{H}_{0}$ is the nonperturbative Hamiltonian expressed as
\begin{align}
        \mathcal{H}_{0} = \int \dfrac{d\vb*{k}}{(2\pi)^{d}}\sum_{a}\epsilon_{\vb*{k}a}c_{\vb*{k}a}^{\dagger}c_{\vb*{k}a}.
\end{align}
Here, $c_{\vb*{k}a}^{\dagger}$ ($c_{\vb*{k}a}$) represents the creation (annihilation) operator for the Bloch state $\ket{\psi_{\vb*{k}a}} = \exp(i\vb*{k} \cdot \vb{r})\ket{u_{\vb*{k}a}}$, where $\ket{u_{\vb*{k}a}}$ is the cell-periodic part of the Bloch state and satisfies the equation
\begin{align}
        H_{0}(\vb*{k})\ket{u_{\vb*{k}a}} = \epsilon_{\vb*{k}a}\ket{u_{\vb*{k}a}}.
\end{align}
According to \cite{ADAMS1959286}, in the infinite volume limit, the position operator in \Eqref{lengthgauge} is expressed in the Bloch basis as
\begin{align}
        \qty[\vb*{r_{k}}]_{ab} = i\nabla_{\vb{k}}\delta_{ab} + \vb*{\xi}_{ab}. \label{position_operator}
\end{align}
Here, $\vb*{\xi}_{ab} = i\mel{u_{\vb*{k}a}}{\nabla_{\vb*{k}}}{u_{\vb*{k}b}}$ is the Berry connection. In the following calculation, $k$-dependence is omitted unless explicitly mentioned.

As we explained in the main text, we can write the von Neumann equation of SPDM $\rho_{ab} = \langle c_{\vb*{k}b}^{\dagger}c_{\vb*{k}a}\rangle $ as 
\begin{align}
        i\pdv{\rho_{ab}(t)}{t} &= \qty[\mathcal{H}, \rho]_{ab} =\qty[\mathcal{H}_{0}, \rho]_{ab} + \qty[\Delta\mathcal{H}(t), \rho]_{ab}.
\end{align}
We define the Fourier transformation as
\begin{align}
        f(t) = \int \dfrac{d\omega}{2\pi}f(\omega)e^{-i(\omega + i\eta)t},
\end{align}
and we obtain the frequency representation of the von Neumann equation,
\begin{align}
       (\omega + i\eta - \epsilon_{ab})\rho_{ab}(\omega) =  \int \dfrac{d\omega_{1}}{2\pi}\qty[\Delta \mathcal{H}(\omega_{1}), \rho(\omega - \omega_{1})]_{ab}.
\end{align}
Here, the infinitesimal parameter $\eta$ is introduced to describe the adiabatic application of the external field and $\epsilon_{ab}=\epsilon_{\bm{k}a}-\epsilon_{\bm{k}b}$.
We can perform a perturbative expansion of this equation by introducing $\rho^{(n)}$, which represents the $n$-th order term with respect to the external field.
\begin{align}
        (\omega + i\eta - \epsilon_{ab})\rho_{ab}^{(n+1)}(\omega) = \int\dfrac{d\omega_{1}}{2\pi}\qty[\Delta \mathcal{H}(\omega_{1}), \rho^{(n)}(\omega - \omega_{1})]_{ab}  \label{vonNeumann_omega}.
\end{align}
We introduce the matrix $d(\omega)$ as 
\begin{align}
        d_{ab}(\omega) = \dfrac{1}{\omega + i\eta - \epsilon_{ab}}
        = \mathcal{P}\dfrac{1}{\omega-\epsilon_{ab}} - i\pi\delta(\omega-\epsilon_{ab}), \label{Dirac_formula}
\end{align}
and by using the Hadamard product $(A \odot B)_{ab} = A_{ab}B_{ab}$ \cite{Ventura2017}, we can rewrite the 
von Neumann equation \Eqref{vonNeumann_omega} as 
\begin{align}
        \rho_{ab}^{(n+1)}(\omega) = \int \dfrac{d\omega_{1}}{2\pi}\qty(d(\omega) \odot [\Delta \mathcal{H}(\omega_{1}), \rho^{(n)}(\omega - \omega_{1})])_{ab}.
\end{align}
In the following calculation, we solve this equation with the condition of $\rho_{ab}^{(0)}(\omega) = 2\pi\delta(\omega)f_{ab}\delta_{ab}$.

Since we focus on the second-order current response to the perturbation, we solve the equation
\begin{align}
        \begin{split}
            \rho_{ab}^{(2)}(\omega) &= \int \dfrac{d\omega_{1}}{2\pi}\qty(d(\omega) \odot [\Delta \mathcal{H}(\omega_{1}), \rho^{(1)}(\omega - \omega_{1})])_{ab} \\
        &= \int \dfrac{d\omega_{1} d\omega_{2}}{(2\pi)^{2}}\qty(d(\omega) \odot [\Delta \mathcal{H}(\omega_{1}), d(\omega - \omega_{1})\odot [\Delta \mathcal{H}(\omega_{2}), \rho^{(0)}(\omega-\omega_{1}-\omega_{2})]])_{ab}.
        \end{split}
\end{align}
Since we consider the perturbation to involve both the light field and the spin dynamics, we can devide $\rho_{ab}^{(2)}(\omega)$ into three components as, 
\begin{align}
        \rho_{ab}^{(2)}(\omega) = \rho_{EE, ab}(\omega) + \rho_{E\mathrm{S}, ab}(\omega) + \rho_{\mathrm{S}\mathrm{S}, ab}(\omega),
\end{align}
where
\begin{align}
        \rho_{EE,ab}(\omega) &= \dfrac{1}{2}\int \dfrac{d\omega_{1} d\omega_{2}}{(2\pi)^{2}}E^{\nu}(\omega_{1})E^{\lambda}(\omega_{2})\qty(d(\omega) \odot [r^{\nu}, d(\omega - \omega_{1})\odot [r^{\lambda}, \rho^{(0)}(\omega-\omega_{1}-\omega_{2})]])_{ab} + [(\nu, \omega_{1}) \leftrightarrow (\lambda, \omega_{2})], \\
            \rho_{E\mathrm{S}, ab}(\omega) &= \dfrac{J}{2}\int \dfrac{d\omega_{1} d\omega_{2}}{(2\pi)^{2}}E^{\nu}(\omega_{1})\mathrm{S}^{\lambda}(\omega_{2})\qty(d(\omega) \odot [r^{\nu}, d(\omega - \omega_{1})\odot [B^{\lambda}, \rho^{(0)}(\omega-\omega_{1}-\omega_{2})]])_{ab}, \\
        \rho_{\mathrm{S}E, ab}(\omega) &= \dfrac{J}{2}\int \dfrac{d\omega_{1} d\omega_{2}}{(2\pi)^{2}}\mathrm{S}^{\lambda}(\omega_{1})E^{\nu}(\omega_{2})\qty(d(\omega) \odot [B^{\lambda}, d(\omega - \omega_{1})\odot [r^{\nu}, \rho^{(0)}(\omega-\omega_{1}-\omega_{2})]])_{ab},\\
        \rho_{\mathrm{S}\mathrm{S}, ab}(\omega) &= \dfrac{J^{2}}{2}\int \dfrac{d\omega_{1} d\omega_{2}}{(2\pi)^{2}}\mathrm{S}^{\nu}(\omega_{1})\mathrm{S}^{\lambda}(\omega_{2})\qty(d(\omega) \odot [B^{\nu}, d(\omega - \omega_{1})\odot [B^{\lambda}, \rho^{(0)}(\omega-\omega_{1}-\omega_{2})]])_{ab} + [(\nu, \omega_{1}) \leftrightarrow (\lambda, \omega_{2})].
\end{align}
Based on this decomposition, we formulate the photocurrent formula in the presence of spin dynamics.

\subsection{Light field induced photocurrent}
Here, we derive the photocurrent induced by the light field based on $\rho_{EE}$. To facilitate the analysis, the position operator in \Eqref{position_operator} is decomposed into intra-band ($\vb*{r}_{i}$) and inter-band ($\vb*{r}_{e}$) component as
\begin{align}
        \begin{split}
            \qty(\vb*{r}_{i})_{ab} &= \delta_{ab}\qty(i\nabla_{\vb{k}} + \bm{\xi}_{aa}), \\
            \qty(\vb*{r}_{e})_{ab} &= \qty(1-\delta_{ab})\bm{\xi}_{ab}.  
        \end{split}    
\end{align}
Based on this decomposition, we can classify $\rho_{EE}(\omega)$ into the following four component, 
\begin{align}
        \rho_{EE,ab}(\omega) = \rho_{EE, ab}^{(ii)} + \rho_{EE,ab}^{(ie)} + \rho_{EE, ab}^{(ei)} + \rho_{EE,ab}^{(ee)},
\end{align}
where each term is explicitly written as,
\begin{align}
        \begin{split}
            \rho_{EE, ab}^{(ii)} &= \dfrac{1}{2}\int \dfrac{d\omega_{1} d\omega_{2}}{(2\pi)^{2}}E^{\nu}(\omega_{1})E^{\lambda}(\omega_{2})\qty(d(\omega) \odot [r_{i}^{\nu}, d(\omega - \omega_{1})\odot [r_{i}^{\lambda}, \rho^{(0)}(\omega-\omega_{1}-\omega_{2})]])_{ab} + [(\nu, \omega_{1}) \leftrightarrow (\lambda, \omega_{2})] \\
            &= -\dfrac{1}{2}\int \dfrac{d\omega_{1}d\omega_{2}}{(2\pi)^{2}}E^{\nu}(\omega_{1})E^{\lambda}(\omega_{2})d_{ab}(\omega)d_{ab}(\omega-\omega_{1})\partial_{\nu}\partial_{\lambda}f(\epsilon_{\vb*{k}a})\delta_{ab}2\pi\delta(\omega-\omega_{1}-\omega_{2}) + [(\nu, \omega_{1}) \leftrightarrow (\lambda, \omega_{2})], 
        \end{split} \\
        \begin{split}
            \rho_{EE, ab}^{(ie)} &= \dfrac{1}{2}\int \dfrac{d\omega_{1} d\omega_{2}}{(2\pi)^{2}}E^{\nu}(\omega_{1})E^{\lambda}(\omega_{2})\qty(d(\omega) \odot [r_{i}^{\nu}, d(\omega - \omega_{1})\odot [r_{e}^{\lambda}, \rho^{(0)}(\omega-\omega_{1}-\omega_{2})]])_{ab} + [(\nu, \omega_{1}) \leftrightarrow (\lambda, \omega_{2})]  \\
            &= -\dfrac{i}{2}\int\dfrac{d\omega_{1}\omega_{1}}{(2\pi)^{2}}E^{\nu}(\omega_{1})E^{\lambda}(\omega_{2})d_{ab}(\omega)\qty[\partial_{\nu}\qty(d_{ab}(\omega-\omega_{1})\xi_{ab}^{\lambda}f_{ab})-i\qty(\xi_{aa}^{\nu} - \xi_{bb}^{\nu})d_{ab}(\omega-\omega_{1})\xi_{ab}^{\lambda}f_{ab}] + [(\nu, \omega_{1}) \leftrightarrow (\lambda, \omega_{2})]
        \end{split},\\
        \begin{split}
            \rho_{EE, ab}^{(ei)} &= \dfrac{1}{2}\int \dfrac{d\omega_{1} d\omega_{2}}{(2\pi)^{2}}E^{\nu}(\omega_{1})E^{\lambda}(\omega_{2})\qty(d(\omega) \odot [r_{e}^{\nu}, d(\omega - \omega_{1})\odot [r_{i}^{\lambda}, \rho^{(0)}(\omega-\omega_{1}-\omega_{2})]])_{ab} + [(\nu, \omega_{1}) \leftrightarrow (\lambda, \omega_{2})]  \\
            &=-\dfrac{i}{2}\int \dfrac{d\omega_{1}d\omega_{2}}{(2\pi)^{2}}E^{\nu}(\omega_{1})E^{\lambda}(\omega_{2})d_{ab}(\omega)d_{aa}(\omega-\omega_{1})\xi_{ab}^{\nu}\partial_{\lambda}f_{ab}2\pi\delta(\omega -\omega_{1} - \omega_{2})+ [(\nu, \omega_{1}) \leftrightarrow (\lambda, \omega_{2})],    
        \end{split},\\
        \begin{split}
            \rho_{EE, ab}^{(ee)} &= \dfrac{1}{2}\int \dfrac{d\omega_{1} d\omega_{2}}{(2\pi)^{2}}E^{\nu}(\omega_{1})E^{\lambda}(\omega_{2})\qty(d(\omega) \odot [r_{e}^{\nu}, d(\omega - \omega_{1})\odot [r_{e}^{\lambda}, \rho^{(0)}(\omega-\omega_{1}-\omega_{2})]])_{ab} + [(\nu, \omega_{1}) \leftrightarrow (\lambda, \omega_{2})] \\
        &= \dfrac{1}{2}\sum_{c}\int \dfrac{d\omega_{1} d\omega_{2}}{(2\pi)^{2}}E^{\nu}(\omega_{1})E^{\lambda}(\omega_{2})d_{ab}(\omega)\qty[d_{cb}(\omega -\omega_{1})\xi_{ac}^{\nu}\xi_{cb}^{\lambda}f_{bc} - d_{ac}(\omega - \omega_{1})\xi_{cb}^{\nu}\xi_{ac}^{\lambda}f_{ca}]2\pi\delta(\omega-\omega_{1}-\omega_{2}) \\
        & \hspace{12cm}+ [(\nu, \omega_{1})\leftrightarrow (\lambda, \omega_{2})].
        \end{split}
\end{align}

Using $\rho_{EE,ab}$, we can obtain the second-order current response as
\begin{align}
        \begin{split}
            J^{\mu}_{EE}(\omega) &= \int \dfrac{d\vb*{k}}{(2\pi)^{d}}\sum_{abc}J_{ab}^{\mu}\rho_{EE, ba}(\omega) \\
        &\eqqcolon  \int\dfrac{d\omega_{1}d\omega_{2}}{(2\pi)^{2}}{\sigma}^{\mu;\nu\lambda}_{EE}(\omega, \omega_{1}, \omega_{2})E^{\nu}(\omega_{1})E^{\lambda}(\omega_{2})2\pi\delta(\omega - \omega_{1} - \omega_{2}).
        \end{split}
\end{align}
As the SPDM can be divided into four contributions, we can also decompose $\sigma_{EE}^{\mu;\nu\lambda}(\omega, \omega_{1}, \omega_{2})$ into the corresponding four components, defined as
\begin{align}
    \sigma_{EE}^{\mu;\nu\lambda} = \sigma_{EE, (ii)}^{\mu;\nu\lambda} + \sigma_{EE, (ie)}^{\mu;\nu\lambda} + \sigma_{EE, (ei)}^{\mu;\nu\lambda} + \sigma_{EE, (ee)}^{\mu;\nu\lambda}.
\end{align}
We can obtain the photocurrent conductivity by setting 
\begin{align}
        \omega =0 , \omega_{1} = -\Omega, \omega_{2} = \Omega.
\end{align}
In the following calculation, we will derive the Drude term, Berry curvature dipole term, injection current term, injection current term, and intrinsic Fermi surface term.
\subsection{Fermi surface effect I: Drude term}
Firstly, we focus on $\sigma_{EE, (ii)}^{\mu;\nu\lambda}$ as
\begin{align}
    \sigma_{EE,(ii)}^{\mu;\nu\lambda}(0;-\Omega,\Omega)&=\frac{1}{2}\int\frac{d\bm{k}}{(2\pi)^d}\sum_{a}-v_{aa}^{\mu}d_{aa}^{0}d_{aa}^{\Omega}\partial_{\nu}\partial_{\lambda}f(\epsilon_{\bm{k}a})+\left[(\nu,-\Omega)\leftrightarrow(\lambda,\Omega)\right],\\
    &=-\frac{1}{2i\eta}\int\frac{d\bm{k}}{(2\pi)^d}\sum_{a}v_{aa}^{\mu}\left(\frac{1}{\Omega+i\eta}+\frac{1}{-\Omega+i\eta}\right)\partial_{\nu}\partial_{\lambda}f(\epsilon_{\bm{k}a}),\\
    &=\frac{1}{\Omega^2+\eta^2}\int\frac{d\bm{k}}{(2\pi)^d}\sum_{a}v_{aa}^{\mu}\partial_{\nu}\partial_{\lambda}f(\epsilon_{\bm{k}a}).
\end{align}
By replacing $\eta$ with $1/\tau$ phenomenologically, we can obtain the Drude term formula as
\begin{align}
    \sigma^{\mu;\nu\lambda}_{\mathrm{D}}(0;-\Omega,\Omega)=\frac{1}{\Omega^2+(1/\tau)^2}\int\frac{d\bm{k}}{(2\pi)^d}\sum_{a}\partial_{\mu}\partial_{\nu}\partial_{\lambda}\epsilon_{\bm{k}a}f(\epsilon_{\bm{k}a}).
\end{align}
This term shows the $\tau^2$-dependence at the peak $\omega=0$. In $\mathcal{T}$-symmetric system, $\sigma^{\mu;\nu\lambda}_{\mathrm{D}}(0;-\Omega,\Omega)$ vanishes because $\partial_{\mu}\partial_{\nu}\partial_{\lambda}\epsilon_{\bm{k}a}f(\epsilon_{\bm{k}a})$ is odd function of $\bm{k}$. 
On the other hand, the breaking of the $\mathcal{P}$ and $\mathcal{T}$-symmetries allows the band dispersion to be antisymmetric between $\bm{k}$, therefore $\sigma^{\mu;\nu\lambda}_{\mathrm{D}}$ can be finite.
The Drude term comes from the diagonal part of the current operator in the output vertex.

\subsection{Fermi surface effect II: Berry curvature dipole}
Secondly, we derive the Berry curvature dipole term from $\sigma_{EE, (ei)}^{\mu;\nu\lambda}$.
\begin{align}
    \sigma_{EE,(ei)}^{\mu;\nu\lambda}(0;-\Omega,\Omega)&=\frac{1}{2(\Omega+i\eta)}\int\frac{d\bm{k}}{(2\pi)^d}\sum_{a\neq b}\xi_{ab}^{\mu}\xi_{ba}^{\nu}\partial_{\lambda}f_{ba}+\left[(\nu,-\Omega)\leftrightarrow(\lambda,\Omega)\right],\notag\\ 
    &=\frac{1}{2(\Omega+i\eta)}\int\frac{d\bm{k}}{(2\pi)^d}\sum_{a\neq b}(\xi_{ba}^{\mu}\xi_{ab}^{\nu}-\xi_{ab}^{\mu}\xi_{ba}^{\nu})\partial_{\lambda}f_{a}+\left[(\nu,-\Omega)\leftrightarrow(\lambda,\Omega)\right],\notag\\
    &=-\frac{i\Omega+\eta}{\Omega^2+\eta^2}\int\frac{d\bm{k}}{(2\pi)^d}\sum_{a\neq b}\operatorname{Im}[\xi_{ab}^{\mu}\xi_{ba}^{\nu}]\partial_{\lambda}f_{a}+\left[(\nu,-\Omega)\leftrightarrow(\lambda,\Omega)\right],\notag\\ 
    &=\frac{i\Omega}{\Omega^2+\eta^2}\int\frac{d\bm{k}}{(2\pi)^d}\sum_{a\neq b}(\partial_{\lambda}\operatorname{Im}[\xi_{ab}^{\mu}\xi_{ba}^{\nu}]-\partial_{\nu}\operatorname{Im}[\xi_{ab}^{\mu}\xi_{ba}^{\lambda}])f_{a}\nonumber\\
    &+\frac{\eta}{\Omega^2+\eta^2}\int\frac{d\bm{k}}{(2\pi)^d}\sum_{a\neq b}(\partial_{\lambda}\operatorname{Im}[\xi_{ab}^{\mu}\xi_{ba}^{\nu}]+\partial_{\nu}\operatorname{Im}[\xi_{ab}^{\mu}\xi_{ba}^{\lambda}])f_{a}.
\end{align}
In the $\mathcal{PT}$-symmetric system, the Berry curvature dipole term vanishes because the Berry curvature vanishes at every wave vector. 
By replacing the $\eta$ with $1/\tau$, we can obtain the formula for the Berry curvature dipole term as
\begin{align}
    \sigma_{\mathrm{BCD;C}}^{\mu;\nu\lambda}(0;-\Omega,\Omega)&=\frac{i\Omega}{\Omega^2+1/\tau^2}\int\frac{d\bm{k}}{(2\pi)^d}\sum_{a\neq b}(\partial_{\lambda}\operatorname{Im}[\xi_{ab}^{\mu}\xi_{ba}^{\nu}]-\partial_{\nu}\operatorname{Im}[\xi_{ab}^{\mu}\xi_{ba}^{\lambda}])f_{a},\\
    \sigma_{\mathrm{BCD;L}}^{\mu;\nu\lambda}(0;-\Omega,\Omega)&=\frac{1/\tau}{\Omega^2+1/\tau^2}\int\frac{d\bm{k}}{(2\pi)^d}\sum_{a\neq b}(\partial_{\lambda}\operatorname{Im}[\xi_{ab}^{\mu}\xi_{ba}^{\nu}]+\partial_{\nu}\operatorname{Im}[\xi_{ab}^{\mu}\xi_{ba}^{\lambda}])f_{a}.
\end{align}
The former term can be induced by circularly polarized light, while the latter term can be induced by non-circularly polarized light and exhibits $\tau$-dependence when $\Omega = 0$. The Berry curvature dipole term originates from the off-diagonal component of the current operator $J^{\mu}$ in the output vertex.

\subsection{Interband effect I: Injection current}
Here, we focus on $\sigma_{EE, (ee)}^{\mu;\nu\lambda}$ with the diagonal component of the current operator in the band basis denoted as $\sigma_{EE, (ee;d)}^{\mu;\nu\lambda}$;
\begin{align}
        \begin{split}
            &\sigma^{\mu;\nu\lambda}_{EE,(ee;d)}(\omega,\omega_{1},\omega_{2}) \\
        &= \dfrac{1}{2}\int \dfrac{d\vb*{k}}{(2\pi)^{d}}\sum_{a\neq c}J_{aa}^{\mu}d_{aa}(\omega)\qty[d_{ca}(\omega-\omega_{1})\xi^{\nu}_{ac}\xi^{\lambda}_{ca}f_{ac} - d_{ac}(\omega-\omega_{1})\xi_{ca}^{\nu}\xi_{ac}^{\lambda}f_{cb}] + [(\nu, \omega_{1}) \leftrightarrow (\lambda, \omega_{2})] \\
        &= \dfrac{1}{2}\dfrac{1}{\omega+i\eta}\int \dfrac{d\vb*{k}}{(2\pi)^{d}}\sum_{a\neq c}(J_{aa}^{\mu} - J_{cc}^{\mu})\qty[d_{ca}(\omega-\omega_{1})\xi^{\nu}_{ac}\xi^{\lambda}_{ca}f_{ac}] + [(\nu, \omega_{1}) \leftrightarrow (\lambda, \omega_{2})] \\
        &= \dfrac{1}{2}\dfrac{1}{\omega+i\eta}\int \dfrac{d\vb*{k}}{(2\pi)^{d}}\sum_{a\neq c}(J_{aa}^{\mu} - J_{cc}^{\mu})\xi^{\nu}_{ac}\xi^{\lambda}_{ca}f_{ac}\qty[d_{ca}(\omega-\omega_{1}) + d_{ac}(\omega-\omega_{2})] \\
        &= \dfrac{1}{2}\dfrac{1}{\omega+i\eta}\int \dfrac{d\vb*{k}}{(2\pi)^{d}}\sum_{a\neq c}\Delta_{ac}^{\mu}\xi^{\nu}_{ac}\xi^{\lambda}_{ca}f_{ac}\qty[d_{ca}(\omega-\omega_{1}) + d_{ac}(\omega-\omega_{2})].
        \end{split}
\end{align}
Here, $\Delta_{ac}^{\mu} = J_{aa}^{\mu} - J_{cc}^{\mu} = \partial_{\mu}\epsilon_{\bm{k}a} - \partial_{\mu}\epsilon_{\bm{k}b}$ is the velocity difference matrix.
In the $\omega \to 0$ limit, the expression diverges because of the factor of $1/(\omega+i\eta)$. To eliminate this unphysical divergence, we introduce the finite sum frequency $\omega_1+\omega_2=\delta$ and will take the limit $\delta\rightarrow0$.
\begin{align}
        \sigma^{\mu;\nu\lambda}_{EE,(ee;d)}(0,-\Omega,\Omega)
    &= \lim_{\delta\rightarrow0}\dfrac{1}{2}\dfrac{1}{\delta+i\eta}\int \dfrac{d\vb*{k}}{(2\pi)^{d}}\sum_{a\neq b}\Delta_{ab}^{\mu}\xi^{\nu}_{ab}\xi^{\lambda}_{ba}f_{ab}\qty[d_{ba}(\Omega) + d_{ab}(-\Omega)],\\ \nonumber
    &=\lim_{\delta\rightarrow0}\dfrac{1}{2}\dfrac{1}{\delta+i\eta}\int \dfrac{d\vb*{k}}{(2\pi)^{d}}\sum_{a\neq b}\Delta_{ab}^{\mu}\xi^{\nu}_{ab}\xi^{\lambda}_{ba}f_{ab}\qty[\frac{1}{\Omega+\frac{\delta}{2}+i\eta-\epsilon_{ba}} + \frac{1}{-\Omega+\frac{\delta}{2}+i\eta+\epsilon_{ba}}].
\end{align}
Here, we perform the Taylor expansion,
\begin{align}
    \frac{1}{\Omega+\frac{\delta}{2}+i\eta-\epsilon_{ba}}+\frac{1}{-\Omega+\frac{\delta}{2}+i\eta+\epsilon_{ba}} =2\operatorname{Im}\frac{1}{\Omega+i\eta-\epsilon_{ba}}-2\operatorname{Re}\frac{\delta/2}{(\Omega+i\eta-\epsilon_{ba})^2}+\mathcal{O}(\delta^2).
\end{align}
By using this method and replacing the $\eta$ with $1/\tau$, we can obtain the formulation of the injection current and intrinsic Fermi surface effect as
\begin{align}
    \sigma^{\mu;\nu\lambda}_{EE,(ee;d)}(0,-\Omega,\Omega)&=\sigma_{EE,\mathrm{inj}}^{\mu;\nu\lambda}(0,-\Omega,\Omega)+\sigma_{EE,\mathrm{IFSI}}^{\mu;\nu\lambda}(0,-\Omega,\Omega),
\end{align}
here,
\begin{align}
        \sigma_{EE,\text{inj}}^{\mu;\nu\lambda}(0,-\Omega,\Omega) 
        &= \pi\tau\int \dfrac{d\vb*{k}}{(2\pi)^{d}}\sum_{a\neq c}\Delta_{ac}^{\mu}\xi^{\nu}_{ac}\xi^{\lambda}_{ca}f_{ac}\delta(\Omega-\epsilon_{ca}),\\
        \sigma_{EE,\mathrm{IFSI}}^{\mu;\nu\lambda}(0,-\Omega,\Omega)&=\frac{1}{2}\int\dfrac{d\vb*{k}}{(2\pi)^{d}}\sum_{a\neq c}\xi^{\nu}_{ac}\xi^{\lambda}_{ca}f_{ac}\partial_{\mu}\mathrm{P}\frac{1}{\Omega-\epsilon_{ca}}.
\end{align}
Here, we decompose the $\sigma_{EE,\text{inj}}^{\mu;\nu\lambda}$ into real and imaginary parts as
\begin{align}
        \sigma_{EE, \text{Inj;M}}^{\mu;\nu\lambda} &= \pi\tau\int \dfrac{d\vb*{k}}{(2\pi)^{d}}\sum_{a\neq c}\Delta_{ac}^{\mu}\operatorname{Re}\qty[\xi^{\nu}_{ac}\xi^{\lambda}_{ca}]f_{ac}\delta(\Omega-\epsilon_{ca}),\\
        \sigma_{EE, \text{Inj;E}}^{\mu;\nu\lambda} &= i\pi\tau\int \dfrac{d\vb*{k}}{(2\pi)^{d}}\sum_{a\neq c}\Delta_{ac}^{\mu}\operatorname{Im}\qty[\xi^{\nu}_{ac}\xi^{\lambda}_{ca}]f_{ac}\delta(\Omega-\epsilon_{ca}).
\end{align}
$\sigma_{\text{Inj;M}}$ ($\sigma_{\text{Inj;E}}$) is called magnetic (electric) injection current, which is induced by linearly (circularly) polarized light. 
We can also decompose the $\sigma_{EE,\text{IFSI}}^{\mu;\nu\lambda}$ into real and imaginary parts as
\begin{align}
    \sigma_{EE,\mathrm{IFSI;M}}^{\mu;\nu\lambda}(0,-\Omega,\Omega)&=\frac{1}{2}\int\dfrac{d\vb*{k}}{(2\pi)^{d}}\sum_{a\neq c}\operatorname{Re}[\xi^{\nu}_{ac}\xi^{\lambda}_{ca}]f_{ac}\partial_{\mu}\mathrm{P}\frac{1}{\Omega-\epsilon_{ca}},\\
    \sigma_{EE,\mathrm{IFSI;E}}^{\mu;\nu\lambda}(0,-\Omega,\Omega)&=\frac{i}{2}\int\dfrac{d\vb*{k}}{(2\pi)^{d}}\sum_{a\neq c}\operatorname{Im}[\xi^{\nu}_{ac}\xi^{\lambda}_{ca}]f_{ac}\partial_{\mu}\mathrm{P}\frac{1}{\Omega-\epsilon_{ca}}.
\end{align}
$\sigma_{\mathrm{IFSI;M}}$ ($\sigma_{\mathrm{IFSI;E}}$) is called magnetic (electric) intrinsic Fermi surface term I, which is induced by linearly (circularly) polarized light. 
The photocurrent responses that we derived in this section come from the diagonal part of the current operator $J^{\mu}$ in the output vertex.

\subsection{Interband effect II: Shift current and intrinsic Fermi surface effect}
Here, we derive the shift current term and intrinsic Fermi surface term based from $\sigma_{EE, (ee)}^{\mu;\nu\lambda}$ and $\sigma_{EE, (ie)}^{\mu;\nu\lambda}$. 
The photocurrent responses that we derive in this section come from the off-diagonal part of the current operator $J^{\mu}$ in the output vertex.
First, we focus on $\sigma_{EE, (ee)}^{\mu;\nu\lambda}$ with the off-diagonal component of the velocity operator defined as $\sigma^{\mu;\nu\lambda}_{EE, (ee;o)}$;
\begin{align}
        \begin{split}
            &\sigma^{\mu;\nu\lambda}_{EE, (ee;o)}(0, -\Omega, \Omega) \\
        &= \dfrac{1}{2}\int \dfrac{d\vb*{k}}{(2\pi)^{d}}\sum_{a\neq b\neq c}J_{ab}^{\mu}d_{ba}(0)\qty[d_{ca}(\Omega)\xi^{\nu}_{bc}\xi^{\lambda}_{ca}f_{ac} - d_{bc}(\Omega)\xi_{ca}^{\nu}\xi_{bc}^{\lambda}f_{cb}] + [(\nu, -\Omega) \leftrightarrow (\lambda, \Omega)] \\
        &= \dfrac{1}{2}\int \dfrac{d\vb*{k}}{(2\pi)^{d}}\sum_{a\neq b\neq c}i\epsilon_{ab}\xi_{ab}^{\mu}d_{ba}(0)\qty[d_{ca}(\Omega)\xi^{\nu}_{bc}\xi^{\lambda}_{ca}f_{ac} - d_{bc}(\Omega)\xi_{ca}^{\nu}\xi_{bc}^{\lambda}f_{cb}] + [(\nu, -\Omega) \leftrightarrow (\lambda, \Omega)] \\
        &= \dfrac{1}{2}\int \dfrac{d\vb*{k}}{(2\pi)^{d}}\sum_{a\neq b \neq c}i\xi_{ab}^{\mu}\qty[d_{ca}(\Omega)\xi^{\nu}_{bc}\xi^{\lambda}_{ca}f_{ac} - d_{bc}(\Omega)\xi_{ca}^{\nu}\xi_{bc}^{\lambda}f_{cb}] + [(\nu, -\Omega) \leftrightarrow (\lambda, \Omega)] \\
        &=  \dfrac{1}{2}\int \dfrac{d\vb*{k}}{(2\pi)^{d}}\sum_{a\neq b \neq c}i(\xi_{ab}^{\mu}\xi_{bc}^{\nu} - \xi_{bc}^{\mu}\xi_{ab}^{\nu})\xi_{ca}^{\lambda}f_{ac}d_{ac}(\Omega) + [(\nu, -\Omega) \leftrightarrow (\lambda, \Omega)] \\
        &= \dfrac{1}{2}\int \dfrac{d\vb*{k}}{(2\pi)^{d}}\sum_{a\neq c}\qty(\qty[D^{\mu}\xi^{\nu}]_{ac}- \qty[D^{\nu}\xi^{\mu}]_{ac})\xi_{ca}^{\lambda}f_{ac}d_{ac}(\Omega)+ [(\nu, -\Omega) \leftrightarrow (\lambda, \Omega)].
        \end{split}
\end{align}
In the last line, we introduced U(1)-covariant derivative 
\begin{align}
        \qty[D^{\mu}O]_{ab} = \pdv{O_{ab}}{k^{\mu}} - i\qty(\xi_{aa}^{\mu} - \xi_{bb}^{\mu})O_{ab},
\end{align}
and used the following formula under the condition of $a \neq c$;
\begin{align}
        \qty[D^{\mu}\xi^{\nu}]_{ac}- \qty[D^{\nu}\xi^{\mu}]_{ac} = \sum_{b}i(\xi_{ab}^{\mu}\xi_{bc}^{\nu} - \xi_{bc}^{\mu}\xi_{ab}^{\nu}).
\end{align}
Next, we focus on the $\sigma_{EE, (ie)}^{\mu;\nu\lambda}$ component;
\begin{align}
        \begin{split}
            &\sigma^{\mu;\nu\lambda}_{EE, (ie)}(0;-\Omega,\Omega) \\
        &= -\dfrac{i}{2}\int \dfrac{d\vb*{k}}{(2\pi)^{d}}\sum_{a\neq b}J_{ab}^{\mu}d_{ba}(\omega) \qty[\pdv{k^{\nu}}\qty(d_{ba}(\Omega)f_{ba}\xi_{ba}^{\lambda} - i \qty(\xi_{aa}^{\nu} - \xi_{bb}^{\nu})d_{ba}(\Omega)f_{ba}\xi_{ba}^{\lambda})] + [(\nu, -\Omega) \leftrightarrow (\lambda, \Omega)] \\
        &= -\dfrac{i}{2}\int \dfrac{d\vb*{k}}{(2\pi)^{d}}\sum_{a\neq b}i\epsilon_{ab}\xi_{ab}^{\mu}d_{ba}(\omega) \qty[\pdv{k^{\nu}}\qty(d_{ba}(\Omega)f_{ba}\xi_{ba}^{\lambda} - i \qty(\xi_{aa}^{\nu} - \xi_{bb}^{\nu})d_{ba}(\Omega)f_{ba}\xi_{ba}^{\lambda})]+ [(\nu, -\Omega) \leftrightarrow (\lambda, \Omega)]  \\
        &= \dfrac{1}{2}\int \dfrac{d\vb*{k}}{(2\pi)^{d}}\sum_{a\neq b}\xi_{ab}^{\mu} \qty[\pdv{k^{\nu}}\qty(d_{ba}(\Omega)f_{ba}\xi_{ba}^{\lambda}) - i \qty(\xi_{aa}^{\nu} - \xi_{bb}^{\nu})d_{ba}(\Omega)f_{ba}\xi^{\lambda}_{ba}] + [(\nu, -\Omega) \leftrightarrow (\lambda, \Omega)] \\
        &= -\dfrac{1}{2}\int \dfrac{d\vb*{k}}{(2\pi)^{d}}\sum_{a\neq b} \qty[D^{\nu}\xi^{\mu}]_{ab} d_{ba}(\Omega)f_{ba}\xi_{ba}^{\lambda}+ [(\nu, -\Omega) \leftrightarrow (\lambda, \Omega)].
        \end{split}
\end{align}
By summing up these contributions, we get
\begin{align}
        \begin{split}
            \sigma^{\mu;\nu\lambda}_{EE, (ee;o)+(ie)} &=  \sigma^{\mu;\nu\lambda}_{EE, (ee;o)}(0, -\Omega, \Omega) + \sigma^{\mu;\nu\lambda}_{EE, (ie)}(0;-\Omega,\Omega) \\
        &= \dfrac{1}{2}\int \dfrac{d\vb*{k}}{(2\pi)^{d}}\sum_{a\neq b}\qty[D^{\mu}\xi^{\nu}]_{ab}\xi_{ba}^{\lambda}f_{ab}d_{ba}(\Omega)+ [(\nu, -\Omega) \leftrightarrow (\lambda, \Omega)] \\
        &= \dfrac{1}{2}\int \dfrac{d\vb*{k}}{(2\pi)^{d}}\sum_{a\neq b}\qty[D^{\mu}\xi^{\nu}]_{ab}\xi_{ba}^{\lambda}f_{ab}d_{ba}(\Omega)+ \qty[D^{\mu}\xi^{\lambda}]_{ab}\xi_{ba}^{\nu}f_{ab}d_{ba}(-\Omega) \\
        &= \dfrac{1}{2}\int \dfrac{d\vb*{k}}{(2\pi)^{d}}\sum_{a\neq b}\qty[\qty[D^{\mu}\xi^{\nu}]_{ab}\xi_{ba}^{\lambda} + \qty[D^{\mu}\xi^{\lambda}]_{ba}\xi_{ab}^{\nu}]f_{ab}\mathcal{P}\dfrac{1}{\Omega-\epsilon_{ba}} \\
        &-\dfrac{i\pi}{2}\int \dfrac{d\vb*{k}}{(2\pi)^{d}}\sum_{a\neq b}\qty[\qty[D^{\mu}\xi^{\nu}]_{ab}\xi_{ba}^{\lambda} - \qty[D^{\mu}\xi^{\lambda}]_{ab}\xi_{ba}^{\nu}]f_{ab}\delta(\Omega-\epsilon_{ba}).
        \end{split}
\end{align}
The absorptive part with $\delta(\Omega-\epsilon_{ba})$ corresponds to the shift current conductivity, and we divide it into the real and imaginary parts as 
\begin{align}
        \sigma_{EE,\text{shift}}^{\mu;\nu\lambda} = \dfrac{\pi}{2}\int \dfrac{d\vb*{k}}{(2\pi)^{d}}\sum_{a\neq b}\operatorname{Im}\qty[\qty[D^{\mu}\xi^{\nu}]_{ab}\xi_{ba}^{\lambda} - \qty[D^{\mu}\xi^{\lambda}]_{ba}\xi_{ab}^{\nu}]f_{ab}\delta(\Omega-\epsilon_{ba}), \\
        \sigma_{EE,\text{gyro}}^{\mu;\nu\lambda} = -\dfrac{i\pi}{2}\int \dfrac{d\vb*{k}}{(2\pi)^{d}}\sum_{a\neq b}\operatorname{Re}\qty[\qty[D^{\mu}\xi^{\nu}]_{ab}\xi_{ba}^{\lambda} - \qty[D^{\mu}\xi^{\lambda}]_{ba}\xi_{ab}^{\nu}]f_{ab}\delta(\Omega-\epsilon_{ba}).
\end{align}
$\sigma_{\text{shift}}$ ($\sigma_{\text{gyro}}$) is called shift (gyration) currents, which is induced by linearly (circularly) polarized light.
On the other hand, the non-absorptive part $\mathrm{P}\frac{1}{\Omega-\epsilon_{ba}}$ corresponds to the intrinsic Fermi surface term II, and we divide it into the real and imaginary parts as
\begin{align}
    \sigma_{EE,\text{IFSI\hspace{-1.2pt}I;M}}^{\mu;\nu\lambda}&=\frac{1}{2}\int \dfrac{d\vb*{k}}{(2\pi)^{d}}\sum_{a\neq b}\partial_{\mu}\operatorname{Re}[\xi^{\nu}_{ab}\xi^{\lambda}_{ba}]f_{ab}\mathrm{P}\frac{1}{\Omega-\epsilon_{ba}},\\
    \sigma_{EE,\text{IFSI\hspace{-1.2pt}I;E}}^{\mu;\nu\lambda}&=\frac{i}{2}\int \dfrac{d\vb*{k}}{(2\pi)^{d}}\sum_{a\neq b}\partial_{\mu}\operatorname{Im}[\xi^{\nu}_{ab}\xi^{\lambda}_{ba}]f_{ab}\mathrm{P}\frac{1}{\Omega-\epsilon_{ba}}.
\end{align}
Here, we used the relation,
\begin{align}
    \qty[D^{\mu}\xi^{\nu}]_{ab}\xi_{ba}^{\lambda} + \qty[D^{\mu}\xi^{\lambda}]_{ba}\xi_{ab}^{\nu}=\partial_{\mu}[\xi_{ab}^{\nu}\xi_{ba}^{\lambda}].
\end{align}

We can get the intrinsic Fermi Surface term $\sigma_{\mathrm{IFS}}$ by combining the $\sigma_{\mathrm{IFSI}}$ and $\sigma_{\mathrm{IFSI\hspace{-1.2pt}I}}$ as
\begin{align}
    \sigma_{EE,\mathrm{IFS}}^{\mu;\nu\lambda}&=\sigma_{\mathrm{IFSI}}^{\mu;\nu\lambda}+\sigma_{\mathrm{IFSI\hspace{-1.2pt}I}}^{\mu;\nu\lambda},\nonumber\\ 
    &=-\frac{1}{2}\int \dfrac{d\vb*{k}}{(2\pi)^{d}}\sum_{a\neq b}\xi^{\nu}_{ab}\xi^{\lambda}_{ba}\partial_{\mu}f_{ab}\mathrm{P}\frac{1}{\Omega-\epsilon_{ba}}.
\end{align}
We can decompose the intrinsic Fermi surface term into the real and imaginary parts as
\begin{align}
    \sigma_{EE,\mathrm{IFS;M}}^{\mu;\nu\lambda}&=-\frac{1}{2}\int \dfrac{d\vb*{k}}{(2\pi)^{d}}\sum_{a\neq b}\operatorname{Re}[\xi^{\nu}_{ab}\xi^{\lambda}_{ba}]\partial_{\mu}f_{ab}\mathrm{P}\frac{1}{\Omega-\epsilon_{ba}},\\
    \sigma_{EE,\mathrm{IFS;E}}^{\mu;\nu\lambda}&=-\frac{i}{2}\int \dfrac{d\vb*{k}}{(2\pi)^{d}}\sum_{a\neq b}\operatorname{Im}[\xi^{\nu}_{ab}\xi^{\lambda}_{ba}]\partial_{\mu}f_{ab}\mathrm{P}\frac{1}{\Omega-\epsilon_{ba}}.
\end{align}

\subsection{Spin dynamics induced photocurrent}
Here, we derived the photocurrent formula related to spin dynamics. Using the SPDM $\rho_{\mathrm{SS}}$, we can express photocurrent response to the spin field as
\begin{align}
        J^{\mu}_{\mathrm{SS}} &= \int \dfrac{d\vb*{k}}{(2\pi)^{d}}\sum_{abc}J_{ab}^{\mu}\rho_{\mathrm{S}\mathrm{S}, ba}^{(2)}(\omega)  \\
        &\eqqcolon \int\dfrac{d\omega_{1}d\omega_{2}}{(2\pi)^{2}}\sigma^{\mu;\nu\lambda}_{\mathrm{SS}}(\omega, \omega_{1}, \omega_{2})\Delta\mathrm{S}^{\nu}(\omega_{1})\Delta\mathrm{S}^{\lambda}(\omega_{2})2\pi\delta(\omega-\omega_{1}-\omega_{2})
\end{align}
In the previous subsection, we derived the photocurrent conductivity by using a perturbative expansion of the von Neumann equation and obtained the formula for the photocurrent.
Similarly, the same formula for the photocurrent can be derived from $\rho_{\mathrm{SS}}$ by replacing the Berry connection term $\xi$, associated with the external light field in $\sigma_{EE}$, with the spin operator $B$.
However, the spin operator $B^{\mu}$ does not contain the derivative term with respect to $\bm{k}$-points.
As a result, the Drude term $\sigma_{\text{D}}^{\mu;\nu\lambda}$ and the Berry curvature dipole term $\sigma_{\text{BCD}}^{\mu;\nu\lambda}$ do not appear in this formula.
Considering these facts, we can get the formula for photocurrent induced by spin dynamics;
\begin{align}
        \sigma_{\mathrm{S}\mathrm{S},\text{shift}}^{\mu;\nu\lambda} &= J^{2}\dfrac{\pi}{2}\int \dfrac{d\vb*{k}}{(2\pi)^{d}}\sum_{a\neq b}\operatorname{Im}\qty[\qty[D^{\mu}B^{\nu}]_{ab}B_{ba}^{\lambda} - \qty[D^{\mu}B^{\lambda}]_{ba}B_{ab}^{\nu}]f_{ab}\delta(\Omega-\epsilon_{ba}), \\
        \sigma_{\mathrm{S}\mathrm{S},\text{gyro}}^{\mu;\nu\lambda} &= -J^{2}\dfrac{i\pi}{2}\int \dfrac{d\vb*{k}}{(2\pi)^{d}}\sum_{a\neq b}\operatorname{Re}\qty[\qty[D^{\mu}B^{\nu}]_{ab}B_{ba}^{\lambda} - \qty[D^{\mu}B^{\lambda}]_{ba}B_{ab}^{\nu}]f_{ab}\delta(\Omega-\epsilon_{ba}), \\
        \sigma_{\mathrm{S}\mathrm{S},\text{Inj;M}}^{\mu;\nu\lambda} &= \pi J^2\tau\int \dfrac{d\vb*{k}}{(2\pi)^{d}}\sum_{a\neq b}\Delta_{ab}^{\mu}\operatorname{Re}\qty[B^{\nu}_{ab}B^{\lambda}_{ba}]f_{ab}\delta(\Omega-\epsilon_{ba}), \\
        \sigma_{\mathrm{S}\mathrm{S}, \text{Inj;E}}^{\mu;\nu\lambda} &= i\pi J^2\tau\int \dfrac{d\vb*{k}}{(2\pi)^{d}}\sum_{a\neq b}\Delta_{ab}^{\mu}\operatorname{Im}\qty[B^{\nu}_{ab}B^{\lambda}_{ba}]f_{ab}\delta(\Omega-\epsilon_{ba}),\\
        \sigma_{\text{SS},\mathrm{IFSI;M}}^{\mu;\nu\lambda}&=\frac{J^2}{2}\int\dfrac{d\vb*{k}}{(2\pi)^{d}}\sum_{a\neq b}\operatorname{Re}[B^{\nu}_{ab}B^{\lambda}_{ba}]f_{ab}\partial_{\mu}\mathrm{P}\frac{1}{\Omega-\epsilon_{ba}},\\
        \sigma_{\text{SS},\mathrm{IFSI;E}}^{\mu;\nu\lambda}&=\frac{iJ^2}{2}\int\dfrac{d\vb*{k}}{(2\pi)^{d}}\sum_{a\neq c}\operatorname{Im}[B^{\nu}_{ab}B^{\lambda}_{ba}]f_{ab}\partial_{\mu}\mathrm{P}\frac{1}{\Omega-\epsilon_{ba}},\\
        \sigma_{\text{SS},\text{IFSI\hspace{-1.2pt}I;M}}^{\mu;\nu\lambda}&=\frac{J^2}{2}\int \dfrac{d\vb*{k}}{(2\pi)^{d}}\sum_{a\neq b}\partial_{\mu}\operatorname{Re}[B^{\nu}_{ab}B^{\lambda}_{ba}]f_{ab}\mathrm{P}\frac{1}{\Omega-\epsilon_{ba}},\\
        \sigma_{\text{SS},\text{IFSI\hspace{-1.2pt}I;E}}^{\mu;\nu\lambda}&=\frac{iJ^2}{2}\int \dfrac{d\vb*{k}}{(2\pi)^{d}}\sum_{a\neq b}\partial_{\mu}\operatorname{Im}[B^{\nu}_{ab}B^{\lambda}_{ba}]f_{ab}\mathrm{P}\frac{1}{\Omega-\epsilon_{ba}}.
\end{align}
The photocurrent response from $\sigma_{\mathrm{SS}}$ does not include any terms exhibiting $\tau$-dependence at $\omega = 0$. The shift current term and the intrinsic Fermi surface term II originate from the off-diagonal components of the current matrix at the output vertex. In contrast, the injection current term and the intrinsic Fermi surface term I arise from the diagonal components of the current matrix at the output vertex.

\subsection{Interference of light field and spin dynamics}
Following the previous subsection, we consider the photocurrent response coming from the interference of the light field and spin dynamics. Using $\rho_{E\mathrm{S}}$ and $\rho_{\mathrm{S}E}$, we can write photocurrent formula as
\begin{align}
        J_{E\mathrm{S}}^{\mu}(\omega) &= \int \dfrac{d\vb*{k}}{(2\pi)^{d}}\sum_{abc}J_{ab}^{\mu}(\rho_{E\mathrm{S},ba}^{(2)}(\omega)+\rho_{\mathrm{S}E,ba}^{(2)}(\omega)) \\
        &\eqqcolon \int\dfrac{d\omega_{1}d\omega_{2}}{(2\pi)^{2}}\left[\sigma_{\text{MD}}^{\mu;\nu\lambda}(\omega, \omega_{1}, \omega_{2})+\tilde{\sigma}_{\mathrm{S}E}^{\mu;\nu\lambda}(\omega, \omega_{1}, \omega_{2})\right]\Delta\mathrm{S}^{\nu}(\omega_{1})E^{\lambda}(\omega_{2})2\pi\delta(\omega-\omega_{1}-\omega_{2}).
\end{align}
Here, the term $\sigma_{\text{MD}}^{\mu;\nu\lambda}(\omega, \omega_{1}, \omega_{2})$ can be expressed as
\begin{align}
    \sigma_{\text{MD}}^{\mu;\nu\lambda} &= \frac{J}{2(\Omega+i/\tau)}\int\frac{d\bm{k}}{(2\pi)^d}\sum_{a\neq b}(\xi_{ba}^{\mu}B_{ab}^{\nu}-\xi_{ab}^{\mu}B_{ba}^{\nu})\partial_{\lambda}f_{a}\notag,\\
    &=\frac{J(1/\tau+i\Omega)}{\Omega^2+1/\tau^2}\int\frac{d\bm{k}}{(2\pi)^d}\sum_{a\neq b}\partial_{\lambda}\operatorname{Im}[\xi_{ab}^{\mu}B_{ba}^{\nu}]f_{a}.
\end{align}
We call this term the mixed dipole term originating from the SPDM $\rho_{\mathrm{S}E}^{(2)}(\omega)$.
We can decompose the mixed dipole term into two components as 
\begin{align}
    \sigma_{\text{MD,C}}^{\mu;\nu\lambda} &=\frac{Ji\Omega}{\Omega^2+1/\tau^2}\int\frac{d\bm{k}}{(2\pi)^d}\sum_{a\neq b}\partial_{\lambda}\operatorname{Im}[\xi_{ab}^{\mu}B_{ba}^{\nu}]f_{a},\\
    \sigma_{\text{MD,L}}^{\mu;\nu\lambda} &=\frac{J/\tau}{\Omega^2+1/\tau^2}\int\frac{d\bm{k}}{(2\pi)^d}\sum_{a\neq b}\partial_{\lambda}\operatorname{Im}[\xi_{ab}^{\mu}B_{ba}^{\nu}]f_{a}.
\end{align}
The former term can be induced by circularly polarized light and the latter term can be induced by non-circularly polarized light.
The latter term exhibits a $\tau$-dependent peak at $\Omega = 0$.
On the other hand, the $\tilde{\sigma}_{\mathrm{S}E}^{\mu;\nu\lambda}$ can be classified into the following eight component.
\begin{align}
        \sigma_{\mathrm{S}E,\text{shift}}^{\mu;\nu\lambda} &= J\dfrac{\pi}{2}\int \dfrac{d\vb*{k}}{(2\pi)^{d}}\sum_{a\neq b}\operatorname{Im}\qty[\qty[D^{\mu}B^{\nu}]_{ab}\xi_{ba}^{\lambda} - \qty[D^{\mu}\xi^{\lambda}]_{ba}B_{ab}^{\nu}]f_{ab}\delta(\Omega-\epsilon_{ba}), \\
        \sigma_{\mathrm{S}E,\text{gyro}}^{\mu;\nu\lambda} &= -J\dfrac{i\pi}{2}\int \dfrac{d\vb*{k}}{(2\pi)^{d}}\sum_{a\neq b}\operatorname{Re}\qty[\qty[D^{\mu}B^{\nu}]_{ab}\xi_{ba}^{\lambda} - \qty[D^{\mu}\xi^{\lambda}]_{ba}B_{ab}^{\nu}]f_{ab}\delta(\Omega-\epsilon_{ba}), \\
        \sigma_{\mathrm{S}E,\text{Inj;M}}^{\mu;\nu\lambda} &= \pi J\tau\int \dfrac{d\vb*{k}}{(2\pi)^{d}}\sum_{a\neq b}\Delta_{ab}^{\mu}\operatorname{Re}\qty[B^{\nu}_{ab}\xi^{\lambda}_{ba}]f_{ab}\delta(\Omega-\epsilon_{ba}),\\
        \sigma_{\mathrm{S}E, \text{Inj;E}}^{\mu;\nu\lambda} &= i\pi J\tau\int \dfrac{d\vb*{k}}{(2\pi)^{d}}\sum_{a\neq b}\Delta_{ab}^{\mu}\operatorname{Im}\qty[B^{\nu}_{ab}\xi^{\lambda}_{ba}]f_{ab}\delta(\Omega-\epsilon_{ba}),\\
        \sigma_{\text{S}E,\mathrm{IFSI;M}}^{\mu;\nu\lambda}&=\frac{J}{2}\int\dfrac{d\vb*{k}}{(2\pi)^{d}}\sum_{a\neq b}\operatorname{Re}[B^{\nu}_{ab}\xi^{\lambda}_{ba}]f_{ab}\partial_{\mu}\mathrm{P}\frac{1}{\Omega-\epsilon_{ba}},\\
        \sigma_{\text{S}E,\mathrm{IFSI;E}}^{\mu;\nu\lambda}&=\frac{iJ}{2}\int\dfrac{d\vb*{k}}{(2\pi)^{d}}\sum_{a\neq c}\operatorname{Im}[B^{\nu}_{ab}\xi^{\lambda}_{ba}]f_{ab}\partial_{\mu}\mathrm{P}\frac{1}{\Omega-\epsilon_{ba}},\\
        \sigma_{\text{S}E,\text{IFSI\hspace{-1.2pt}I;M}}^{\mu;\nu\lambda}&=\frac{J}{2}\int \dfrac{d\vb*{k}}{(2\pi)^{d}}\sum_{a\neq b}\partial_{\mu}\operatorname{Re}[B^{\nu}_{ab}\xi^{\lambda}_{ba}]f_{ab}\mathrm{P}\frac{1}{\Omega-\epsilon_{ba}},\\
        \sigma_{\text{S}E,\text{IFSI\hspace{-1.2pt}I;E}}^{\mu;\nu\lambda}&=\frac{iJ}{2}\int \dfrac{d\vb*{k}}{(2\pi)^{d}}\sum_{a\neq b}\partial_{\mu}\operatorname{Im}[B^{\nu}_{ab}\xi^{\lambda}_{ba}]f_{ab}\mathrm{P}\frac{1}{\Omega-\epsilon_{ba}}.
\end{align}
It should be noted that the mixed dipole term, $\sigma_{\text{MD}}^{\mu;\nu\lambda}$, arises solely from the $\rho_{\mathrm{S}E}^{(2)}(\omega)$ component, as the spin operator does not contain any $\bm{k}$-derivative terms. 
The mixed dipole term, shift current term, and intrinsic Fermi surface term II originate from the off-diagonal components of the current matrix at the output vertex. In contrast, the injection current term and intrinsic Fermi surface term I arise from the diagonal components of the current matrix at the output vertex.

\section{U(2) gauge description of photocurrent response in $\mathcal{PT}$-symmetric systems}\label{U(2)_formula}
The Bloch states have at least U(2)-gauge degree of freedom due to the PT-symmetry-endowed double degeneracy.
In the previous section, we formulated the photocurrent responses in spinless systems or $\mathcal{PT}$-violated spinful systems.
The formulas for the photocurrent responses in the previous section are not invariant under the U(2)-gauge transformation. Thus, we rewrite the formulas for the photocurrent responses in the U(2) invariant form \cite{Watanabe2021}.
More general treatment can be found in Ref. \cite{avdoshkin2024} where the projection operator is utilized in order to retain the gauge invariance.
First, we can decompose the Berry connection as
\begin{align}
    \xi_{ab}^{\mu}=\alpha_{ab}^{\mu}+\mathcal{A}_{ab}^{\mu}
\end{align}
where the $\alpha_{ab}^{\mu}$ is the intraband component and the $\mathcal{A}_{ab}^{\mu}$ is the interband component of the Berry connection. With this decomposition of the Berry connection, the intraband component of the position operator $r^{\mu}_i$ is modified as 
\begin{align}
    (r^{\mu}_i)_{ab}=i\partial_{\mu}\delta_{ab}+\alpha_{ab}^{\mu},
\end{align}
and the interband component of the position operator is given by $(r_{e}^{\mu})_{ab}=\mathcal{A}_{ab}^{\mu}$.
Using the intraband component of the position operator, the U(2)-gauge-covariant derivative is defined by $\mathfrak{D}_{\mu}=-ir_{i}^{\mu}$.
The derivative acts on the physical quantities in the Bloch representation $O_{ab}$ as
\begin{align}
    [\mathfrak{D}_{\mu}O]_{ab}=\partial_{\mu}O_{ab}-i\left(\sum_{c}\alpha_{ac}^{\mu}O_{cb}-\sum_{c}O_{ac}\alpha_{cb}^{\mu}\right).
\end{align}
We can check that $[\mathfrak{D}_{\mu}O]_{ab}$ is U(2) covariant by taking the U(2)-gauge transformation $\ket{u_a(\bm{k})}\rightarrow\ket{u_b(\bm{k})}U_{ba}$, where the summation of the band index is taken over the Kramers pair.
Using this U(2)-gauge covariant derivative, we show the formulas for the photocurrent response in $\mathcal{PT}$-symmetric spinful systems. In the same manner as the Berry connection, we can decompose the spin operator as
\begin{align}
    B_{ab}^{\mu}=s_{ab}^{\mu}+\mathcal{S}_{ab}^{\mu},
\end{align}
where the $s_{ab}^{\mu}$ is the intraband component and the $\mathcal{S}_{ab}^{\mu}$ is the interband component of the spin operator.
By using these methods, we show the formulas for the photocurrent responses in $\mathcal{PT}$-symmetric spinfull systems in the following subsections.

\subsection{Light field induced photocurrent in $\mathcal{PT}$-symmetric spinful systems}
Here, we show the formulas for the photocurrent responses from the light field in $\mathcal{PT}$-symmetric systems.
In the previous section, we can derive the formulas for the photocurrent responses in spinless systems.
In the same manner, we can get the formulas for the photocurrent responses in $\mathcal{PT}$-symmetric spinful systems by replacing the Berry connection $\xi$ with $\mathcal{A}$ and the U(1)-covariant derivative $D_{\mu}$ with the U(2)-covariant derivative $\mathfrak{D}_{\mu}$.
The formulas are expressed as
\begin{align}
        \sigma^{\mu;\nu\lambda}_{EE,\mathrm{D}}&=\frac{1}{\Omega^2+1/\tau^2}\int\frac{d\bm{k}}{(2\pi)^d}\sum_{a}\partial_{\mu}\partial_{\nu}\partial_{\lambda}\epsilon_{\bm{k}a}f(\epsilon_{\bm{k}a}),\\
        \sigma_{EE,\mathrm{BCD;C}}^{\mu;\nu\lambda}&=\frac{i\Omega}{\Omega^2+1/\tau^2}\int\frac{d\bm{k}}{(2\pi)^d}\sum_{a\neq b}(\partial_{\lambda}\operatorname{Im}[\mathcal{A}_{ab}^{\mu}\mathcal{A}_{ba}^{\nu}]-\partial_{\nu}\operatorname{Im}[\mathcal{A}_{ab}^{\mu}\mathcal{A}_{ba}^{\lambda}])f_{a},\\
        \sigma_{EE,\mathrm{BCD;L}}^{\mu;\nu\lambda}&=\frac{1/\tau}{\Omega^2+1/\tau^2}\int\frac{d\bm{k}}{(2\pi)^d}\sum_{a\neq b}(\partial_{\lambda}\operatorname{Im}[\mathcal{A}_{ab}^{\mu}\mathcal{A}_{ba}^{\nu}]+\partial_{\nu}\operatorname{Im}[\mathcal{A}_{ab}^{\mu}\mathcal{A}_{ba}^{\lambda}])f_{a},\\
        \sigma_{EE, \text{shift}}^{\mu;\nu\lambda} &= \dfrac{\pi}{2}\int \dfrac{dk}{(2\pi)^d}\sum_{a\neq b}\operatorname{Im}\qty[\qty[\mathfrak{D}^{\mu}\mathcal{A}^{\nu}]_{ab}\mathcal{A}_{ba}^{\lambda} - \qty[\mathfrak{D}^{\mu}\mathcal{A}^{\lambda}]_{ba}\mathcal{A}_{ab}^{\nu}]f_{ab}\delta(\Omega-\epsilon_{ab}), \\
        \sigma_{EE, \text{gyro}}^{\mu;\nu\lambda} &= -\dfrac{i\pi}{2}\int \dfrac{dk}{(2\pi)^d}\sum_{a\neq b}\operatorname{Re}\qty[\qty[\mathfrak{D}^{\mu}\mathcal{A}^{\nu}]_{ab}\mathcal{A}_{ba}^{\lambda} - \qty[\mathfrak{D}^{\mu}\mathcal{A}^{\lambda}]_{ba}\mathcal{A}_{ab}^{\nu}]f_{ab}\delta(\Omega-\epsilon_{ab}),\\
        \sigma_{EE, \text{Inj;M}}^{\mu;\nu\lambda} &= \pi\tau\int \dfrac{dk}{(2\pi)^d}\sum_{a\neq b}\Delta_{ab}^{\mu}\operatorname{Re}\qty[\mathcal{A}^{\nu}_{ab}\mathcal{A}^{\lambda}_{ba}]f_{ab}\delta(\Omega-\epsilon_{ba}),\\
        \sigma_{EE, \text{Inj;E}}^{\mu;\nu\lambda} &= i\pi\tau\int \dfrac{dk}{(2\pi)^d}\sum_{a\neq b}\Delta_{ab}^{\mu}\operatorname{Im}\qty[\mathcal{A}^{\nu}_{ab}\mathcal{A}^{\lambda}_{ba}]f_{ab}\delta(\Omega-\epsilon_{ba}),\\
        \sigma_{EE,\mathrm{IFSI;M}}^{\mu;\nu\lambda}&=\frac{1}{2}\int\dfrac{d\vb*{k}}{(2\pi)^{d}}\sum_{a\neq b}\operatorname{Re}[\mathcal{A}^{\nu}_{ab}\mathcal{A}^{\lambda}_{ba}]f_{ab}\partial_{\mu}\mathrm{P}\frac{1}{\Omega-\epsilon_{ba}},\\
        \sigma_{EE,\mathrm{IFSI;E}}^{\mu;\nu\lambda}&=\frac{i}{2}\int\dfrac{d\vb*{k}}{(2\pi)^{d}}\sum_{a\neq c}\operatorname{Im}[\mathcal{A}^{\nu}_{ab}\mathcal{A}^{\lambda}_{ba}]f_{ab}\partial_{\mu}\mathrm{P}\frac{1}{\Omega-\epsilon_{ba}},\\
        \sigma_{EE,\text{IFSI\hspace{-1.2pt}I;M}}^{\mu;\nu\lambda}&=\frac{1}{2}\int \dfrac{d\vb*{k}}{(2\pi)^{d}}\sum_{a\neq b}\partial_{\mu}\operatorname{Re}[\mathcal{A}^{\nu}_{ab}\mathcal{A}^{\lambda}_{ba}]f_{ab}\mathrm{P}\frac{1}{\Omega-\epsilon_{ba}},\\
        \sigma_{EE,\text{IFSI\hspace{-1.2pt}I;E}}^{\mu;\nu\lambda}&=\frac{i}{2}\int \dfrac{d\vb*{k}}{(2\pi)^{d}}\sum_{a\neq b}\partial_{\mu}\operatorname{Im}[\mathcal{A}^{\nu}_{ab}\mathcal{A}^{\lambda}_{ba}]f_{ab}\mathrm{P}\frac{1}{\Omega-\epsilon_{ba}}.
\end{align}

\subsection{Spin dynamics induced photocurrent in $\mathcal{PT}$-symmetric spinful systems}
Here, we show the formulas for the photocurrent responses from the spin dynamics in $\mathcal{PT}$-symmetric systems.
In the previous section, we derived the formulas for the photocurrent responses from the light field in $\mathcal{PT}$-symmetric spinful systems.
In the same manner, we can obtain the formulas for the photocurrent responses from the localized spin dynamics by replacing the Berry connection $\mathcal{A}$ with the interband component of the spin operators $\mathcal{S}$.
The formulas for the photocurrent conductivity solely from the spin dynamics are written as
\begin{align}
        \sigma_{\mathrm{S}\mathrm{S},\text{shift}}^{\mu;\nu\lambda} &= J^{2}\dfrac{\pi}{2}\int \dfrac{d\vb*{k}}{(2\pi)^{d}}\sum_{a\neq b}\operatorname{Im}\qty[\qty[\mathfrak{D}^{\mu}\mathcal{S}^{\nu}]_{ab}\mathcal{S}_{ba}^{\lambda} - \qty[\mathfrak{D}^{\mu}\mathcal{S}^{\lambda}]_{ba}\mathcal{S}_{ab}^{\nu}]f_{ab}\delta(\Omega-\epsilon_{ba}), \\
        \sigma_{\mathrm{S}\mathrm{S},\text{gyro}}^{\mu;\nu\lambda} &= -J^{2}\dfrac{i\pi}{2}\int \dfrac{d\vb*{k}}{(2\pi)^{d}}\sum_{a\neq b}\operatorname{Re}\qty[\qty[\mathfrak{D}^{\mu}\mathcal{S}^{\nu}]_{ab}\mathcal{S}_{ba}^{\lambda} - \qty[\mathfrak{D}^{\mu}\mathcal{S}^{\lambda}]_{ba}\mathcal{S}_{ab}^{\nu}]f_{ab}\delta(\Omega-\epsilon_{ba}), \\
        \sigma_{\mathrm{S}\mathrm{S},\text{Inj;M}}^{\mu;\nu\lambda} &= J^{2}\pi\tau\int \dfrac{d\vb*{k}}{(2\pi)^{d}}\sum_{a\neq b}\Delta_{ab}^{\mu}\operatorname{Re}\qty[\mathcal{S}^{\nu}_{ab}\mathcal{S}^{\lambda}_{ba}]f_{ab}\delta(\Omega-\epsilon_{ba}), \\
        \sigma_{\mathrm{S}\mathrm{S}, \text{Inj;E}}^{\mu;\nu\lambda} &= J^{2}i\pi\tau\int \dfrac{d\vb*{k}}{(2\pi)^{d}}\sum_{a\neq b}\Delta_{ab}^{\mu}\operatorname{Im}\qty[\mathcal{S}^{\nu}_{ab}\mathcal{S}^{\lambda}_{ba}]f_{ab}\delta(\Omega-\epsilon_{ba}),\\
        \sigma_{\text{SS},\mathrm{IFSI;M}}^{\mu;\nu\lambda}&=\frac{J^2}{2}\int\dfrac{d\vb*{k}}{(2\pi)^{d}}\sum_{a\neq b}\operatorname{Re}[\mathcal{S}^{\nu}_{ab}\mathcal{S}^{\lambda}_{ba}]f_{ab}\partial_{\mu}\mathrm{P}\frac{1}{\Omega-\epsilon_{ba}},\\
        \sigma_{\text{SS},\mathrm{IFSI;E}}^{\mu;\nu\lambda}&=\frac{iJ^2}{2}\int\dfrac{d\vb*{k}}{(2\pi)^{d}}\sum_{a\neq c}\operatorname{Im}[\mathcal{S}^{\nu}_{ab}\mathcal{S}^{\lambda}_{ba}]f_{ab}\partial_{\mu}\mathrm{P}\frac{1}{\Omega-\epsilon_{ba}},\\
        \sigma_{\text{SS},\text{IFSI\hspace{-1.2pt}I;M}}^{\mu;\nu\lambda}&=\frac{J^2}{2}\int \dfrac{d\vb*{k}}{(2\pi)^{d}}\sum_{a\neq b}\partial_{\mu}\operatorname{Re}[\mathcal{S}^{\nu}_{ab}\mathcal{S}^{\lambda}_{ba}]f_{ab}\mathrm{P}\frac{1}{\Omega-\epsilon_{ba}},\\
        \sigma_{\text{SS},\text{IFSI\hspace{-1.2pt}I;E}}^{\mu;\nu\lambda}&=\frac{iJ^2}{2}\int \dfrac{d\vb*{k}}{(2\pi)^{d}}\sum_{a\neq b}\partial_{\mu}\operatorname{Im}[\mathcal{S}^{\nu}_{ab}\mathcal{S}^{\lambda}_{ba}]f_{ab}\mathrm{P}\frac{1}{\Omega-\epsilon_{ba}}.
\end{align}

\subsection{Interference of light and spin dynamics in $\mathcal{PT}$-symmetric spinful systems}
Here, we show the formulas for the photocurrent responses stemming from the interference of the light and spin dynamics in $\mathcal{PT}$-symmetric spinful systems.
First, the mixed dipole term can be expressed in the spinful system as
\begin{align}
    \sigma_{\text{MD}}^{\mu;\nu\lambda} &= \frac{J}{2(\Omega+i/\tau)}\int\frac{d\bm{k}}{(2\pi)^d}\sum_{a\neq b}(\mathcal{A}_{ba}^{\mu}\mathcal{S}_{ab}^{\nu}-\mathcal{A}_{ab}^{\mu}\mathcal{S}_{ba}^{\nu})\partial_{\lambda}f_{a}\notag,\\
    &=\frac{J(1/\tau+i\Omega)}{\Omega^2+1/\tau^2}\int\frac{d\bm{k}}{(2\pi)^d}\sum_{a\neq b}\partial_{\lambda}\operatorname{Im}[\mathcal{A}_{ab}^{\mu}\mathcal{S}_{ba}^{\nu}]f_{a}.
\end{align}
We can separate the mixed dipole term into two components as
\begin{align}
    \sigma_{\text{MD,C}}^{\mu;\nu\lambda} &=\frac{Ji\Omega}{\Omega^2+1/\tau^2}\int\frac{d\bm{k}}{(2\pi)^d}\sum_{a\neq b}\partial_{\lambda}\operatorname{Im}[\mathcal{A}_{ab}^{\mu}\mathcal{S}_{ba}^{\nu}]f_{a},\\
    \sigma_{\text{MD,L}}^{\mu;\nu\lambda} &=\frac{J/\tau}{\Omega^2+1/\tau^2}\int\frac{d\bm{k}}{(2\pi)^d}\sum_{a\neq b}\partial_{\lambda}\operatorname{Im}[\mathcal{A}_{ab}^{\mu}\mathcal{S}_{ba}^{\nu}]f_{a}.
\end{align}
Second, the other terms can be written as
\begin{align}
        \sigma_{\mathrm{S}E,\text{shift}}^{\mu;\nu\lambda} &= J\dfrac{\pi}{2}\int \dfrac{d\vb*{k}}{(2\pi)^{d}}\sum_{a\neq b}\operatorname{Im}\qty[\qty[\mathfrak{D}^{\mu}\mathcal{S}^{\nu}]_{ab}\mathcal{A}_{ba}^{\lambda} - \qty[\mathfrak{D}^{\mu}\mathcal{A}^{\lambda}]_{ba}\mathcal{S}_{ab}^{\nu}]f_{ab}\delta(\Omega-\epsilon_{ba}), \\
        \sigma_{\mathrm{S}E,\text{gyro}}^{\mu;\nu\lambda} &= -J\dfrac{i\pi}{2}\int \dfrac{d\vb*{k}}{(2\pi)^{d}}\sum_{a\neq b}\operatorname{Re}\qty[\qty[\mathfrak{D}^{\mu}\mathcal{S}^{\nu}]_{ab}\mathcal{A}_{ba}^{\lambda} - \qty[\mathfrak{D}^{\mu}\mathcal{A}^{\lambda}]_{ba}\mathcal{S}_{ab}^{\nu}]f_{ab}\delta(\Omega-\epsilon_{ba}), \\
        \sigma_{\mathrm{S}E,\text{Inj;M}}^{\mu;\nu\lambda} &= J\pi\tau\int \dfrac{d\vb*{k}}{(2\pi)^{d}}\sum_{a\neq b}\Delta_{ab}^{\mu}\operatorname{Re}\qty[\mathcal{S}^{\nu}_{ab}\mathcal{A}^{\lambda}_{ba}]f_{ab}\delta(\Omega-\epsilon_{ba}),\\
        \sigma_{\mathrm{S}E, \text{Inj;E}}^{\mu;\nu\lambda} &= Ji\pi\tau\int \dfrac{d\vb*{k}}{(2\pi)^{d}}\sum_{a\neq b}\Delta_{ab}^{\mu}\operatorname{Im}\qty[\mathcal{S}^{\nu}_{ab}\mathcal{A}^{\lambda}_{ba}]f_{ab}\delta(\Omega-\epsilon_{ba}),\\
        \sigma_{\text{S}E,\mathrm{IFSI;M}}^{\mu;\nu\lambda}&=\frac{J}{2}\int\dfrac{d\vb*{k}}{(2\pi)^{d}}\sum_{a\neq b}\operatorname{Re}[\mathcal{S}^{\nu}_{ab}\mathcal{A}^{\lambda}_{ba}]f_{ab}\partial_{\mu}\mathrm{P}\frac{1}{\Omega-\epsilon_{ba}},\\
        \sigma_{\text{S}E,\mathrm{IFSI;E}}^{\mu;\nu\lambda}&=\frac{iJ}{2}\int\dfrac{d\vb*{k}}{(2\pi)^{d}}\sum_{a\neq c}\operatorname{Im}[\mathcal{S}^{\nu}_{ab}\mathcal{A}^{\lambda}_{ba}]f_{ab}\partial_{\mu}\mathrm{P}\frac{1}{\Omega-\epsilon_{ba}},\\
        \sigma_{\text{S}E,\text{IFSI\hspace{-1.2pt}I;M}}^{\mu;\nu\lambda}&=\frac{J}{2}\int \dfrac{d\vb*{k}}{(2\pi)^{d}}\sum_{a\neq b}\partial_{\mu}\operatorname{Re}[\mathcal{S}^{\nu}_{ab}\mathcal{A}^{\lambda}_{ba}]f_{ab}\mathrm{P}\frac{1}{\Omega-\epsilon_{ba}},\\
        \sigma_{\text{S}E,\text{IFSI\hspace{-1.2pt}I;E}}^{\mu;\nu\lambda}&=\frac{iJ}{2}\int \dfrac{d\vb*{k}}{(2\pi)^{d}}\sum_{a\neq b}\partial_{\mu}\operatorname{Im}[\mathcal{S}^{\nu}_{ab}\mathcal{A}^{\lambda}_{ba}]f_{ab}\mathrm{P}\frac{1}{\Omega-\epsilon_{ba}}.
\end{align}

\section{Symmetry analysis of the photocurrent}\label{sym_analysis_photocurrent}
In this section, we present the symmetry analysis of photocurrent responses including what stems from spin dynamics. The $\mathcal{PT}$ symmetry and phase matching between the different fields restrict the photocurrent generation. In the $\mathcal{PT}$-symmetric spinful system, the eigenenergy of the electronic system is doubly degenerate by the Kramers theorem. Thus, in $\mathcal{PT}$-symmetric systems, the spin-up state $\ket{u_{+}(\bm{k})}$ and the spin-down state $\ket{u_{-}(\bm{k})}$ are related to each other as
\begin{align}
    \mathcal{PT}\ket{u_{\pm}(\bm{k})}=\pm e^{-i\theta(\bm{k})}\ket{u_{\mp}(\bm{k})}.
\end{align}
where the $\theta(\bm{k})$ is a real-valued function.
In other words, we can express the transform property of the wave function as 
\begin{align}
    \mathcal{PT}\ket{u_{a}(\bm{k})}=\ket{u_{b}(\bm{k})}w_{ba}(\bm{k}),
\end{align}
where the index $a,b$ denotes the spin degree of the freedom and we introduce $w(\bm{k})=-i\sigma_ye^{-i\theta}$. 
Using this transformation property of the wave function, we derive the transformation property of the physical quantity.
First, we prove the formulas
\begin{align}
    \mathcal{A}_{ab}^{\mu}(\bm{k})\mathcal{A}_{ba}^{\nu}(\bm{k})&=\mathcal{A}_{\bar{b}\bar{a}}^{\mu}(\bm{k})\mathcal{A}_{\bar{a}\bar{b}}^{\nu}(\bm{k}),\label{Quantum_geometry}\\
    \mathcal{A}_{ab}^{\mu}(\bm{k})\mathcal{S}_{ba}^{\nu}(\bm{k})&=-\sigma_{\mathcal{S}^{\nu}}\mathcal{A}_{\bar{b}\bar{a}}^{\mu}(\bm{k})\mathcal{S}_{\bar{a}\bar{b}}^{\nu}(\bm{k}),\label{ES_geometry}\\
    \mathcal{S}_{ab}^{\mu}(\bm{k})\mathcal{S}_{ba}^{\nu}(\bm{k})&=\sigma_{\mathcal{S}^{\mu}}\sigma_{\mathcal{S}^{\nu}}\mathcal{S}_{\bar{b}\bar{a}}^{\mu}(\bm{k})\mathcal{S}_{\bar{a}\bar{b}}^{\nu}(\bm{k}).\label{SS_geometry}
\end{align}
where the $\sigma_{\mathcal{S}^{\mu}}$ is the sign under the $\mathcal{PT}$ operation and $(s,\bar{s})$ labels a Kramers pair.
In $\mathcal{PT}$-symmetric systems, the signs satisfy the following relations:
\begin{align}\label{PT_symmetry}
    \sigma_{\mathcal{S}^{\mathrm{L}^x}}=-1,\sigma_{\mathcal{S}^{\mathrm{M}^z}}=1.
\end{align}
Owing to the $(\mathcal{PT})^{-1}\mathcal{PT}=1$, we can obtain the formula
\begin{align}
    \braket*{u_{a\sigma}(\bm{k})}{u_{b\tau}(\bm{k})}&=i\braket*{\mathcal{PT}(u_{b\tau}(\bm{k}))}{\mathcal{PT}(u_{a\sigma}(\bm{k}))},\nonumber\\
    &=\braket*{u_{b\sigma}(\bm{k})}{u_{a\tau}(\bm{k})}.
\end{align}
By using this relation, the transformation property of the Berry connection between the Kramers doublet under the $\mathcal{PT}$ symmetry is expressed as
\begin{align}
    \xi_{a\sigma;b\tau}^{\mu}(\bm{k})&=i\braket*{\mathcal{PT}(\partial_{\mu}u_{b\tau}(\bm{k}))}{\mathcal{PT}(u_{a\sigma}(\bm{k}))},\notag\\
    &=\left\{\ket*{\partial_{\mu}u_{b\tau'}(\bm{k})}w_{\tau'\tau}(\bm{k})+\ket*{u_{b\tau'}(\bm{k})}\partial_{\mu}w_{\tau'\tau}(\bm{k})\right\}^*\ket*{u_{a\sigma'}(\bm{k})}w_{\sigma'\sigma}(\bm{k}),\notag\\
    &=\left[i\braket*{\partial_{\mu}u_{b\tau'}(\bm{k})}{u_{a\sigma'}(\bm{k})}-\braket*{u_{b\tau'}(\bm{k})}{u_{a\sigma'}(\bm{k})}\partial_{\mu}\theta(\bm{k})\right](i\sigma_{y})^{\dag}_{\tau\tau'}(i\sigma_{y})_{\sigma\sigma'},\notag\\
    &=(-\xi_{b\tau';a\sigma'}^{\mu}(\bm{k})-\partial_{\mu}\theta(\bm{k})\delta_{ab}\delta_{\sigma'\tau'})(i\sigma_{y})^{\dag}_{\tau\tau'}(i\sigma_{y})_{\sigma\sigma'}.\label{transform_of_connection}
\end{align}
On the other hand, the transformation property of the spin operator $\mathcal{S}^{\mu}_{ab}$ is expressed as
\begin{align}
    \mathcal{S}_{a\sigma;b\tau}^{\mu}(\bm{k})&=i\braket*{\mathcal{PT}(u_{b\tau}(\bm{k}))}{\mathcal{PT}[(\mathcal{S}^{\mu})^{\dag}u_{a\sigma}(\bm{k})]},\notag\\
    &=\left\{\ket*{u_{b\tau'}(\bm{k})}w_{\tau'\tau}(\bm{k})\right\}^*\ket*{\mathcal{PT}\mathcal{S}^{\mu}u_{a\sigma}(\bm{k})},\notag\\
    &=\matrixel{u_{b\tau'}(\bm{k})}{\mathcal{PT}\mathcal{S}^{\mu}(\mathcal{PT})^{-1}}{u_{a\sigma'}(\bm{k})}(i\sigma_{y})^{\dag}_{\tau\tau'}(i\sigma_{y})_{\sigma\sigma'},\notag\\
    &=\sigma_{\mathcal{S^{\mu}}}\mathcal{S}_{b\tau';a\sigma'}^{\mu}(\bm{k})(i\sigma_{y})^{\dag}_{\tau\tau'}(i\sigma_{y})_{\sigma\sigma'}.\label{transform_of_spin}
\end{align}

Taking different energy band indices $a\neq b$ and applying \Eqref{transform_of_connection} and \Eqref{transform_of_spin} to the product of $\mathcal{A}_{ab}^{\mu}\mathcal{A}_{ba}^{\nu}$, we can obtain the relation \Eqref{Quantum_geometry}, \Eqref{ES_geometry} and \Eqref{SS_geometry}.

Similarly, we can derive the relation 
\begin{align}
    [\mathfrak{D}_{\mu}(\bm{k})\mathcal{A}^{\nu}(\bm{k})]_{ab}\mathcal{A}^{\lambda}_{ba}(\bm{k})&=[\mathfrak{D}_{\mu}(\bm{k})\mathcal{A}^{\nu}(\bm{k})]_{\bar{b}\bar{a}}\mathcal{A}^{\lambda}_{\bar{a}\bar{b}}(\bm{k}),\label{trans_covariant_EE}\\
    [\mathfrak{D}_{\mu}(\bm{k})\mathcal{A}^{\nu}(\bm{k})]_{ab}\mathcal{S}^{\lambda}_{ba}(\bm{k})&=-\sigma_{\mathcal{S}^{\lambda}}[\mathfrak{D}_{\mu}(\bm{k})\mathcal{A}^{\nu}(\bm{k})]_{\bar{b}\bar{a}}\mathcal{S}^{\lambda}_{\bar{a}\bar{b}}(\bm{k}),\label{trans_covariant_ES}\\
    [\mathfrak{D}_{\mu}(\bm{k})\mathcal{S}^{\nu}(\bm{k})]_{ab}\mathcal{A}^{\lambda}_{ba}(\bm{k})&=-\sigma_{\mathcal{S}^{\nu}}[\mathfrak{D}_{\mu}(\bm{k})\mathcal{S}^{\nu}(\bm{k})]_{\bar{b}\bar{a}}\mathcal{A}^{\lambda}_{\bar{a}\bar{b}}(\bm{k}),\label{trans_covariant_SE}\\
    [\mathfrak{D}_{\mu}(\bm{k})\mathcal{S}^{\nu}(\bm{k})]_{ab}\mathcal{S}^{\lambda}_{ba}(\bm{k})&=\sigma_{\mathcal{S}^{\nu}}\sigma_{\mathcal{S}^{\lambda}}[\mathfrak{D}_{\mu}(\bm{k})\mathcal{S}^{\nu}(\bm{k})]_{\bar{b}\bar{a}}\mathcal{S}^{\lambda}_{\bar{a}\bar{b}}(\bm{k}).\label{trans_covariant_SS}
\end{align}
in which $\mathfrak{D}_{\mu}$ indicates the $\mathrm{U(2)}$ gauge covariant derivative. For the different energy band indices $a,b$, the covariant derivative of the Berry connection satisfies the relation as
\begin{align}
    [\mathfrak{D}_{\mu}(\bm{k})\mathcal{A}^{\nu}(\bm{k})]_{a\sigma;b\tau}&=\partial_{\mu}\xi^{\nu}_{a\sigma;b\tau}-i(\xi^{\mu}_{a\sigma;a\sigma}-\xi^{\mu}_{b\tau;b\tau})\xi^{\mu}_{a\sigma;b\tau}-i(\xi^{\mu}_{a\sigma;a\bar{\sigma}}\xi^{\nu}_{a\bar{\sigma};b\tau}-\xi^{\nu}_{a\sigma;b\bar{\tau}}\xi^{\mu}_{b\bar{\tau};b\tau}),\notag\\
    &=\left[-\partial_{\mu}\xi^{\nu}_{b\bar{\tau};a\bar{\sigma}}-i(\xi^{\mu}_{a\bar{\sigma};a\bar{\sigma}}+\partial_{\mu}\theta-\xi^{\mu}_{b\tau;b\tau}-\partial_{\mu}\theta)\xi^{\nu}_{b\bar{\tau};a\bar{\sigma}}-i(\xi^{\mu}_{a\sigma;a\bar{\sigma}}\xi^{\nu}_{b\bar{\tau};a\sigma}-\xi^{\nu}_{b\tau;a\bar{\sigma}}\xi^{\mu}_{b\bar{\tau};b\tau})\right](i\sigma_{y})^{\dag}_{\tau\tau'}(i\sigma_{y})_{\sigma\sigma'},\notag\\
    &=-[\mathfrak{D}_{\mu}\xi^{\nu}]_{b\bar{\tau};a\bar{\sigma}}(i\sigma_{y})^{\dag}_{\tau\tau'}(i\sigma_{y})_{\sigma\sigma'}.
\end{align}
On the other hand, we can derive the relation of the covariant derivative of the spin operator as
\begin{align}
    [\mathfrak{D}_{\mu}(\bm{k})\mathcal{S}^{\nu}(\bm{k})]_{a\sigma;b\tau}&=\partial_{\mu}\mathcal{S}^{\nu}_{a\sigma;b\tau}-i(\xi^{\mu}_{a\sigma;a\sigma}-\xi^{\mu}_{b\tau;b\tau})\mathcal{S}^{\mu}_{a\sigma;b\tau}-i(\xi^{\mu}_{a\sigma;a\bar{\sigma}}\mathcal{S}^{\nu}_{a\bar{\sigma};b\tau}-\mathcal{S}^{\nu}_{a\sigma;b\bar{\tau}}\xi^{\mu}_{b\bar{\tau};b\tau}),\notag\\
    &=\left[\partial_{\mu}\mathcal{S}^{\nu}_{b\bar{\tau};a\bar{\sigma}}-i(\xi^{\mu}_{a\bar{\sigma};a\bar{\sigma}}+\partial_{\mu}\theta-\xi^{\mu}_{b\tau;b\tau}-\partial_{\mu}\theta)\mathcal{S}^{\nu}_{b\bar{\tau};a\bar{\sigma}}-i(\xi^{\mu}_{a\sigma;a\bar{\sigma}}\mathcal{S}^{\nu}_{b\bar{\tau};a\sigma}-\mathcal{S}^{\nu}_{b\tau;a\bar{\sigma}}\xi^{\mu}_{b\bar{\tau};b\tau})\right]\sigma_{\mathcal{S^{\mu}}}(i\sigma_{y})^{\dag}_{\tau\tau'}(i\sigma_{y})_{\sigma\sigma'},\notag\\
    &=\sigma_{\mathcal{S^{\nu}}}[\mathfrak{D}_{\mu}\mathcal{S}^{\nu}]_{b\bar{\tau};a\bar{\sigma}}(i\sigma_{y})^{\dag}_{\tau\tau'}(i\sigma_{y})_{\sigma\sigma'}.
\end{align}
Combining this equation with \Eqref{transform_of_connection}, we can get the relations \Eqref{trans_covariant_EE}, \Eqref{trans_covariant_ES}, \Eqref{trans_covariant_SE} and \Eqref{trans_covariant_SS}. Using these relations, we classify the photocurrent response in terms of the $\mathcal{PT}$ symmetry.

\subsection{Light field induced photocurrent}
As drawn in the previous subsection, the photocurrent along the $y$ direction induced by the light field along the $x$ direction is expressed as
\begin{align}
        J_{EE}^{y} = \int \dfrac{d\Omega}{2\pi}\sigma_{EE}^{y;xx}(0;-\Omega,\Omega)E^{x}(-\Omega)E^{x}(\Omega),\label{J_PVE_independent particle approximation}
\end{align} 
where $\sigma^{y;xx}$ can be classified into following eight contributions,
\begin{align}
        \sigma^{y;xx}_{EE,\mathrm{D}}&=\frac{1}{\Omega^2+1/\tau^2}\int\frac{d\bm{k}}{(2\pi)^d}\sum_{a}\partial_{y}\partial_{x}\partial_{x}\epsilon_{\bm{k}a}f(\epsilon_{\bm{k}a}),\\
        \sigma_{EE,\mathrm{BCD;C}}^{y;xx}&=\frac{i\Omega}{\Omega^2+1/\tau^2}\int\frac{d\bm{k}}{(2\pi)^d}\sum_{a\neq b}(\partial_{x}\operatorname{Im}[\mathcal{A}_{ab}^{y}\mathcal{A}_{ba}^{x}]-\partial_{x}\operatorname{Im}[\mathcal{A}_{ab}^{y}\mathcal{A}_{ba}^{x}])f_{a},\\
        \sigma_{EE,\mathrm{BCD;L}}^{y;xx}&=\frac{1/\tau}{\Omega^2+1/\tau^2}\int\frac{d\bm{k}}{(2\pi)^d}\sum_{a\neq b}(\partial_{x}\operatorname{Im}[\mathcal{A}_{ab}^{y}\mathcal{A}_{ba}^{x}]+\partial_{x}\operatorname{Im}[\mathcal{A}_{ab}^{y}\mathcal{A}_{ba}^{x}])f_{a},\\
        \sigma_{EE, \text{shift}}^{y;xx} &= \dfrac{\pi}{2}\int \dfrac{dk}{2\pi}\sum_{a\neq b}\operatorname{Im}\qty[\qty[\mathfrak{D}^{y}\mathcal{A}^{x}]_{ab}\mathcal{A}_{ba}^{x} - \qty[\mathfrak{D}^{y}\mathcal{A}^{x}]_{ba}\mathcal{A}_{ab}^{x}]f_{ab}\delta(\Omega-\epsilon_{ab}), \\
        \sigma_{EE, \text{gyro}}^{y;xx} &= -\dfrac{i\pi}{2}\int \dfrac{dk}{2\pi}\sum_{a\neq b}\operatorname{Re}\qty[\qty[\mathfrak{D}^{y}\mathcal{A}^{x}]_{ab}\mathcal{A}_{ba}^{x} - \qty[\mathfrak{D}^{y}\mathcal{A}^{x}]_{ba}\mathcal{A}_{ab}^{x}]f_{ab}\delta(\Omega-\epsilon_{ab}),\\
        \sigma_{EE, \text{Inj;M}}^{y;xx} &= \pi\tau\int \dfrac{dk}{2\pi}\sum_{a\neq b}\Delta_{ab}^{y}\operatorname{Re}\qty[\mathcal{A}^{x}_{ab}\mathcal{A}^{x}_{ba}]f_{ab}\delta(\Omega-\epsilon_{ba}),\\
        \sigma_{EE, \text{Inj;E}}^{y;xx} &= i\pi\tau\int \dfrac{dk}{2\pi}\sum_{a\neq b}\Delta_{ab}^{y}\operatorname{Im}\qty[\mathcal{A}^{x}_{ab}\mathcal{A}^{x}_{ba}]f_{ab}\delta(\Omega-\epsilon_{ba})=0,\\
        \sigma_{EE,\mathrm{IFSI;M}}^{y;xx}&=\frac{1}{2}\int\dfrac{d\vb*{k}}{(2\pi)^{d}}\sum_{a\neq b}\operatorname{Re}[\mathcal{A}^{x}_{ab}\mathcal{A}^{x}_{ba}]f_{ab}\partial_{y}\mathrm{P}\frac{1}{\Omega-\epsilon_{ba}},\\
        \sigma_{EE,\mathrm{IFSI;E}}^{y;xx}&=\frac{i}{2}\int\dfrac{d\vb*{k}}{(2\pi)^{d}}\sum_{a\neq c}\operatorname{Im}[\mathcal{A}^{x}_{ab}\mathcal{A}^{x}_{ba}]f_{ab}\partial_{y}\mathrm{P}\frac{1}{\Omega-\epsilon_{ba}}=0,\\
        \sigma_{EE,\text{IFSI\hspace{-1.2pt}I;M}}^{y;xx}&=\frac{1}{2}\int \dfrac{d\vb*{k}}{(2\pi)^{d}}\sum_{a\neq b}\partial_{y}\operatorname{Re}[\mathcal{A}^{x}_{ab}\mathcal{A}^{x}_{ba}]f_{ab}\mathrm{P}\frac{1}{\Omega-\epsilon_{ba}},\\
        \sigma_{EE,\text{IFSI\hspace{-1.2pt}I;E}}^{y;xx}&=\frac{i}{2}\int \dfrac{d\vb*{k}}{(2\pi)^{d}}\sum_{a\neq b}\partial_{y}\operatorname{Im}[\mathcal{A}^{x}_{ab}\mathcal{A}^{x}_{ba}]f_{ab}\mathrm{P}\frac{1}{\Omega-\epsilon_{ba}}=0.
\end{align}
Owing to $\operatorname{Im}[\mathcal{A}^{x}_{ab}\mathcal{A}^{x}_{ba}]=0$, the photocurrent conductivity $\sigma_{EE, \text{Inj;E}}^{y;xx}$, $\sigma_{EE,\mathrm{IFSI;E}}^{y;xx}$ and $\sigma_{EE,\text{IFSI\hspace{-1.2pt}I;E}}^{y;xx}$ vanish.
In the $\mathcal{PT}$-symmetric system, the Drude term $\sigma_{EE,\mathrm{D}}^{y;xx}$ can be finite because the $\mathcal{PT}$-symmetry does not forbid the asymmetric band structure.
In \Eqref{J_PVE_independent particle approximation}, since the left hand side is real, and $E^{x}(\Omega)\qty[E^{x}(\Omega)]^{\ast}$ is also real, $\sigma^{y;xx}_{EE}(\Omega)$ should be real.
Therefore only $\sigma_{\text{BCD;L}}^{y;xx}, \sigma_{\text{shift}}^{y;xx}, \sigma_{\text{Inj;M}}^{y;xx}, \sigma_{\text{IFSI;M}}^{y;xx}, \sigma_{\text{IFSI\hspace{-1.2pt}I;M}}^{y;xx}$ can contribute to the photocurrent generation.
Moreover, by using the $\mathcal{PT}$ symmetry in the system, we can derive the following relations owing to the double degeneracy of the electronic bands.
\begin{align}
    &\sum_{a\neq b}\mathcal{A}_{ba}^{\mu}\mathcal{A}_{ab}^{\nu}\partial_{\lambda}f_{a}=\frac{1}{2}\sum_{a\neq b}[\mathcal{A}_{ba}^{\mu}\mathcal{A}_{ab}^{\nu}+\mathcal{A}_{\bar{b}\bar{a}}^{\mu}\mathcal{A}_{\bar{a}\bar{b}}^{\nu}]\partial_{\lambda}f_{a},\\
        &\sum_{a\neq b}\qty[\qty[\mathfrak{D}^{y}\mathcal{A}^{x}]_{ab}\mathcal{A}_{ba}^{x} - \qty[\mathfrak{D}^{y}\mathcal{A}^{x}]_{ba}\mathcal{A}_{ab}^{x}]f_{ab}\delta(\Omega-\epsilon_{ba})\nonumber\\
        &=\frac{1}{2}\sum_{a\neq b}\qty[\qty[\mathfrak{D}^{y}\mathcal{A}^{x}]_{ab}\mathcal{A}_{ba}^{x}+\qty[\mathfrak{D}^{y}\mathcal{A}^{x}]_{\bar{a}\bar{b}}\mathcal{A}_{\bar{b}\bar{a}}^{x}- \qty[\mathfrak{D}^{y}\mathcal{A}^{x}]_{ba}\mathcal{A}_{ab}^{x} - \qty[\mathfrak{D}^{y}\mathcal{A}^{x}]_{\bar{b}\bar{a}}\mathcal{A}_{\bar{a}\bar{b}}^{x}]f_{ab}\delta(\Omega-\epsilon_{ba}),\\
    &\sum_{a\neq b}[\mathcal{A}_{ab}^{x}\mathcal{A}_{ba}^{x}]f_{ab}\delta(\Omega-\epsilon_{ba})=\frac{1}{2}\sum_{a\neq b}[\mathcal{A}_{ab}^{x}\mathcal{A}_{ba}^{x}+\mathcal{A}_{\bar{a}\bar{b}}^{x}\mathcal{A}_{\bar{b}\bar{a}}^{x}]f_{ab}\delta(\Omega-\epsilon_{ba}),\\
    &\sum_{a\neq b}[\mathcal{A}_{ab}^{x}\mathcal{A}_{ba}^{x}]f_{ab}\mathrm{P}\frac{1}{\Omega-\epsilon_{ba}}=\frac{1}{2}\sum_{a\neq b}[\mathcal{A}_{ab}^{x}\mathcal{A}_{ba}^{x}+\mathcal{A}_{\bar{a}\bar{b}}^{x}\mathcal{A}_{\bar{b}\bar{a}}^{x}]f_{ab}\mathrm{P}\frac{1}{\Omega-\epsilon_{ba}},
\end{align}
where the band index $\bar{a} (\bar{b})$ represents the Kramers pair corresponding to $a (b)$.
In the derivation of these relations, the summation over the degenerated band indices can be computed by putting aside the energy-related terms, such as $\partial_{\lambda}f_a$, $f_{ab}\delta(\Omega-\epsilon_{ba})$, and $f_{ab}\mathrm{P}\frac{1}{\Omega-\epsilon_{ba}}$.
By considering the $\mathcal{PT}$ operation, the matrix elements in each photocurrent conductivity satisfy the following relations.
\begin{align}
    &\sum_{a\neq b}\operatorname{Im}[\mathcal{A}_{ba}^{\mu}\mathcal{A}_{ab}^{\nu}]\partial_{\lambda}f_{a}=\frac{1+\sigma_{\mathcal{A}^{\nu}}\sigma_{\mathcal{A}^{\lambda}}}{4i}\sum_{a\neq b}[\mathcal{A}_{ba}^{\mu}\mathcal{A}_{ab}^{\nu}-\mathcal{A}_{ab}^{\mu}\mathcal{A}_{ba}^{\nu}]\partial_{\lambda}f_{a},\\
        &\sum_{a\neq b}\operatorname{Re}\qty[\qty[\mathfrak{D}^{y}\mathcal{A}^{x}]_{ab}\mathcal{A}_{ba}^{x} - \qty[\mathfrak{D}^{y}\mathcal{A}^{x}]_{ba}\mathcal{A}_{ab}^{x}]f_{ab}\delta(\Omega-\epsilon_{ba})\nonumber\\
        &=\frac{1+\sigma_{\mathcal{A}^{x}}\sigma_{\mathcal{A}^{x}}}{4}\sum_{a\neq b}\qty[\qty[\mathfrak{D}^{y}\mathcal{A}^{x}]_{ab}\mathcal{A}_{ba}^{x}+\qty[\mathfrak{D}^{y}\mathcal{A}^{x}]_{ba}\mathcal{A}_{ab}^{x}- \qty[\mathfrak{D}^{y}\mathcal{A}^{x}]_{ba}\mathcal{A}_{ab}^{x} - \qty[\mathfrak{D}^{y}\mathcal{A}^{x}]_{ab}\mathcal{A}_{ba}^{x}]f_{ab}\delta(\Omega-\epsilon_{ba}),\\
        &\sum_{a\neq b}\operatorname{Im}\qty[\qty[\mathfrak{D}^{y}\mathcal{A}^{x}]_{ab}\mathcal{A}_{ba}^{x} - \qty[\mathfrak{D}^{y}\mathcal{A}^{x}]_{ba}\mathcal{A}_{ab}^{x}]f_{ab}\delta(\Omega-\epsilon_{ba})\nonumber\\
        &=\frac{1-\sigma_{\mathcal{A}^{x}}\sigma_{\mathcal{A}^{x}}}{4i}\sum_{a\neq b}\qty[\qty[\mathfrak{D}^{y}\mathcal{A}^{x}]_{ab}\mathcal{A}_{ba}^{x}-\qty[\mathfrak{D}^{y}\mathcal{A}^{x}]_{ba}\mathcal{A}_{ab}^{x}- \qty[\mathfrak{D}^{y}\mathcal{A}^{x}]_{ba}\mathcal{A}_{ab}^{x} + \qty[\mathfrak{D}^{y}\mathcal{A}^{x}]_{ab}\mathcal{A}_{ba}^{x}]f_{ab}\delta(\Omega-\epsilon_{ba}),\\
    &\sum_{a\neq b}\operatorname{Re}[\mathcal{A}_{ab}^{x}\mathcal{A}_{ba}^{x}]f_{ab}\delta(\Omega-\epsilon_{ba})=\frac{1+\sigma_{\mathcal{A}^{x}}\sigma_{\mathcal{A}^{x}}}{4}\sum_{a\neq b}[\mathcal{A}_{ab}^{x}\mathcal{A}_{ba}^{x}+\mathcal{A}_{ba}^{x}\mathcal{A}_{ab}^{x}]f_{ab}\delta(\Omega-\epsilon_{ba}),\\
    &\sum_{a\neq b}\operatorname{Re}[\mathcal{A}_{ab}^{x}\mathcal{A}_{ba}^{x}]f_{ab}\mathrm{P}\frac{1}{\Omega-\epsilon_{ba}}=\frac{1+\sigma_{\mathcal{A}^{x}}\sigma_{\mathcal{A}^{x}}}{4}\sum_{a\neq b}[\mathcal{A}_{ab}^{x}\mathcal{A}_{ba}^{x}+\mathcal{A}_{ba}^{x}\mathcal{A}_{ab}^{x}]f_{ab}\mathrm{P}\frac{1}{\Omega-\epsilon_{ba}},
\end{align}
where the $\sigma_{\mathcal{A}^{\mu}}$ is the parity under the $\mathcal{PT}$ operator.
In $\mathcal{PT}$-symmetric systems, the $\sigma_{EE,\mathrm{BCD;L}}^{y;xx}$ and $\sigma_{EE,\mathrm{shift}}^{y;xx}$ vanish because $\sigma_{\mathcal{A}^x}=\sigma_{\mathcal{A}^y}=-1$.
Therefore, the response comes from the Drude term $\sigma_{EE,\mathrm{D}}^{y;xx}$, injection current $\sigma_{EE,\mathrm{inj}}^{y;xx}$, and intrinsic Fermi surface term $\sigma_{EE,\mathrm{IFS;M}}^{y;xx}$.

\subsection{Spin dynamics induced photocurrent}
Photocurrent induced solely by localized spin dynamics can be described as 
\begin{align}
        \begin{split}
            J^{y}_{SS} &= \int\dfrac{d\Omega}{2\pi}\sigma_{\mathrm{SS}}^{y;\nu\lambda}(0;-\Omega,\Omega)\Delta \mathrm{S}^{\nu}(-\Omega)\Delta \mathrm{S}^{\lambda}(\Omega) \\
        &= \int\dfrac{d\Omega}{2\pi}\sigma_{\mathrm{SS}}^{y;\nu\lambda}(0;-\Omega,\Omega)\qty[\Delta \mathrm{S}^{\nu}(\Omega)]^{\ast}\Delta \mathrm{S}^{\lambda}(\Omega). \label{spindynamics_photocurrent}
        \end{split}
\end{align}
Here, $\sigma_{\mathrm{SS}}$ can be classified into the following eight components
\begin{align}
        \sigma_{\mathrm{S}\mathrm{S},\text{shift}}^{y;\nu\lambda} &= J^{2}\dfrac{\pi}{2}\int \dfrac{d\vb*{k}}{(2\pi)^{d}}\sum_{a\neq b}\operatorname{Im}\qty[\qty[\mathfrak{D}^{y}\mathcal{S}^{\nu}]_{ab}\mathcal{S}_{ba}^{\lambda} - \qty[\mathfrak{D}^{y}\mathcal{S}^{\lambda}]_{ba}\mathcal{S}_{ab}^{\nu}]f_{ab}\delta(\Omega-\epsilon_{ba}), \\
        \sigma_{\mathrm{S}\mathrm{S},\text{gyro}}^{y;\nu\lambda} &= -J^{2}\dfrac{i\pi}{2}\int \dfrac{d\vb*{k}}{(2\pi)^{d}}\sum_{a\neq b}\operatorname{Re}\qty[\qty[\mathfrak{D}^{y}\mathcal{S}^{\nu}]_{ab}\mathcal{S}_{ba}^{\lambda} - \qty[\mathfrak{D}^{y}\mathcal{S}^{\lambda}]_{ba}\mathcal{S}_{ab}^{\nu}]f_{ab}\delta(\Omega-\epsilon_{ba}), \\
        \sigma_{\mathrm{S}\mathrm{S},\text{Inj;M}}^{y;\nu\lambda} &= J^{2}\pi\tau\int \dfrac{d\vb*{k}}{(2\pi)^{d}}\sum_{a\neq b}\Delta_{ab}^{y}\operatorname{Re}\qty[\mathcal{S}^{\nu}_{ab}\mathcal{S}^{\lambda}_{ba}]f_{ab}\delta(\Omega-\epsilon_{ba}), \\
        \sigma_{\mathrm{S}\mathrm{S}, \text{Inj;E}}^{y;\nu\lambda} &= J^{2}i\pi\tau\int \dfrac{d\vb*{k}}{(2\pi)^{d}}\sum_{a\neq b}\Delta_{ab}^{y}\operatorname{Im}\qty[\mathcal{S}^{\nu}_{ab}\mathcal{S}^{\lambda}_{ba}]f_{ab}\delta(\Omega-\epsilon_{ba}),\\
        \sigma_{\text{SS},\mathrm{IFSI;M}}^{y;\nu\lambda}&=\frac{J^2}{2}\int\dfrac{d\vb*{k}}{(2\pi)^{d}}\sum_{a\neq b}\operatorname{Re}[\mathcal{S}^{\nu}_{ab}\mathcal{S}^{\lambda}_{ba}]f_{ab}\partial_{y}\mathrm{P}\frac{1}{\Omega-\epsilon_{ba}},\\
        \sigma_{\text{SS},\mathrm{IFSI;E}}^{y;\nu\lambda}&=\frac{iJ^2}{2}\int\dfrac{d\vb*{k}}{(2\pi)^{d}}\sum_{a\neq c}\operatorname{Im}[\mathcal{S}^{\nu}_{ab}\mathcal{S}^{\lambda}_{ba}]f_{ab}\partial_{y}\mathrm{P}\frac{1}{\Omega-\epsilon_{ba}},\\
        \sigma_{\text{SS},\text{IFSI\hspace{-1.2pt}I;M}}^{y;\nu\lambda}&=\frac{J^2}{2}\int \dfrac{d\vb*{k}}{(2\pi)^{d}}\sum_{a\neq b}\partial_{y}\operatorname{Re}[\mathcal{S}^{\nu}_{ab}\mathcal{S}^{\lambda}_{ba}]f_{ab}\mathrm{P}\frac{1}{\Omega-\epsilon_{ba}},\\
        \sigma_{\text{SS},\text{IFSI\hspace{-1.2pt}I;E}}^{y;\nu\lambda}&=\frac{iJ^2}{2}\int \dfrac{d\vb*{k}}{(2\pi)^{d}}\sum_{a\neq b}\partial_{y}\operatorname{Im}[\mathcal{S}^{\nu}_{ab}\mathcal{S}^{\lambda}_{ba}]f_{ab}\mathrm{P}\frac{1}{\Omega-\epsilon_{ba}}.
\end{align}
The following relations can be derived by using the double degeneracy of the energy bands.
\begin{align}
        &\sum_{a\neq b}\qty[\qty[\mathfrak{D}^{y}\mathcal{S}^{\nu}]_{ab}\mathcal{S}_{ba}^{\lambda} - \qty[\mathfrak{D}^{y}\mathcal{S}^{\lambda}]_{ba}\mathcal{S}_{ab}^{\nu}]f_{ab}\delta(\Omega-\epsilon_{ba})\nonumber\\
        &=\frac{1}{2}\sum_{a\neq b}\qty[\qty[\mathfrak{D}^{y}\mathcal{S}^{\nu}]_{ab}\mathcal{S}_{ba}^{\lambda}+\qty[\mathfrak{D}^{y}\mathcal{S}^{\nu}]_{\bar{a}\bar{b}}\mathcal{S}_{\bar{b}\bar{a}}^{\lambda}- \qty[\mathfrak{D}^{y}\mathcal{S}^{\lambda}]_{ba}\mathcal{S}_{ab}^{\nu} - \qty[\mathfrak{D}^{y}\mathcal{S}^{\lambda}]_{\bar{b}\bar{a}}\mathcal{S}_{\bar{a}\bar{b}}^{\nu}]f_{ab}\delta(\Omega-\epsilon_{ba}),\\
    &\sum_{a\neq b}[\mathcal{S}_{ab}^{\nu}\mathcal{S}_{ba}^{\lambda}]f_{ab}\delta(\Omega-\epsilon_{ba})=\frac{1}{2}\sum_{a\neq b}[\mathcal{S}_{ab}^{\nu}\mathcal{S}_{ba}^{\lambda}+\mathcal{S}_{\bar{a}\bar{b}}^{\nu}\mathcal{S}_{\bar{b}\bar{a}}^{\lambda}]f_{ab}\delta(\Omega-\epsilon_{ba}),\\
    &\sum_{a\neq b}[\mathcal{S}_{ab}^{\nu}\mathcal{S}_{ba}^{\lambda}]f_{ab}\mathrm{P}\frac{1}{\Omega-\epsilon_{ba}}=\frac{1}{2}\sum_{a\neq b}[\mathcal{S}_{ab}^{\nu}\mathcal{S}_{ba}^{\lambda}+\mathcal{S}_{\bar{a}\bar{b}}^{\nu}\mathcal{S}_{\bar{b}\bar{a}}^{\lambda}]f_{ab}\mathrm{P}\frac{1}{\Omega-\epsilon_{ba}}.
\end{align}

Taking into account the $\mathcal{PT}$ symmetry, the matrix elements in each photocurrent conductivity obey the following relations.
\begin{align}
        &\sum_{a\neq b}\operatorname{Re}\qty[\qty[\mathfrak{D}^{y}\mathcal{S}^{\nu}]_{ab}\mathcal{S}_{ba}^{\lambda} - \qty[\mathfrak{D}^{y}\mathcal{S}^{\lambda}]_{ba}\mathcal{S}_{ab}^{\nu}]f_{ab}\delta(\Omega-\epsilon_{ba})\nonumber\\
        &=\frac{1+\sigma_{\mathcal{S}^{\nu}}\sigma_{\mathcal{S}^{\lambda}}}{4}\sum_{a\neq b}\qty[\qty[\mathfrak{D}^{y}\mathcal{S}^{\nu}]_{ab}\mathcal{S}_{ba}^{\lambda}+\qty[\mathfrak{D}^{y}\mathcal{S}^{\nu}]_{ba}\mathcal{S}_{ab}^{\lambda}- \qty[\mathfrak{D}^{y}\mathcal{S}^{\lambda}]_{ba}\mathcal{S}_{ab}^{\nu} - \qty[\mathfrak{D}^{y}\mathcal{S}^{\lambda}]_{ab}\mathcal{S}_{ba}^{\nu}]f_{ab}\delta(\Omega-\epsilon_{ba}),\\
        &\sum_{a\neq b}\operatorname{Im}\qty[\qty[\mathfrak{D}^{y}\mathcal{S}^{\nu}]_{ab}\mathcal{S}_{ba}^{\lambda} - \qty[\mathfrak{D}^{y}\mathcal{S}^{\lambda}]_{ba}\mathcal{S}_{ab}^{\nu}]f_{ab}\delta(\Omega-\epsilon_{ba})\nonumber\\
        &=\frac{1-\sigma_{\mathcal{S}^{\nu}}\sigma_{\mathcal{S}^{\lambda}}}{4i}\sum_{a\neq b}\qty[\qty[\mathfrak{D}^{y}\mathcal{S}^{\nu}]_{ab}\mathcal{S}_{ba}^{\lambda}-\qty[\mathfrak{D}^{y}\mathcal{S}^{\nu}]_{ba}\mathcal{S}_{ab}^{\lambda}- \qty[\mathfrak{D}^{y}\mathcal{S}^{\lambda}]_{ba}\mathcal{S}_{ab}^{\nu} + \qty[\mathfrak{D}^{y}\mathcal{S}^{\lambda}]_{ab}\mathcal{S}_{ba}^{\nu}]f_{ab}\delta(\Omega-\epsilon_{ba}),\\
    &\sum_{a\neq b}\operatorname{Re}[\mathcal{S}_{ab}^{\nu}\mathcal{S}_{ba}^{\lambda}]f_{ab}\delta(\Omega-\epsilon_{ba})=\frac{1+\sigma_{\mathcal{S}^{\nu}}\sigma_{\mathcal{S}^{\lambda}}}{4}\sum_{a\neq b}[\mathcal{S}_{ab}^{\nu}\mathcal{S}_{ba}^{\lambda}+\mathcal{S}_{ba}^{\nu}\mathcal{S}_{ab}^{\lambda}]f_{ab}\delta(\Omega-\epsilon_{ba}),\\
    &\sum_{a\neq b}\operatorname{Im}[\mathcal{S}_{ab}^{\nu}\mathcal{S}_{ba}^{\lambda}]f_{ab}\delta(\Omega-\epsilon_{ba})=\frac{1-\sigma_{\mathcal{S}^{\nu}}\sigma_{\mathcal{S}^{\lambda}}}{4i}\sum_{a\neq b}[\mathcal{S}_{ab}^{\nu}\mathcal{S}_{ba}^{\lambda}-\mathcal{S}_{ba}^{\nu}\mathcal{S}_{ab}^{\lambda}]f_{ab}\delta(\Omega-\epsilon_{ba}),\\
    &\sum_{a\neq b}\operatorname{Re}[\mathcal{S}_{ab}^{\nu}\mathcal{S}_{ba}^{\lambda}]f_{ab}\mathrm{P}\frac{1}{\Omega-\epsilon_{ba}}=\frac{1+\sigma_{\mathcal{S}^{\nu}}\sigma_{\mathcal{S}^{\lambda}}}{4}\sum_{a\neq b}[\mathcal{S}_{ab}^{\nu}\mathcal{S}_{ba}^{\lambda}+\mathcal{S}_{ba}^{\nu}\mathcal{S}_{ab}^{\lambda}]f_{ab}\mathrm{P}\frac{1}{\Omega-\epsilon_{ba}},\\
    &\sum_{a\neq b}\operatorname{Im}[\mathcal{S}_{ab}^{\nu}\mathcal{S}_{ba}^{\lambda}]f_{ab}\mathrm{P}\frac{1}{\Omega-\epsilon_{ba}}=\frac{1-\sigma_{\mathcal{S}^{\nu}}\sigma_{\mathcal{S}^{\lambda}}}{4i}\sum_{a\neq b}[\mathcal{S}_{ab}^{\nu}\mathcal{S}_{ba}^{\lambda}-\mathcal{S}_{ba}^{\nu}\mathcal{S}_{ab}^{\lambda}]f_{ab}\mathrm{P}\frac{1}{\Omega-\epsilon_{ba}}.
\end{align}
In \Eqref{spindynamics_photocurrent}, when $\mathrm{S}^{\nu}$ and $\mathrm{S}^{\lambda}$ are in-phase (out-of-phase), $\qty[\Delta \mathrm{S}^{\nu}(\Omega)]^{\ast}\Delta \mathrm{S}^{\lambda}(\Omega)$ becomes real (pure-imaginary).
When the two fields are in-phase, the photocurrent response is characterized by the real part of $\sigma_{\mathrm{SS}}$, such as $\sigma_{\mathrm{SS, shift}}$, $\sigma_{\mathrm{SS, Inj;M}}$, and $\sigma_{\mathrm{SS, IFS;M}}$. 
On the other hand, the photocurrent induced by two fields that are out-of-phase with each other can be characterized by the imaginary part of $\sigma_{\mathrm{SS}}$, such as $\sigma_{\mathrm{SS, gyro}}$, $\sigma_{\mathrm{SS, Inj;E}}$, and $\sigma_{\mathrm{SS, IFS;E}}$.
Combining these correspondences with the $\mathcal{PT}$ symmetry constraints in \Eqref{PT_symmetry}, we identify the photocurrent contributions as summarized in \tabref{photocurrent_classification_intra} and \tabref{photocurrent_classification_inter}. 
Although we focused on the photocurrent response to the linearly polarized light, $\sigma_{\mathrm{SS}}$ includes $\sigma_{\text{Inj;E}}$, which corresponds to the photocurrent induced by circularly polarized light.
Moreover, the photocurrent generated by spin dynamics can include contributions from the shift current, specifically $\sigma_{\text{shift}}$ and $\sigma_{\text{gyro}}$, which is less torelant of disorder effects \cite{Hatada2020}.

\subsection{Interference of light field and spin dynamics}
Photocurrent arising from the interference of light field and spin dynamics can be expressed as follows.
\begin{align}\label{J_ES}
        J_{E\mathrm{S}}^{y}(\omega) &= \int \dfrac{d\vb*{k}}{(2\pi)^{d}}\sum_{abc}J_{ab}^{y}(\rho_{E\mathrm{S},ba}^{(2)}(\omega)+\rho_{\mathrm{S}E,ba}^{(2)}(\omega))\nonumber \\
        &\eqqcolon \int\dfrac{d\omega_{1}d\omega_{2}}{(2\pi)^{2}}\left[\sigma_{\text{MD}}^{y;\nu x}(\omega, \omega_{1}, \omega_{2})+\tilde{\sigma}_{\mathrm{S}E}^{y;\nu x}(\omega, \omega_{1}, \omega_{2})\right]\Delta\mathrm{S}^{\nu}(\omega_{1})E^{x}(\omega_{2})2\pi\delta(\omega-\omega_{1}-\omega_{2}).
\end{align}
Here, the mixed dipole term $\sigma_{\text{MD}}^{y;\nu x}(\omega, \omega_{1}, \omega_{2})$ can be expressed as
\begin{align}
    \sigma_{\text{MD}}^{y;\nu x} &= \frac{J}{2(\Omega+i/\tau)}\int\frac{d\bm{k}}{(2\pi)^d}\sum_{a\neq b}(\mathcal{A}_{ba}^{y}\mathcal{S}_{ab}^{\nu}-\mathcal{A}_{ab}^{y}\mathcal{S}_{ba}^{\nu})\partial_{x}f_{a}\notag,\\
    &=\frac{J(1/\tau+i\Omega)}{\Omega^2+1/\tau^2}\int\frac{d\bm{k}}{(2\pi)^d}\sum_{a\neq b}\partial_{x}\operatorname{Im}[\mathcal{A}_{ab}^{y}\mathcal{S}_{ba}^{\nu}]f_{a}.
\end{align}
We can decompose the mixed dipole term into two components as 
\begin{align}
    \sigma_{\text{MD,C}}^{y;\nu x} &=\frac{Ji\Omega}{\Omega^2+1/\tau^2}\int\frac{d\bm{k}}{(2\pi)^d}\sum_{a\neq b}\partial_{x}\operatorname{Im}[\mathcal{A}_{ab}^{y}\mathcal{S}_{ba}^{\nu}]f_{a},\\
    \sigma_{\text{MD,L}}^{y;\nu x} &=\frac{J/\tau}{\Omega^2+1/\tau^2}\int\frac{d\bm{k}}{(2\pi)^d}\sum_{a\neq b}\partial_{x}\operatorname{Im}[\mathcal{A}_{ab}^{y}\mathcal{S}_{ba}^{\nu}]f_{a}.
\end{align}
On the other hand, $\tilde{\sigma}_{\mathrm{S}E}^{y;\nu x}$ can be classified into the following eight components.
\begin{align}
        \sigma_{\mathrm{S}E,\text{shift}}^{y;\nu x} &= J\dfrac{\pi}{2}\int \dfrac{d\vb*{k}}{(2\pi)^{d}}\sum_{a\neq b}\operatorname{Im}\qty[\qty[\mathfrak{D}^{y}\mathcal{S}^{\nu}]_{ab}\mathcal{A}_{ba}^{x} - \qty[\mathfrak{D}^{y}\mathcal{A}^{x}]_{ba}\mathcal{S}_{ab}^{\nu}]f_{ab}\delta(\Omega-\epsilon_{ba}), \\
        \sigma_{\mathrm{S}E,\text{gyro}}^{y;\nu x} &= -J\dfrac{i\pi}{2}\int \dfrac{d\vb*{k}}{(2\pi)^{d}}\sum_{a\neq b}\operatorname{Re}\qty[\qty[\mathfrak{D}^{y}\mathcal{S}^{\nu}]_{ab}\mathcal{A}_{ba}^{x} - \qty[\mathfrak{D}^{y}\mathcal{A}^{x}]_{ba}\mathcal{S}_{ab}^{\nu}]f_{ab}\delta(\Omega-\epsilon_{ba}), \\
        \sigma_{\mathrm{S}E,\text{Inj;M}}^{y;\nu x} &= J\pi\tau\int \dfrac{d\vb*{k}}{(2\pi)^{d}}\sum_{a\neq b}\Delta_{ab}^{y}\operatorname{Re}\qty[\mathcal{S}^{\nu}_{ab}\mathcal{A}^{x}_{ba}]f_{ab}\delta(\Omega-\epsilon_{ba}),\\
        \sigma_{\mathrm{S}E, \text{Inj;E}}^{y;\nu x} &= Ji\pi\tau\int \dfrac{d\vb*{k}}{(2\pi)^{d}}\sum_{a\neq b}\Delta_{ab}^{y}\operatorname{Im}\qty[\mathcal{S}^{\nu}_{ab}\mathcal{A}^{x}_{ba}]f_{ab}\delta(\Omega-\epsilon_{ba}),\\
        \sigma_{\text{S}E,\mathrm{IFSI;M}}^{y;\nu x}&=\frac{J}{2}\int\dfrac{d\vb*{k}}{(2\pi)^{d}}\sum_{a\neq b}\operatorname{Re}[\mathcal{S}^{\nu}_{ab}\mathcal{A}^{x}_{ba}]f_{ab}\partial_{y}\mathrm{P}\frac{1}{\Omega-\epsilon_{ba}},\\
        \sigma_{\text{S}E,\mathrm{IFSI;E}}^{y;\nu x}&=\frac{iJ}{2}\int\dfrac{d\vb*{k}}{(2\pi)^{d}}\sum_{a\neq c}\operatorname{Im}[\mathcal{S}^{\nu}_{ab}\mathcal{A}^{x}_{ba}]f_{ab}\partial_{y}\mathrm{P}\frac{1}{\Omega-\epsilon_{ba}},\\
        \sigma_{\text{S}E,\text{IFSI\hspace{-1.2pt}I;M}}^{y;\nu x}&=\frac{J}{2}\int \dfrac{d\vb*{k}}{(2\pi)^{d}}\sum_{a\neq b}\partial_{y}\operatorname{Re}[\mathcal{S}^{\nu}_{ab}\mathcal{A}^{x}_{ba}]f_{ab}\mathrm{P}\frac{1}{\Omega-\epsilon_{ba}},\\
        \sigma_{\text{S}E,\text{IFSI\hspace{-1.2pt}I;E}}^{y;\nu x}&=\frac{iJ}{2}\int \dfrac{d\vb*{k}}{(2\pi)^{d}}\sum_{a\neq b}\partial_{y}\operatorname{Im}[\mathcal{S}^{\nu}_{ab}\mathcal{A}^{x}_{ba}]f_{ab}\mathrm{P}\frac{1}{\Omega-\epsilon_{ba}}.
\end{align}
Under the $\mathcal{PT}$ operation, the matrix element related to photocurrent generation satisfies the following relations owing to the double degeneracy of the energy bands.
\begin{align}
    &\sum_{a\neq b}\partial_{x}[\mathcal{A}_{ab}^{y}\mathcal{S}_{ba}^{\nu}]f_{a}=\frac{1}{2}\sum_{a\neq b}\partial_{x}[\mathcal{A}_{ab}^{y}\mathcal{S}_{ba}^{\nu}+\mathcal{A}_{\bar{a}\bar{b}}^{y}\mathcal{S}_{\bar{b}\bar{a}}^{\nu}]f_{a},\\
        &\sum_{a\neq b}\qty[\qty[\mathfrak{D}^{y}\mathcal{S}^{\nu}]_{ab}\mathcal{A}_{ba}^{x} - \qty[\mathfrak{D}^{y}\mathcal{A}^{x}]_{ba}\mathcal{S}_{ab}^{\nu}]f_{ab}\delta(\Omega-\epsilon_{ba})\nonumber\\
        &=\frac{1}{2}\sum_{a\neq b}\qty[\qty[\mathfrak{D}^{y}\mathcal{S}^{\nu}]_{ab}\mathcal{A}_{ba}^{x}+\qty[\mathfrak{D}^{y}\mathcal{S}^{\nu}]_{\bar{a}\bar{b}}\mathcal{A}_{\bar{b}\bar{a}}^{x}- \qty[\mathfrak{D}^{y}\mathcal{A}^{x}]_{ba}\mathcal{S}_{ab}^{\nu} - \qty[\mathfrak{D}^{y}\mathcal{A}^{x}]_{\bar{b}\bar{a}}\mathcal{S}_{\bar{a}\bar{b}}^{\nu}]f_{ab}\delta(\Omega-\epsilon_{ba}),\\
    &\sum_{a\neq b}[\mathcal{S}_{ab}^{\nu}\mathcal{A}_{ba}^{x}]f_{ab}\delta(\Omega-\epsilon_{ba})=\frac{1}{2}\sum_{a\neq b}[\mathcal{S}_{ab}^{\nu}\mathcal{A}_{ba}^{x}+\mathcal{S}_{\bar{a}\bar{b}}^{\nu}\mathcal{A}_{\bar{b}\bar{a}}^{x}]f_{ab}\delta(\Omega-\epsilon_{ba}),\\
    &\sum_{a\neq b}[\mathcal{S}_{ab}^{\nu}\mathcal{A}_{ba}^{x}]f_{ab}\mathrm{P}\frac{1}{\Omega-\epsilon_{ba}}=\frac{1}{2}\sum_{a\neq b}[\mathcal{S}_{ab}^{\nu}\mathcal{A}_{ba}^{x}+\mathcal{S}_{\bar{a}\bar{b}}^{\nu}\mathcal{A}_{\bar{b}\bar{a}}^{x}]f_{ab}\mathrm{P}\frac{1}{\Omega-\epsilon_{ba}}.
\end{align}
By considering the $\mathcal{PT}$ operation, the matrix elements in each photocurrent conductivity satisfy the following relations.
\begin{align}
    &\sum_{a\neq b}\partial_{x}\operatorname{Im}[\mathcal{A}_{ab}^{y}\mathcal{S}_{ba}^{\nu}]f_{a}=\frac{1+\sigma_{\mathcal{S}^{\nu}}}{4i}\sum_{a\neq b}\partial_{x}[\mathcal{A}_{ab}^{y}\mathcal{S}_{ba}^{\nu}-\mathcal{A}_{ba}^{y}\mathcal{S}_{ab}^{\nu}]f_{a},\\
        &\sum_{a\neq b}\operatorname{Re}\qty[\qty[\mathfrak{D}^{y}\mathcal{S}^{\nu}]_{ab}\mathcal{A}_{ba}^{x} - \qty[\mathfrak{D}^{y}\mathcal{A}^{x}]_{ba}\mathcal{S}_{ab}^{\nu}]f_{ab}\delta(\Omega-\epsilon_{ba})\nonumber\\
        &=\frac{1+\sigma_{\mathcal{S}^{\nu}}}{2}\sum_{a\neq b}\qty[\qty[\mathfrak{D}^{y}\mathcal{S}^{\nu}]_{ab}\mathcal{A}_{ba}^{x}+\qty[\mathfrak{D}^{y}\mathcal{S}^{\nu}]_{ba}\mathcal{A}_{ab}^{x}- \qty[\mathfrak{D}^{y}\mathcal{A}^{x}]_{ba}\mathcal{S}_{ab}^{\nu} - \qty[\mathfrak{D}^{y}\mathcal{A}^{x}]_{ab}\mathcal{S}_{ba}^{x}]f_{ab}\delta(\Omega-\epsilon_{ba}),\\
        &\sum_{a\neq b}\operatorname{Im}\qty[\qty[\mathfrak{D}^{y}\mathcal{S}^{\nu}]_{ab}\mathcal{A}_{ba}^{x} - \qty[\mathfrak{D}^{y}\mathcal{A}^{x}]_{ba}\mathcal{S}_{ab}^{\nu}]f_{ab}\delta(\Omega-\epsilon_{ba})\nonumber\\
        &=\frac{1-\sigma_{\mathcal{S}^{\nu}}}{4i}\sum_{a\neq b}\qty[\qty[\mathfrak{D}^{y}\mathcal{S}^{\nu}]_{ab}\mathcal{A}_{ba}^{x}-\qty[\mathfrak{D}^{y}\mathcal{S}^{\nu}]_{ba}\mathcal{A}_{ab}^{x}- \qty[\mathfrak{D}^{y}\mathcal{A}^{x}]_{ba}\mathcal{S}_{ab}^{\nu} + \qty[\mathfrak{D}^{y}\mathcal{A}^{x}]_{ab}\mathcal{S}_{ba}^{x}]f_{ab}\delta(\Omega-\epsilon_{ba}),\\
    &\sum_{a\neq b}\operatorname{Re}[\mathcal{S}_{ab}^{\nu}\mathcal{A}_{ba}^{x}]f_{ab}\delta(\Omega-\epsilon_{ba})=\frac{1+\sigma_{\mathcal{S}^{\nu}}}{4}\sum_{a\neq b}[\mathcal{S}_{ab}^{\nu}\mathcal{A}_{ba}^{x}+\mathcal{S}_{ba}^{x}\mathcal{A}_{ab}^{x}]f_{ab}\delta(\Omega-\epsilon_{ba}),\\
    &\sum_{a\neq b}\operatorname{Im}[\mathcal{S}_{ab}^{\nu}\mathcal{A}_{ba}^{x}]f_{ab}\delta(\Omega-\epsilon_{ba})=\frac{1-\sigma_{\mathcal{S}^{\nu}}}{4i}\sum_{a\neq b}[\mathcal{S}_{ab}^{\nu}\mathcal{A}_{ba}^{x}-\mathcal{S}_{ba}^{x}\mathcal{A}_{ab}^{x}]f_{ab}\delta(\Omega-\epsilon_{ba}),\\
    &\sum_{a\neq b}\operatorname{Re}[\mathcal{S}_{ab}^{\nu}\mathcal{A}_{ba}^{x}]f_{ab}\mathrm{P}\frac{1}{\Omega-\epsilon_{ba}}=\frac{1+\sigma_{\mathcal{S}^{\nu}}}{4}\sum_{a\neq b}[\mathcal{S}_{ab}^{\nu}\mathcal{A}_{ba}^{x}+\mathcal{S}_{ba}^{x}\mathcal{A}_{ab}^{x}]f_{ab}\mathrm{P}\frac{1}{\Omega-\epsilon_{ba}},\\
    &\sum_{a\neq b}\operatorname{Im}[\mathcal{S}_{ab}^{\nu}\mathcal{A}_{ba}^{x}]f_{ab}\mathrm{P}\frac{1}{\Omega-\epsilon_{ba}}=\frac{1-\sigma_{\mathcal{S}^{\nu}}}{4i}\sum_{a\neq b}[\mathcal{S}_{ab}^{\nu}\mathcal{A}_{ba}^{x}-\mathcal{S}_{ba}^{x}\mathcal{A}_{ab}^{x}]f_{ab}\mathrm{P}\frac{1}{\Omega-\epsilon_{ba}}.
\end{align}
In addition to the $\mathcal{PT}$ symmetry restriction, phase degrees of freedom between the light field and fictitious spin field play an important role in the photocurrent response. 
Rewriting the \Eqref{J_ES} with the electromagnetic susceptibility $\chi_{\mathrm{S}E^x}$ of the light field, we obtain
\begin{align}
        J_{ES}^{y} &= \int \dfrac{d\Omega}{2\pi}[\sigma_{\mathrm{MD}}^{y;\nu x}(0;-\Omega,\Omega)+\tilde{\sigma}_{\mathrm{S}E}^{y;\nu x}(0;-\Omega,\Omega)]\chi_{\mathrm{S}^{\nu}E^x}(-\Omega)E^{x}(-\Omega)E^{x}(\Omega), \nonumber\\
        &= \int \dfrac{d\Omega}{2\pi}[\sigma_{\mathrm{MD}}^{y;\nu x}(0;-\Omega,\Omega)+\tilde{\sigma}_{\mathrm{S}E}^{y;\nu x}(0;-\Omega,\Omega)]\qty[\operatorname{Re}\chi_{\mathrm{S}^{\nu}E^{x}}(-\Omega) + i\operatorname{Im}\chi_{\mathrm{S}^{\nu}E^{x}}(-\Omega)]\qty[E^{x}(\Omega)]^{\ast}E^{x}(\Omega),\nonumber \\
        &=\int\dfrac{d\Omega}{2\pi}\left\{\operatorname{Re}[\sigma_{\mathrm{MD}}^{y;\nu x}(0;-\Omega,\Omega)+\tilde{\sigma}_{\mathrm{S}E}^{y;\nu x}(0;-\Omega,\Omega)]\operatorname{Re}\chi_{\mathrm{S}^{\nu}E^x}(-\Omega)\right. \nonumber\\
        &\left.\qquad\qquad - \operatorname{Im}[\sigma_{\mathrm{MD}}^{y;\nu x}(0;-\Omega,\Omega)+\tilde{\sigma}_{\mathrm{S}E}^{y;\nu x}(0;-\Omega,\Omega)]\operatorname{Im}\chi_{\mathrm{S}^{\nu}E^x}(-\Omega)\right\}[E^{x}(\Omega)]^{\ast}E^{x}(\Omega).
\end{align}
Here, we utilized the fact that the left-hand side is real, and that $\qty[E^{z}(\Omega)]^{\ast}E^{z}(\Omega)$ is also a real quantity.
Considering symmetry constraints, we summarize the photocurrent generation by the interference of light field and spin dynamics in \tabref{photocurrent_classification_intra} and \tabref{photocurrent_classification_inter}.
In contrast to the case of independent particle approximation, the various mechanisms for photocurrent generation are found in the $\sigma_{\text{MD}}$ and $\sigma_{\text{S}E}$.

\section{Relaxation time dependence of photocurrent response}\label{freq_dep}
In this section, we discuss the relaxation time dependence of the photocurrent response, especially the injection current, shift current, and intrinsic Fermi surface term from the interband transition of the electrons.
\subsection{Injection current}
First, we discuss the relaxation time dependence of the injection current. We can express the injection current in general as
\begin{align}
    \sigma_{\text{Inj}}^{\mu;\nu\lambda} 
        &= \pi\tau\int \dfrac{d\vb*{k}}{(2\pi)^{d}}\sum_{a\neq b}\Delta_{ab}^{\mu}X^{\nu}_{ab}X^{\lambda}_{ba}f_{ab}\delta(\Omega-\epsilon_{ba}),
\end{align}
where the matrix $X_{ab}$ is defined as the operator, such as the Berry connection or spin operator. Considering the effect of the phenomenological relaxation, we can replace the delta function $\delta(\omega)$ with the Lorentz function, which is defined as
\begin{align}
    \mathcal{L}(\omega)=\frac{\tau^{-1}}{\pi}\frac{1}{\Omega^2+(1/\tau)^2}.
\end{align}
By using this method, we can rewrite the formula for the injection current as
\begin{align}
    \sigma_{\text{Inj}}^{\mu;\nu\lambda} 
        &= \pi\tau\int \dfrac{d\vb*{k}}{(2\pi)^{d}}\sum_{a\neq b}\Delta_{ab}^{\mu}X^{\nu}_{ab}X^{\lambda}_{ba}f_{ab}\frac{\tau^{-1}}{\pi}\frac{1}{(\Omega-\epsilon_{ba})^2+(1/\tau)^2}.
\end{align}
In the low-frequency regime, in which the frequency of the light satisfies the relation $\left\lvert\omega-\epsilon_{ba}\right\rvert\gg \tau^{-1}$, we can approximate the formula for the injection current as
\begin{align}
    \sigma_{\text{Inj}}^{\mu;\nu\lambda} 
        &\simeq \int \dfrac{d\vb*{k}}{(2\pi)^{d}}\sum_{a\neq b}\Delta_{ab}^{\mu}X^{\nu}_{ab}X^{\lambda}_{ba}f_{ab}\frac{1}{(\Omega-\epsilon_{ba})^2}.
\end{align}
Considering this formula for the injection current, we find the injection current does not show the relaxation time dependence and shows frequency dependence as 
\begin{align}
    \sigma_{\text{Inj}}^{\mu;\nu\lambda}\propto\frac{1}{(\Omega-\epsilon_g)^2},
\end{align}
where the $\epsilon_g$ reflects the property of the optical gap of the electronic system.

\subsection{Shift current}
Second, we focus on the relaxation time dependence of the shift current. We can discuss the behavior of the shift current in the same manner as the injection current case. In general, the shift current term can be expressed as
\begin{align}
    \sigma_{\text{shift}}^{\mu;\nu\lambda} =-\dfrac{i\pi}{2}\int \dfrac{d\vb*{k}}{(2\pi)^{d}}\sum_{a\neq b}\qty[\qty[D^{\mu}X^{\nu}]_{ab}X_{ba}^{\lambda} - \qty[D^{\mu}X^{\lambda}]_{ab}X_{ba}^{\nu}]f_{ab}\delta(\Omega-\epsilon_{ba}).
\end{align}
By replacing the delta function with the Lorentz function, we can rewrite the formula for the shift current as
\begin{align}
    \sigma_{\text{shift}}^{\mu;\nu\lambda} =-\dfrac{i\pi}{2}\int \dfrac{d\vb*{k}}{(2\pi)^{d}}\sum_{a\neq b}\qty[\qty[D^{\mu}X^{\nu}]_{ab}X_{ba}^{\lambda} - \qty[D^{\mu}X^{\lambda}]_{ab}X_{ba}^{\nu}]f_{ab}\frac{\tau^{-1}}{\pi}\frac{1}{(\Omega-\epsilon_{ba})^2+(1/\tau)^2}.
\end{align}
In the low-frequency regime, we can approximate the formula for the shift current as
\begin{align}
    \sigma_{\text{shift}}^{\mu;\nu\lambda} \simeq-\dfrac{i\tau^{-1}}{2}\int \dfrac{d\vb*{k}}{(2\pi)^{d}}\sum_{a\neq b}\qty[\qty[D^{\mu}X^{\nu}]_{ab}X_{ba}^{\lambda} - \qty[D^{\mu}X^{\lambda}]_{ab}X_{ba}^{\nu}]f_{ab}\frac{1}{(\Omega-\epsilon_{ba})^2}.
\end{align}
By considering this formula, we can find the relaxation time dependence of the shift current below
\begin{align}
    \sigma_{\mathrm{shift}}^{\mu;\nu\lambda}\propto\frac{\tau^{-1}}{(\Omega-\epsilon_g)^2}.
\end{align}

\subsection{Intrinsic Fermi surface term}
Finally, we discuss the relaxation time dependence of the intrinsic Fermi surface term. In general, we can express the formula for the intrinsic Fermi surface term as
\begin{align}
    \sigma_{\text{IFS}}^{\mu;\nu\lambda} &= -\dfrac{1}{2}\int \dfrac{d\vb*{k}}{(2\pi)^{d}}\sum_{a\neq b}X^{\nu}_{ab}X^{\lambda}_{ba}\partial_{\mu}f_{ab}\mathcal{P}\frac{1}{\Omega-\epsilon_{ba}}.
\end{align}
Considering the effect of the relaxation time, we can replace the principle value as
\begin{align}
    \mathrm{P}\frac{1}{\Omega-\epsilon_{ba}}\to\frac{\Omega-\epsilon_{ba}}{(\Omega-\epsilon_{ba})^2+(1/\tau)^2}.
\end{align}
In the low-frequency regime, we can approximate the formula for the intrinsic Fermi surface effect as
\begin{align}
    \sigma_{\text{IFS}}^{\mu;\nu\lambda} &= -\dfrac{1}{2}\int \dfrac{d\vb*{k}}{(2\pi)^{d}}\sum_{a\neq b}X^{\nu}_{ab}X^{\lambda}_{ba}\partial_{\mu}f_{ab}\frac{\Omega-\epsilon_{ba}}{(\Omega-\epsilon_{ba})^2+(1/\tau)^2},\notag\\ 
    &\simeq -\dfrac{1}{2}\int \dfrac{d\vb*{k}}{(2\pi)^{d}}\sum_{a\neq b}X^{\nu}_{ab}X^{\lambda}_{ba}\partial_{\mu}f_{ab}\frac{1}{\Omega-\epsilon_{ba}}.
\end{align}
Considering this formula, we can find the frequency dependence of the intrinsic Fermi surface effect below
\begin{align}
    \sigma_{\mathrm{IFS}}^{\mu;\nu\lambda}\propto\frac{1}{\Omega-\epsilon_g}.
\end{align}
Therefore, the intrinsic Fermi surface term shows no $\tau$-dependence in the low-frequency regime.

\twocolumngrid

\bibliography{refs}

\end{document}